\documentclass[aps,pra,reprint,superscriptaddress]{revtex4-1}
\usepackage[pdftex]{graphicx}

\usepackage[pdftex,
            colorlinks=true,
            citecolor=blue,
            linkcolor=blue]{hyperref}
\usepackage{braket}
\usepackage{amsmath,amssymb,times,multirow,amsfonts,bm,xspace,ulem,pifont}
\begin{document}
\newcommand{\Eqn}[1]{&\hspace{-0.5em}#1\hspace{-0.5em}&}
\newcommand{\simg}{\stackrel{>}{_\sim}}
\newcommand{\siml}{\stackrel{<}{_\sim}}
\newcommand{\rr}{{\bm r}}
\newcommand{\q}{{\bm q}}
\renewcommand{\k}{{\bm k}}

\newcommand*\wien    {\textsc{wien}2k\xspace}
\newcommand*\wtow    {\textsc{wien2wannier}\xspace}
\newcommand*\wannier {\textsc{wannier90}\xspace}
\newcommand*\JI[1]{\textcolor{red}{#1}}
\newcommand*\JJ[1]{\textcolor{blue}{#1}}
\newcommand*\YY[1]{\textcolor{red}{#1}}
%\preprint{}

\title{Odd-parity multipole fluctuation and unconventional superconductivity in locally noncentrosymmetric crystal}


%\author{Jun Ishizuka$^1$, Takemi Yamada$^2$, Yuki Yanagi$^3$, Yoshiaki \=Ono$^1$}
%\affiliation{$^1$Department of Physics, Niigata University, Ikarashi, Niigata 950-2181, Japan \\
%$^2$ Department of Physics, Faculty of Science and Technology, Tokyo University of Science, Noda 278-8510, Japan \\
%$^3$ Department of Physics, School of Science and Technology, Meiji University, Kawasaki 214-8571, Japan} 

 \author{Jun Ishizuka}
 \email[]{ishizuka.jun.8c@kyoto-u.ac.jp}
 \affiliation{Department of Physics, Graduate School of Science, Kyoto University, Kyoto 606-8502, Japan}

\author{Youichi Yanase}
%\email[]{yanase@scphys.kyoto-u.ac.jp}
\affiliation{Department of Physics, Graduate School of Science, Kyoto University, Kyoto 606-8502, Japan}


\date{\today}

\begin{abstract}
A microscopic calculation and symmetry argument reveal superconductivity in the vicinity of parity-violating magnetic order.
In a crystal structure lacking local space inversion symmetry, an augmented cluster magnetic multipole order may break global inversion symmetry, and therefore it is classified into an odd-parity multipole order.
We investigate unconventional superconductivity induced by an odd-parity magnetic multipole fluctuation in a two-dimensional two-sublattice Hubbard model motivated by Sr$_2$IrO$_4$.
We find that even-parity superconductivity is more significantly suppressed by spin-orbit coupling than that in a globally noncentrosymmetric system.
Consequently, two odd-parity superconducting states are stabilized by magnetic multipole fluctuations in a large spin-orbit coupling region. Both of them are identified as $Z_2$ topological superconducting states. 
The obtained gap function of inter-sublattice pairing shows a gapped or nodal structure protected by nonsymmorphic symmetry.
Our finding implies a family of odd-parity topological superconductors. 
Candidate materials are discussed.
\end{abstract}

%\keywords{}

\maketitle


\section{introduction}
\label{sec:intro}

Recent intensive research has clarified the intriguing effects of spin-orbit coupling in locally noncentrosymmetric (NCS) crystals \cite{Sigrist_LNCS,Maruyama_LNCS,Fischer_noncentro,Nakosai,Yoshida_PDW,Yoshida_CS,Yoshida_TCSC,Yoshida_CeCoIn5_Z8}. 
The locally NCS crystal preserves global inversion symmetry in the crystal structure, although inversion symmetry on a local atomic site is lacking.
Sublattice-dependent antisymmetric spin-orbit coupling (ASOC) appears in locally NCS crystals, and it may induce exotic superconductivity distinct from well-studied globally NCS superconductivity \cite{Bauer_Sigrist,Settai_review,Edelstein_2,Gorkov,Frigeri_CePt3Si,Samokhin,Yanase_CePt3Si_1,Yanase_CePt3Si_2,Yokoyama,Tada,Agterberg_helical,Fujimoto_1,Fujimoto_2,Brydon_YPtBi,Bauer_CePt3Si_1,Sugitani_CeIrSi3}. For instance, singlet-triplet mixing \cite{Fischer_noncentro}, a topological crystalline superconductivity of a pair density wave state \cite{Yoshida_PDW,Yoshida_TCSC,Yoshida_CeCoIn5_Z8}, and an antiferromagnetic (AFM) Edelstein effect \cite{Yanase_zigzag,Zelezny_Mn2Au,Wadley,Watanabe_mpole} have been studied.
Motivated by these works, we investigate exotic superconductivity induced by magnetic fluctuation in locally NCS crystals. 
From the viewpoint of materials, many unconventional superconductors in the vicinity of the AFM critical point, such as iron-based superconductors \cite{Kuroki_1,Kamihara_2} and CeCoIn$_5$ superlattices \cite{Mizukami_CeCoIn5,Maruyama_LNCS}, are classified into locally NCS superconductors. Thus, it is interesting to clarify the effects of local parity violation on AFM-fluctuation-induced  superconductivity from a microscopic point of view. 

Another topic of recent interest in locally NCS crystals is an odd-parity electromagnetic multipole order \cite{Yanase_zigzag,Sumita_zigzag,Watanabe_mpole,Fu_Multipole,Hitomi_Sr3Ru2O7,Hitomi_bilayer,Hayami_Toroidal,Hayami_valley,Hayami_zigzag,Sumita_Ir,Matteo_Sr2IrO4,Hayami_AOsO4,Yanagi_Co4Nb2O9,Hayami_mpole_hyb}, which spontaneously breaks global inversion symmetry through an anisotropic spin and charge distribution.
Although previous studies provided a profound understanding of even-parity multipole order in strongly correlated electron systems \cite{Kuramoto_multipole,Haule_URu2Si2,Kusunose_URu2Si2,Ikeda_URu2Si2}, it has been recently recognized that odd-parity electromagnetic multipole order is ubiquitous in materials. For instance, BaMn$_2$As$_2$ \cite{Singh_BaMn2As2,Watanabe_mpole}, Sr$_2$IrO$_4$ \cite{Zhao_Sr2IrO4,Matteo_Sr2IrO4,Sumita_Ir,B.J.Kim_Sr2IrO4_2008,Y.K.Kim_Sr2IrO4_2014,Yan_Sr2IrO4,Battisti,Y.K.Kim_Sr2IrO4_2016,Meng,B.J.Kim_Sr2IrO4_2009,Boseggia,Clancy,Chetan,Watanabe_Ir,Huang_Sr2IrO4,Crawford}, Cd$_2$ReO$_7$ \cite{Steven_Cd2Re2O7,Harter_Cd2Re2O7,Fu_Multipole,Matteo_Cd2Re2O7,Harima_Cd2Re2O7,Yamaura_Cd2Re2O7_2017,Yamaura_Cd2Re2O7_2002,Hiroi_Cd2Re2O7,Hanawa_Cd2Re2O7,Sakai_Cd2Re2O7} , and SrTiO$_3$ \cite{Bednorz_STO,Schooley,Rischau_STO} have been studied from the viewpoint of odd-parity multipole order.
More recently, more than 110 AFM compounds have been identified as odd-parity magnetic multipole states by a group-theoretical analysis \cite{Watanabe_mpole_2}. For those compounds, a multipole moment in the unit cell (augmented cluster multipole) has an odd parity and leads to parity violation.

Superconductivity near the odd-parity electromagnetic multipole order invokes an unconventional pairing mechanism induced by {\it an odd-parity multipole fluctuation}.
However, theoretical studies  based on microscopic models have not been conducted except for a few works on electric multipole fluctuation \cite{Kozii,Wang_Cd2Re2O7,Edge}.
Because the AFM order in locally NCS crystals with sublattice-dependent ASOC realizes odd-parity magnetic multipole order \cite{Watanabe_mpole_2}, our study of fluctuation-induced superconductivity naturally reveals superconductivity due to the magnetic odd-parity multipole fluctuation. 
%the superconductivity induced by the fluctuation may be closely related to the ASOC. 
The pairing interaction and the resulting superconducting state may be different from those of conventional magnetic-fluctuation-induced superconductivity. Therefore, a different platform of topological superconductivity may be found in this study.

Previous theories based on the random phase approximation (RPA) have investigated the superconductivity induced by AFM fluctuation in globally NCS crystals \cite{Yanase_CePt3Si_1,Yanase_CePt3Si_2,Yokoyama,Takimoto_CePt3Si_2,Tada}.
In this paper, we clarify a peculiar superconducting state and magnetic multipole fluctuation in a locally NCS crystal with the same approximation.
To be specific, we analyze a two-sublattice Hubbard model with sublattice-dependent ASOC.
The crystallographic point group is centrosymmetric $D_{4h}$ and the local site symmetry is $D_{2d}$ lacking inversion symmetry.
This is a minimal model taking account of the locally NCS structure, spin-orbit coupling, and odd-parity magnetic multipole fluctuation.
For instance, BaMn$_2$As$_2$ and Sr$_2$IrO$_4$ are captured by this model from the viewpoint of symmetry.  

The seemingly conventional $G$-type AFM order in our model shows unbroken translation symmetry, because of the two-sublattice structure peculiar to locally NCS crystals. The magnetic propagation vector is indeed $\q=\bm 0$. Instead of the translation symmetry, the space inversion symmetry is broken. Therefore, the AFM order is regarded as an odd-parity magnetic order.  
From the group-theoretical study \cite{Watanabe_mpole}, the magnetoelectric multipole moment has been classified based on the point group $D_{4h}$. The magnetic multipole order in the AFM state with $\bm m\parallel c$ belongs to the $B_{2u}$ irreducible representation (IR).
The candidates of the order parameter are identified as the magnetic quadrupole moment and the hexadecapole moment.
On the other hand, the $G$-type AFM order with $\bm m\perp c$ corresponds to the magnetic quadrupole and toroidal order.
Our calculation takes into account all these magnetic multipole fluctuations.
In this paper, we will perform a microscopic study of unconventional superconductivity induced by odd-parity magnetic fluctuation.

This paper is constructed as follows. In Secs.~\ref{sec:gfg^T} and \ref{sec:gapstruct}, symmetry operations for pair amplitudes with sublattice degrees of freedom are revealed.
We clarify the symmetry properties in the present crystal structure.
Since our model preserves nonsymmorphic crystal symmetry, the pair amplitudes have peculiar structures. 
%which elucidate the microscopic origin of the node/gap structure protected by nonsymmorphic crystalline symmetry \cite{Norman,Micklitz,Yanase_UPt3_Weyl,Kobayashi_TSC_2016,Nomoto_UPt3,Nomoto_UCoGe,Sumita_UPt3}.
Section \ref{sec:model} introduces a two-sublattice Hubbard model with spin-orbit coupling and provides the formulation of the microscopic calculation based on the RPA and Eliashberg equation. Numerical results are shown in Secs.~\ref{sec:chis} and \ref{sec:sc}.
In Sec.~\ref{sec:chis}, we show the odd-parity magnetic fluctuation and its anisotropy. Effects of the ASOC on the magnetic fluctuations are discussed. 
In Sec.~\ref{sec:sc}, we identify four stable pairing states, which are distinguished by symmetry.
Effects of ASOC in locally NCS crystals are compared with those in globally NCS crystals.
It is demonstrated that local parity violation prefers odd-parity superconductivity. Therefore, the $Z_2$-topological odd-parity superconductivity in the DIII class is stabilized in a large ASOC region.  
A brief summary and discussion are given in Sec.~\ref{sec:sum}.

\section{Symmetry of Superconductivity in Multisublattice Systems}
\label{sec:gfg^T}

For the classification of pair amplitudes in multicomponent superconductors, we need to take into account the internal degrees of freedom of electrons, which were neglected in classical theories summarized by Sigrist and Ueda \cite{Sigrist_Ueda}.
For instance, multiorbital systems have been analyzed in Ref.~\onlinecite{Nomoto_MSC}.
We here classify the systems with sublattice degrees of freedom.

To study locally NCS superconductors, it is important to clarify the intersublattice and intrasublattice pairing amplitudes. 
A complete classification is given by introducing the permutation of sites.
We study a single-orbital model for simplicity.
An extension to multiorbital and multisublattice systems is straightforward by considering the permutation of local orbitals.

A creation operator of a Bloch state $c^\dag_{\k ms}$ with spin $s$ on sublattice $m$ is transformed by a space-group operation,
\begin{align}
g c^\dag_{\k ms} g^{-1} & = \sum_{\bm R} g c^\dag_{s} (\bm R+\rr_m) g^{-1} e^{-i\k \cdot \bm R} \nonumber\\
& = \sum_{\bm R s'} c^\dag_{s'} (p(\bm R+\rr_m) + \bm a) D^{(1/2)}_{s's} (p) e^{-i\k \cdot \bm R} \nonumber\\
& = e^{ip\k\bm a} \sum_{s'} c^\dag_{p\k gms'} D^{(1/2)}_{s's} (p) e^{-ip\k (\rr_{gm} - p\rr_m)}.
\end{align}
Here, $\bm R$ is a basic lattice vector and $\rr_m$ is a relative coordinate of $m$th sublattice in a unit cell.
The operation $g=\{p | \bm{a}\}$ is defined by a conventional Seitz space-group symbol with a point-group operation $p$ and a translation $\bm{a}$.
By choosing a representation matrix indicating the permutation of sites as 
\begin{equation}
D^{({\rm perm})}_{m'm} (g;\k)=e^{-ip\k (\rr_{gm} - p\rr_m)}\delta_{m',gm},
\label{eq:c_D^perm}
\end{equation}
the transformation is simply represented as
\begin{align}
g c^\dag_{\k ms} g^{-1} = e^{ip\k\bm a} \sum_{m',s'} c^\dag_{p\k m's'} D^{(1/2)}_{s's} (p) D^{({\rm perm})}_{m'm} (g;\k).
\label{eq:gcg-1}
\end{align}

In a superconducting state, a pair amplitude is defined as 
\begin{equation}
F_{msm's'}(\k)=\langle c_{\k ms}c_{-\k m's'}\rangle,
\label{eq:f}
\end{equation}
where $\langle \cdots \rangle$ denotes the thermal average. The fermion antisymmetry gives
\begin{equation}
F_{msm's'} (\k) = -F_{m's'ms} (-\k).
\label{eq:fermion_antisymmetry}
\end{equation}
From Eqs.~(\ref{eq:gcg-1}) and (\ref{eq:f}), the pair amplitude is translated as 
\begin{align}
 g & F^{\Gamma_i}_{msm's'} (\k) g^{\rm -1} = \hat D (g;\k) \hat F^{\Gamma_i} (p\k) \hat D^{\rm T} (g;\k)\nonumber \\
& = \sum_{\substack{m_1,m_2\\ s_1,s_2}} F^{\Gamma_i}_{m_1s_1m_2s_2} (p\k) \nonumber \\
& \times D^{(1/2)}_{s_1s_2ss'} (p) D^{({\rm perm})}_{m_1m_2mm'} (g;\k) D^{\phi_\k} (g;\Gamma_i).
\label{eq:gfg^T}
\end{align}
Here, the corresponding representation matrix is
\begin{align}
D^{({\rm perm})}_{m_1m_2mm'} (g;\k)& =e^{ - ip\k (\rr_{gm} - p\rr_m)}e^{ ip\k (\rr_{gm'} - p\rr_{m'})} \nonumber \\
& \times\delta_{m_1,gm}\delta_{m_2,gm'},
\label{eq:f_D^perm}
\end{align}
and $D^{\phi_\k} (g;\Gamma_i)$ is the representation matrix of the $\Gamma_i$ IR of the gap function, whose characters are explicitly given in Table~\ref{tb:character_D_4h} for the $D_{4h}$ point group. 
Note that $D^{({\rm perm})}$ is unity for $m=m'$.

When the total Hamiltonian commutes with the space inversion operator $I$,
the pair amplitude possesses an even parity (odd parity), namely,
$I F^{\Gamma_{g(u)}}_{msm's'} (\k) I^{ -1} = + (-) F^{\Gamma_{g(u)}}_{ImsIm's'} (-\k)$.
When the local symmetry on each sublattice has inversion symmetry, $Im=m$, the spin-singlet and spin-triplet pairing states are distinguished by the intrasublattice pair amplitude $I F^{\Gamma_{g(u)}}_{msms'} (\k) I^{ -1} = - (+) F^{\Gamma_{g(u)}}_{ms'ms} (\k)$.
In the absence of global inversion symmetry, the space inversion parity is not a good quantum number, and parity mixing between the singlet and triplet channels occurs.
On the other hand, in locally NCS superconductors the parity mixing appears in a different way. Consequently, the superconducting gap function shows a nontrivial and symmetry-protected structure.
We focus on this case in the next section.
Hereafter, we assume even-frequency pairing, which is thermodynamically stable \cite{Heid} in the usual cases.

\section{Pair amplitude in Locally Noncentrosymmetric Crystal}
\label{sec:gapstruct}
\begin{figure}[ht]
\begin{center}
\includegraphics[width=82mm]{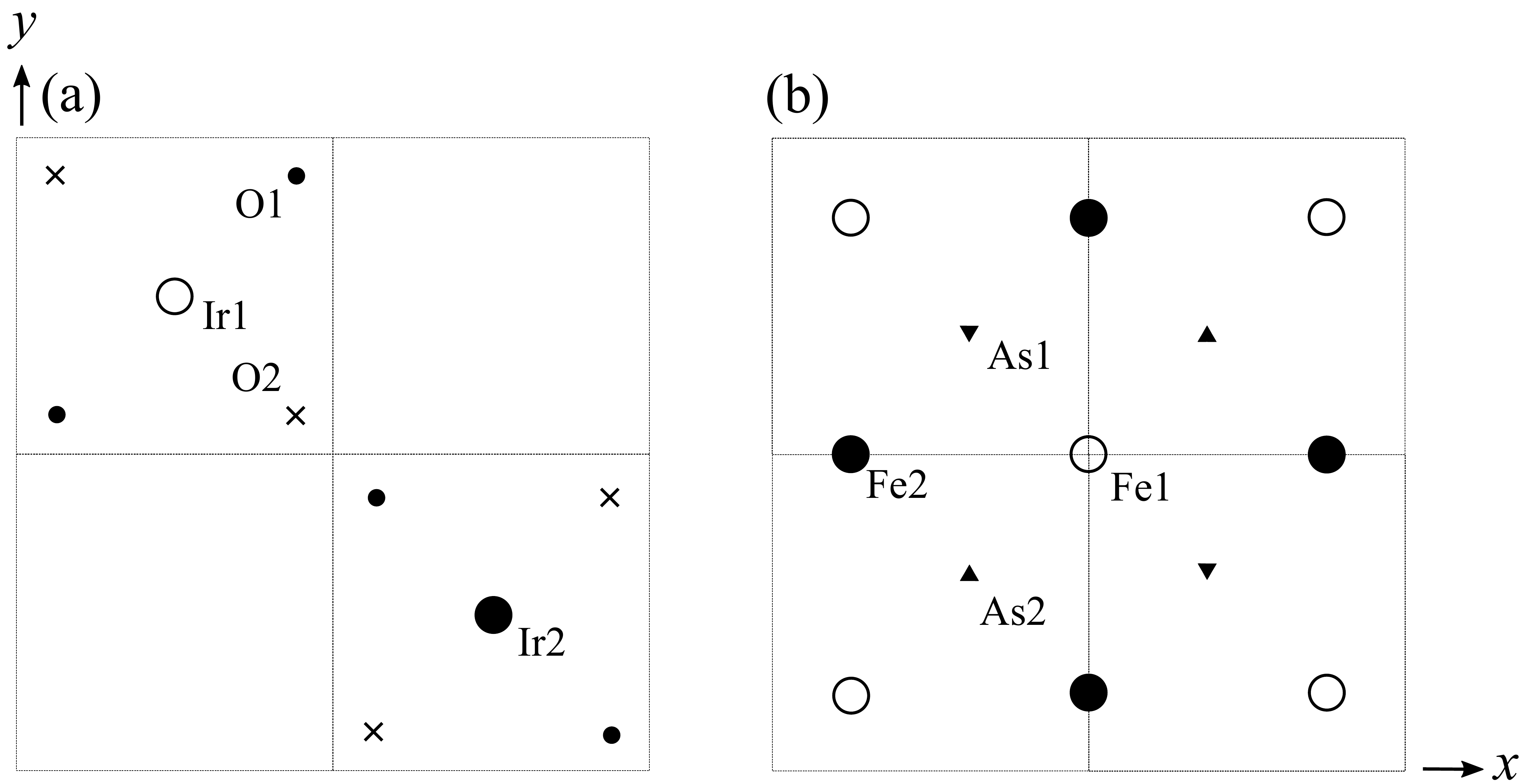}	
\caption{
Crystal structure of (a) Sr$_2$IrO$_4$ and (b) iron-based superconductors.
Two-dimensional Ir and Fe layers with ligands, which break $\sigma_v$ mirror symmetry, are plotted.
A $D_{2d}$-type ASOC appears in these structures.
\label{fig:struct}}
\end{center}
\end{figure}

For a demonstration, we introduce a typical crystal structure lacking local inversion symmetry and examine the symmetry properties of the pair amplitudes. 
The example considered throughout this paper is a tetragonal crystal lattice with two sublattices, each of which lacks local inversion symmetry.
In a crystal structure of Sr$_2$IrO$_4$, which is depicted in Fig.~\ref{fig:struct}(a), oxygen ions out of the Ir-O layer violate inversion symmetry at an Ir ion.
A similar local parity violation appears in iron-based superconductors [Fig.~\ref{fig:struct}(b)], whose zigzag structure of the pnictogen or chalcogen ions also breaks local inversion symmetry, owing to the absence of $\sigma_v$ mirror symmetry.
Hereafter, we study the crystal structure in Fig.~\ref{fig:struct}(a).
A coset decomposition is given by
\begin{align}
 G &= \{E | \bm{0}\} T + \{I | \bm{0}\} T + \{2_z | \bm{\tau}_x + \bm{\tau}_y\} T \nonumber \\
 & +  \{2_x | \bm{\tau}_y\} T +  \{2_y | \bm{\tau}_x\} T +  \{4^+_z | \bm{\tau}_x\} T +  \{4^-_z | \bm{\tau}_y\} T \nonumber \\
 & + \{\sigma_x | \bm{\tau}_y\} T + \{\sigma_y | \bm{\tau}_x\} T + \{\sigma_z | \bm{\tau}_x + \bm{\tau}_y\} T \nonumber \\
& + \{\sigma_{110} | \bm{\tau}_x + \bm{\tau}_y\} T + \{\sigma_{1-10} | \bm{0}\} T \nonumber \\
& + \{2_{110} | \bm{0}\} T + \{2_{1-10} | \bm{\bm{\tau}_x + \bm{\tau}_y}\} T \nonumber \\
& + \{I4^+_{z} | \bm{\tau_x}\} T + \{I4^-_{z} | \bm{\tau_y}\} T, \label{eq:SG_2site}
\end{align}
where the translation group $T$ defines a Bravais Lattice, and $\bm{\tau}_x = \frac{a}{2} \bm{e}_a$, $\bm{\tau}_y = \frac{a}{2} \bm{e}_b$ are non-primitive translation vectors.

Let us consider the glide reflection $G_{y}=\{\sigma_y | \bm{\tau}_x\}$. A Bloch state $c^{\dag}_{\k ms}$ is transformed as
$(k_x,k_y,k_z) \rightarrow (k_x,-k_y,k_z)$, $(s_x,s_y,s_z) \rightarrow (-s_x,s_y,-s_z)$, and sublattice indices $(a,b) \rightarrow (b,a)$,
and then the representation matrices are given by $D^{(1/2)}_{s's} (\sigma_y) = i \sigma_y^{\rm T} = {\scriptsize \begin{pmatrix}0 & -1 \\ 1 & 0\end{pmatrix}}$ 
and $D^{({\rm perm})}_{m'm} (\{\sigma_y | \bm{\tau}_x\}; \k) = {\scriptsize \begin{pmatrix}0 & e^{-i(-k_x/2-k_y)} \\ e^{-i(k_x/2-k_y)} & 0\end{pmatrix}}$.
Therefore, $G_{y}$ gives a relative phase factor $e^{ik_x}$ between the two sublattices. Intersublattice hybridization is forbidden at the zone face $k_{x}=\pm\pi$ owing to this phase factor.
To prove it, we need to consider combined inversion-glide symmetry, which commutes with the Hamiltonian $[H, G_yI]=0$. This imposes a constraint for intersublattice hybridization
\begin{align}
\varepsilon_1(\k) & = \langle \k as | H | \k bs \rangle \nonumber \\
& = \langle \k as | (G_yI)^{-1} G_yI H (G_yI)^{-1} G_yI | \k bs \rangle \nonumber \\
& = \left(\langle (-k_x, k_y) as | e^{i(k_x/2-k_y)}\right)  H \nonumber \\
&\times \left( e^{-i(-k_x/2-k_y)} | (-k_x, k_y) bs \rangle \right) \nonumber \\
& = e^{ik_x} \langle (-k_x, k_y) as |  H | (-k_x, k_y) bs \rangle \nonumber \\
& = e^{ik_x} \varepsilon_1(-k_x, k_y).
\end{align}
Then, the phase factor is $e^{ik_x}=-1$ on the zone face $k_x=\pm\pi$, and therefore the intersublattice hybridization must be zero.

The space group $G$ in Eq.~(\ref{eq:SG_2site}) can be reduced to a subgroup $G_{\rm intra}$ by restricting to sublattice-conserving operations (see Table~\ref{tb:permutation_2site}) 
\begin{align}
 G_{\rm intra} &= \{E | \bm{0}\} T + \{2_z | \bm{\tau}_x + \bm{\tau}_y\} T \nonumber \\
& +  \{2_x | \bm{\tau}_y\} T +  \{2_y | \bm{\tau}_x\} T \nonumber \\
 & + \{\sigma_{110} | \bm{\tau}_x + \bm{\tau}_y\} T + \{\sigma_{1-10} | \bm{0}\} T \nonumber \\
& + \{I4^+_{z} | \bm{\tau_x}\} T + \{I4^-_{z} | \bm{\tau_y}\} T. \label{eq:SG_intra_2site}
\end{align}
From $G_{\rm intra}$, we notice that the local site symmetry is $D_{2d}$.
Thus, parity mixing in intrasublattice pair amplitudes is allowed by the symmetry reduction $D_{4h}\rightarrow D_{2d}$, which is determined by the compatibility relation  shown in Table~\ref{tb:D4h_D2d}.
Note that this simple rule is only applicable to the intrasublattice components.

\begin{table}[t]
\caption{List of characters for the IRs of the $D_{4h}$ point group. The last column shows basis functions. }
\label{tb:character_D_4h}
\footnotesize
\renewcommand{\arraystretch}{1.3}
%\scalebox{0.92}[0.92]{
\begin{tabular*}{1.0\columnwidth}{@{\extracolsep{\fill}}lrrrrrrrrrrc}
%\begin{tabular}{lrrrrrrrrrrc}
\hline\hline 
 &\\[-10pt] & $E$ & $2C_4$ & $C_2$ & $2 C_2'$ & $2 C_2''$ & $I$ & $2 S_4$ & $\sigma_h$ & $2 \sigma_v$ & $2 \sigma_d$ & Basis functions\\[-10pt] & \\ 
\hline 
$A_{1g}$ & $1$ & $1$ & $1$ & $1$ & $1$ & $1$ & $1$ & $1$ & $1$ & $1$ & $k_z^2$ \\
$A_{2g}$ & $1$ & $1$ & $1$ & $-1$ &$-1$ & $1$ & $1$ & $1$ &$-1$ &$-1$ & $  k_xk_y (k_x^2 - k_y^2)$ \\
$B_{1g}$ & $1$ &$-1$ & $1$ & $1$ &$-1$ & $1$ &$-1$ & $1$ & $1$ &$-1$ & $k_x^2-k_y^2$\\
$B_{2g}$ & $1$ &$-1$ & $1$ &$-1$ & $1$ & $1$ &$-1$ & $1$ &$-1$ & $1$ & $k_xk_y$\\
$E_{g}$ & 2 & 0 & $-2$ & 0 & 0 & 2 & 0 &$-2$ & 0 & 0 & $k_zk_x, k_yk_z$\\
$A_{1u}$ & $1$ & $1$ & $1$ & $1$ & $1$ &$-1$ &$-1$ &$-1$ &$-1$ &$-1$ & $k_x \hat {\bm x} + k_y \hat{\bm y}$\\
$A_{2u}$ & $1$ & $1$ & $1$ &$-1$ &$-1$ &$-1$ &$-1$ &$-1$ & $1$ & $1$ & $k_y \hat{\bm x} - k_x \hat{\bm y}$ \\
$B_{1u}$ & $1$ &$-1$ & $1$ & $1$ &$-1$ &$-1$ & $1$ &$-1$ &$-1$ & $1$ & $k_x \hat{\bm x} -k_y \hat{\bm y}$ \\
$B_{2u}$ & $1$ &$-1$ & $1$ &$-1$ & $1$ &$-1$ & $1$ &$-1$ & $1$ &$-1$ & $k_y \hat{\bm x} + k_x \hat{\bm y}$\\
$E_{u}$ & 2 & 0 & $-2$ & 0 & 0 &$-2$ & 0 & 2 & 0 & 0 & $ k_x \hat{\bm z}, k_y \hat{\bm z}$ \\
\hline\hline 
%\end{tabular}
\end{tabular*}
%}
\end{table}

\begin{table}[htbp]
\caption{List of permutation for each symmetry operation. The sublattice-conserving operations are labeled by a check mark (\ding{51}).}
% in the last column.}
\label{tb:permutation_2site}
\begin{tabular*}{1.0\columnwidth}{@{\extracolsep{\fill}}lccc}
%\begin{tabular}{cccc}
\hline\hline 
 &\\[-10pt] & $a$ & $b$ & Sublattice conservation \\[-10pt] &\\ \hline 
$\{E | \bm{0}\}$ & $a$ & $b$  & \ding{51} \\
$\{I | \bm{0}\}$ & $b$ & $a$ & \ding{55} \\
$\{2_z | \bm{\tau}_x + \bm{\tau}_y\}$ & $a$ & $b$ & \ding{51} \\
$\{2_x | \bm{\tau}_y\}$ & $a$ & $b$ & \ding{51} \\
$\{2_y | \bm{\tau}_x\}$ & $a$ & $b$ & \ding{51} \\
$\{4^+_z | \bm{\tau}_x\}$ & $b$ & $a$ & \ding{55} \\
$\{4^-_z | \bm{\tau}_y\}$ & $b$ & $a$ & \ding{55} \\
$\{\sigma_x | \bm{\tau}_y\}$ & $b$ & $a$ & \ding{55} \\
$\{\sigma_y | \bm{\tau}_x\}$ & $b$ & $a$ & \ding{55} \\
$\{\sigma_z | \bm{\tau}_x + \bm{\tau}_y\}$ & $b$ & $a$ & \ding{55} \\
$\{\sigma_{110} | \bm{\tau}_x + \bm{\tau}_y\}$ & $a$ & $b$ & \ding{51} \\
$\{\sigma_{1-10} | \bm{0}\}$ & $a$ & $b$ & \ding{51} \\
$\{2_{110} | \bm{0}\}$ & $b$ & $a$ & \ding{55} \\
$\{2_{1-10} | \bm{\tau}_x + \bm{\tau}_y\}$ & $b$ & $a$ & \ding{55} \\
$\{I4^+_{z} | \bm{\tau_x}\}$ & $a$ & $b$ & \ding{51} \\
$\{I4^-_{z} | \bm{\tau_y}\}$ & $a$ & $b$ & \ding{51} \\
\hline\hline 
%\end{tabular}
\end{tabular*}
\end{table}

\begin{table}[htbp]
 \caption{Reduction of IRs $ D_{4h} \rightarrow D_{2d}$. The two fold rotational symmetry axes of $D_{2d}$ are the $x$/$y$ axes of $D_{4h}$. }
\label{tb:D4h_D2d}
\begin{tabular*}{1.0\columnwidth}{@{\extracolsep{\fill}}lcccccccccc}
\hline\hline 
  &\\[-10pt] $D_{4h}$&$  A_{1g}  $&$  A_{2g}  $&$  B_{1g}  $&$ B _{2g}  $&$ E _{g}  $&$  A_{1u}  $&$  A_{2u}  $&$   B_{1u} 	  $&$ B _{2u}  $&$ E _{u}$  \\[-10pt] & \\ \hline
 $D_{4h} \downarrow D_{2d}$&$  A_{1}  $&$  A_{2}  $&$  B_{1}  $&$ B _{2}  $&$ E   $&$  B_{1}  $&$  B_{2}  $&$   A_{1}   $&$ A _{2}  $&$ E   $ \\ \hline\hline
\end{tabular*}
 \end{table}

For instance, the $B_{2g}$ IR mixes with the $A_{2u}$ IR because they are reduced to the same $B_2$ IR in $D_{2d}$.
The admixture is, however, different from that in the globally NCS superconductors: (1) One of the admixed components has a staggered form between sublattices for an intrasublattice pairing, and (2) parity mixing in intersublattice components is forbidden, to preserve the global inversion symmetry.
These properties of the pair amplitude in locally NCS superconductors are derived from Eq.~(\ref{eq:gfg^T});
the inversion symmetry represented by $D^{(1/2)}_{s_1s_2ss'} (I) = \hat1_{4\times4}$ and $D^{({\rm perm})}_{m_1m_2mm'} (\{I|\bm 0\})={\scriptsize \begin{pmatrix}0 & \tau_x \\ \tau_x & 0\end{pmatrix}}$ imposes a constraint for intra- and inter-sublattice pair amplitudes
\begin{align}
I F^{\Gamma_{g(u)}}_{asas'} (\k) I^{ -1} & = +(-) F^{\Gamma_{g(u)}}_{bsbs'} (-\k) = -(+) F^{\Gamma_{g(u)}}_{bs'bs} (\k), \label{eq:IFI^T_aa} \\
I F^{\Gamma_{g(u)}}_{asbs'} (\k) I^{ -1} & = +(-) F^{\Gamma_{g(u)}}_{bsas'} (-\k) = -(+) F^{\Gamma_{g(u)}}_{as'bs} (\k), \label{eq:IFI^T_ab}
\end{align}
for even-parity (odd-parity) superconductivity.

For example, let us consider the even-parity superconductivity, for which a part of Eqs. (\ref{eq:IFI^T_aa}) and (\ref{eq:IFI^T_ab}) are explicitly described as
\begin{align}
I F^{\Gamma_{g}}_{a\uparrow a\uparrow} (\k) I^{ -1} & = F^{\Gamma_{g}}_{b\uparrow b\uparrow} (-\k) = - F^{\Gamma_{g}}_{b\uparrow b\uparrow} (\k), \label{eq:IFI^T_aupaup} \\
I F^{\Gamma_{g}}_{a\uparrow a\downarrow} (\k) I^{ -1} & = F^{\Gamma_{g}}_{b\uparrow b\downarrow} (-\k) = - F^{\Gamma_{g}}_{b\downarrow b\uparrow} (\k), \label{eq:IFI^T_adnadn} \\
I F^{\Gamma_{g}}_{a\uparrow b\uparrow} (\k) I^{ -1} & = F^{\Gamma_{g}}_{b\uparrow a\uparrow} (-\k) = - F^{\Gamma_{g}}_{a\uparrow b\uparrow} (\k),
\label{eq:IFI^T_aupbup} \\
I F^{\Gamma_{g}}_{a\uparrow b\downarrow} (\k) I^{ -1} & = F^{\Gamma_{g}}_{b\uparrow a\downarrow} (-\k) = - F^{\Gamma_{g}}_{a\downarrow b\uparrow} (\k). \label{eq:IFI^T_aupbdn}
\end{align}
Equation (\ref{eq:IFI^T_aupaup}) shows a symmetry property of the $S_z=1$ intrasublattice Cooper pair, namely, parity-mixed spin-triplet pairing.
This equation means that the exchange of sublattice indices $(a\rightarrow b)$ gives a sign change of pair amplitude $F^{\Gamma_g}_{a\uparrow a\uparrow} (\k) = - F^{\Gamma_g}_{b\uparrow b\uparrow} (\k)$.
Thus, the parity-mixed intrasublattice component has a staggered form.
Equation (\ref{eq:IFI^T_aupbup}) indicates a relation of the parity-mixed intersublattice pairing $F^{\Gamma_g}_{a\uparrow b\uparrow} (\k) = - F^{\Gamma_g}_{a\uparrow b\uparrow} (\k) = 0$.
We immediately find that the parity-mixed intersublattice component must be zero.

Recent progress on the group-theoretical analysis of superconductivity has shown unusual nodal/gapped structures ensured by nonsymmorphic symmetry \cite{Yanase_UPt3_Weyl,Nomoto_UPt3,Nomoto_UCoGe,Norman,Micklitz,Kobayashi_TSC_2016,Sumita_UPt3,Kobayashi_TSC_2018}. 
Similarly, we show peculiar structures of the pair amplitude. The space-group symmetry of our model (\ref{eq:SG_2site}) is nonsymmorphic since it contains the glide symmetry.  

Let us consider the glide symmetry $G_{x}=\{\sigma_x | \bm{\tau}_y\}$ represented by
\begin{align}
D^{(1/2)}_{s_1s_2ss'} (\sigma_x) & = \begin{pmatrix} & -\sigma_x \\ -\sigma_x & \end{pmatrix},\\
D^{({\rm perm})}_{m_1m_2mm'} (\{\sigma_x | \bm{\tau}_y\}) & = \begin{pmatrix} & & & 1 \\ & & e^{ik_y} & \\  & e^{-ik_y} & & \\ 1 & & & \end{pmatrix}.
\end{align}
% and imposes a constraint,
%\begin{align}
%G_{x} F^{\Gamma_i}_{a\uparrow b\downarrow} (\k) G_{x}^{ -1} & = - D^{\phi_\k} (\sigma_x;\Gamma_i) F^{\Gamma_i}_{b\downarrow a\uparrow} (-k_x,k_y,k_z) e^{-ik_y} \nonumber \\
%& = D^{\phi_\k} (\sigma_x;\Gamma_i) F^{\Gamma_i}_{a\uparrow b\downarrow} (k_x,-k_y,-k_z) e^{-ik_y}.
%\label{eq:GxFGx^T_antisym} 
%\end{align}
We especially focus on intersublattice Cooper pairs on the zone face $k_y=\pm\pi$ at $k_z=0$. On this high-symmetry line, $G_x$ imposes a constraint, \begin{align}
G_{x} F^{\Gamma_i}_{a\uparrow b\downarrow} (k_x,\pi,0) G_{x}^{ -1} & = e^{-i\pi} D^{\phi_\k} (\sigma_x;\Gamma_i) F^{\Gamma_i}_{a\uparrow b\downarrow} (k_x,-\pi,0).
\label{eq:GxFGx^T_ky_pi} 
\end{align}
Therefore, it is indicated that intersublattice gap functions must be zero for glide-even superconducting states.
For instance, we find that the intersublattice pair amplitudes in the $A_{1g}$ IR ($s$-wave state) and $B_{1g}$ IR ($d_{x^2-y^2}$-wave state) have nodal lines at $k_z=0$ and $k_{x,y}=\pm\pi$. On the other hand, the pair amplitude is finite for glide-odd IRs such as $B_{2g}$ IR ($d_{xy}$-wave state). These features are opposite to those expected from a group-theoretical analysis of symmorphic superconductors \cite{Sigrist_Ueda}.

 In the following sections, we study superconductivity in a model preserving the space group symmetry (\ref{eq:SG_2site}). The gap functions obtained from numerical calculations satisfy the symmetry constraints, which have been revealed in this section.

\section{Model and Method}
\label{sec:model}
In this section, we introduce a two-dimensional two-sublattice Hubbard model which was adopted for Sr$_2$IrO$_4$ \cite{Sumita_Ir}. 
We consider a two-dimensional IrO$_2$ plane of quasi-two-dimensional Sr$_2$IrO$_4$.
The crystal structure has been illustrated in Fig.~\ref{fig:struct}(a).
We do not restrict our discussions to Sr$_2$IrO$_4$ and later propose some other candidate materials. 
However, it is significant to study a well-studied model for Sr$_2$IrO$_4$ as a typical example and to illustrate the effects of spin-orbit coupling and multipole fluctuations.

The total Hamiltonian is written as
$H=H_0+H_{\rm int}+H_{\rm ASOC}$.
The Hamiltonian of kinetic energy terms is 
\begin{align}
H_{0} & = \sum_{\k s} \sum_{m\neq m'}(\varepsilon_1(\k) c^{\dag}_{\k ms} c_{\k m's} + {\rm H.c.}) \nonumber \\
& + \sum_{\k ms} \varepsilon_2(\k) c^{\dag}_{\k ms} c_{\k ms},
\label{eq:H_0}
\end{align}
where $c^{(\dag)}_{\k ms}$ is the annihilation (creation) operator of an Ir-5$d$ electron with pseudospin $s$ on sublattice $m=(a,b)$.
The pseudospin corresponds to the $j_{\rm eff}=1/2$ doublet states formed by a strong spin-orbit coupling \cite{B.J.Kim_Sr2IrO4_2008}.
The single-electron kinetic energy is described by taking into account the nearest- and next-nearest-neighbor hoppings, 
\begin{align}
 \varepsilon_1(\k) &= - t_1 (1 + e^{ik_x}) (1 + e^{-ik_y}), \\
 \varepsilon_2(\k) &= - 2 t_2 (\cos k_x + \cos k_y).
\label{eq:epsilon_k}
\end{align}
The on-site Coulomb interaction on an Ir site is given by
\begin{equation}
H_{\rm int}= U \sum_{im} n_{im\uparrow} n_{im\downarrow}.
\label{eq:H_int}
\end{equation}
%
%---------------------------------------------
%The anti-symmetric spin-orbit coupling term is written as
%\begin{equation}
%H_{\rm ASOC}= \sum_{\k,m,m',s,s'} \alpha_m \bm g_m(\k) \cdot \bm\sigma_{ss'} c^{\dag}_{\k ms} c_{\k m's'}.
%\label{eq:H_ASOC}
%\end{equation}
%\begin{equation}
%H_{\rm ASOC}= H_{\rm ASOC1} (\k) + H_{\rm ASOC2} (\k).
%\label{eq:H_ASOC}
%\end{equation}
%The anti-symmetric spin-orbit coupling appears due to the locally inversion symmetry breaking at Ir site, which is given by
%\begin{align}
% \hat{H}_{\text{ASOC1}} (\k) &= i \alpha_1 \cos\frac{k_x}{2} \cos\frac{k_y}{2} \bm 0times i \hat{\sigma}_y^{\text{(sl)}} \bm 0times \hat{\sigma}_z^{\text{(spin)}}, \\
% \hat{H}_{\text{ASOC2}} (\k) &= \alpha_2 \bm 0times \hat{\sigma}_z^{\text{(sl)}} \bm 0times ( \sin k_x \cos k_y \hat{\sigma}_x^{\text{(spin)}} - \sin k_y \cos k_x \hat{\sigma}_y^{\text{(spin)}} ).
%\label{eq:H_ASOC_k}
%\end{align}
%---------------------------------------------
The ASOC term is written as
\begin{equation}
H_{\rm ASOC} = \alpha \sum_{\substack{\k ss' \\ mm'}} \bm g (\k) \cdot \bm\sigma_{ss'} c^{\dag}_{\k ms} c_{\k m's'} \zeta^z_{mm'},
\label{eq:H_ASOC}
\end{equation}
where $\zeta^\mu$ is a Pauli matrix for sublattice degrees of freedom.
We consider the staggered ASOC arising from the spin-dependent intrasublattice hopping, 
\begin{equation}
\bm g (\k) = \sin k_x \cos k_y \hat x - \sin k_y \cos k_x \hat y.
\label{eq:g_vector}
\end{equation}
Superconductivity in this model is investigated by solving the  linearized Eliashberg equation, 
 \begin{align}
 \lambda \Delta_{\xi\xi'}(k)&=-\frac{T}{N}\sum_{k'}
 \sum_{\xi_1\xi_2\xi_3\xi_4}V_{\xi \xi_1,\xi_2\xi'}(k-k') \nonumber \\
 &\times G_{\xi_3\xi_1}(-k')\Delta_{\xi_3\xi_4}(k') G_{\xi_4\xi_2}(k'),
\label{eq:gap}
 \end{align}
% \begin{align}
% \lambda \Delta_{ll'}(k)&=-\frac{T}{N}\sum_{k'}
% \sum_{l_1l_2l_3l_4}V_{ll_1,l_2l'}(k-k') \nonumber \\
% &\times G_{l_3l_1}(-k')\Delta_{l_3l_4}(k') G_{l_4l_2}(k'),
%\label{eq:gap}
% \end{align}
where $\hat G (k) = [(i\varepsilon_m-\mu)\hat 1 - \hat H (\k)]^{-1}$ and $i \varepsilon_m = i(2m+1)\pi T$ is the fermionic Matsubara frequency.
Here, we use abbreviated notations $k=(\k, i \varepsilon_m)$ and $\xi=(m,s)$.
In the RPA, effective pairing interaction is described by the generalized susceptibility in the $8\times8$ matrix,
\begin{equation}
\hat V(q) = - \hat\Gamma^0 \hat\chi(q) \hat\Gamma^0 - \hat\Gamma^0.
 \label{eq:veff}
\end{equation}
In the two-sublattice single-orbital model, the bare irreducible vertex is obtained as
\begin{align}
\Gamma^0_{ms_1ms_2,m's_3m's_4} &= \frac{1}{2} \Gamma^s_{mm'}\bm \sigma_{s_1s_2} \cdot \bm \sigma_{s_4s_3} \nonumber \\
& - \frac{1}{2} \Gamma^c_{mm'} \delta_{s_1s_2} \delta_{s_4s_3}, \label{eq:vertex}
\end{align}
and $\hat \Gamma^{s(c)}_{mm'} = U \delta_{mm'} $.
The RPA susceptibility is given by
\begin{equation}
\hat \chi (q)  = \hat \chi^0 (q) \left[\hat 1 - \hat \Gamma^0 \hat \chi^0 (q) \right]^{-1},
\label{eq:chi}
\end{equation}
where the irreducible susceptibility is defined as $\hat \chi^0(q) = - (T/N) \sum_{k} \hat G (k+q) \hat G (k)$.
Now we introduce the magnetic susceptibilities
\begin{equation}
\chi^{\mu\nu}_{mm'} (q)  = \sum_{s_1s_2s_3s_4}\sigma^{\mu}_{s_1s_2} \chi_{ms_1ms_2,m's_3m's_4} (q) \sigma^{\nu}_{s_4s_3},
\label{eq:chi_m}
\end{equation}
where $\mu,\nu=x,y,z$.
The magnetic fluctuation parallel (perpendicular) to the $c$ axis 
$\chi^{\parallel}_{mm'} (q)$ [$\chi^{\perp}_{mm'} (q)$]
 is characterized by $\chi^{\parallel}_{mm'} (q) \equiv \chi^{zz}_{mm'}(q)$ [$\chi^{\perp}_{mm'} (q) \equiv \left(\chi^{xx}_{mm'} (q) + \chi^{yy}_{mm'} (q)\right)/2$].

We define the band filling $n$ as the number of electrons per unit cell (e.g., $n=4$ for full filling).
The doping level $x$ is related to the band filling as $n=2+2x$.
A variational Monte Carlo study \cite{Watanabe_Ir} for Sr$_2$IrO$_4$ shows that the $d$-wave superconducting state is stable near the doping level $x=0.2$.
Thus, we study $x=0.2$ unless mentioned otherwise. We also discuss the result in the undoped case $x=0$. We set $(t_1,t_2)=(1.0,0.26)$, $T=0.02$, $64\times 64$ ${\bm k}$-point meshes, and 1024 Matsubara frequencies in the numerical calculations.

\begin{figure*}[ht]
\begin{center}
\includegraphics[width=145mm]{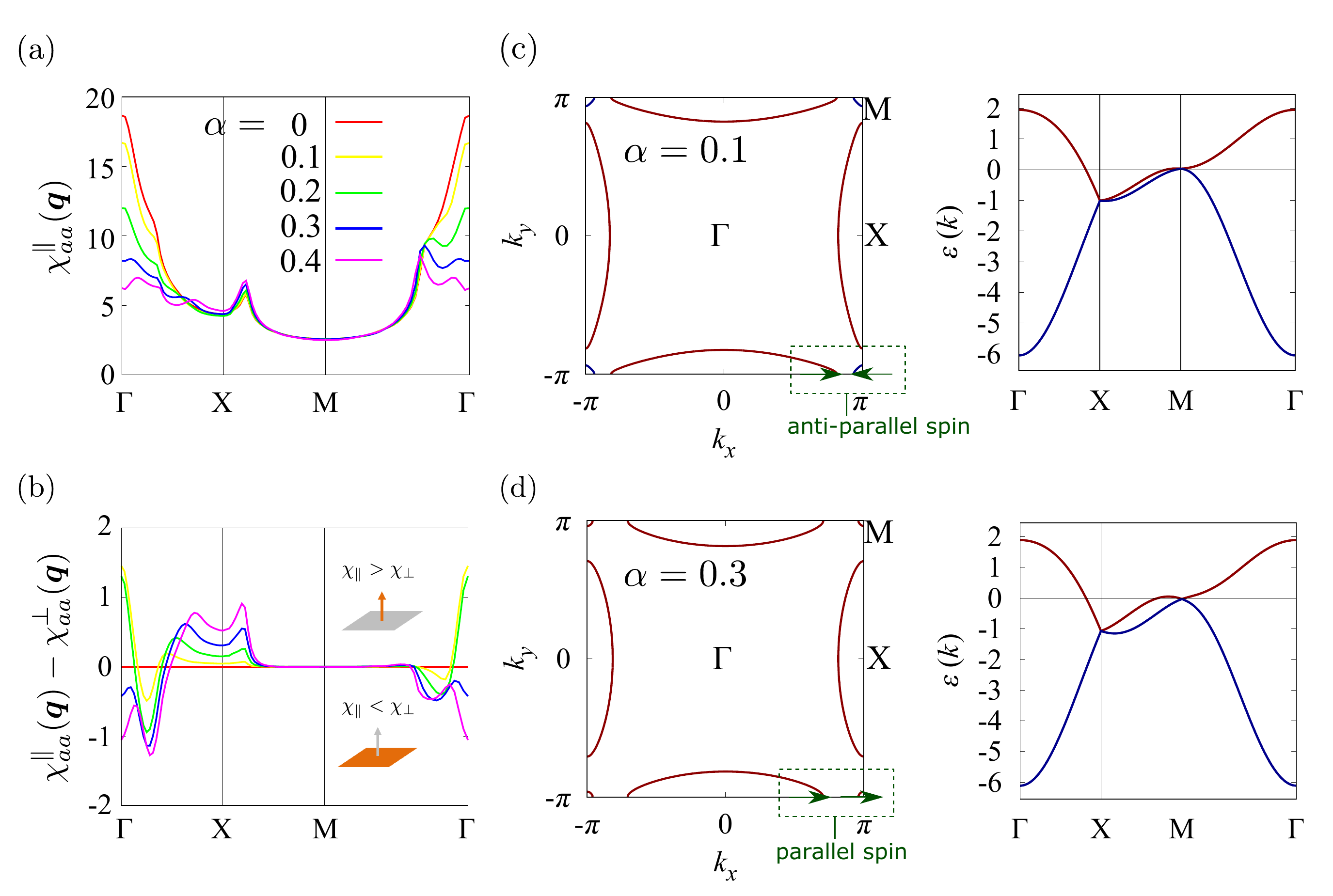}
\caption{
(a), (b) Momentum dependence of the magnetic susceptibility. A matrix element of the static susceptibility $\chi^{\parallel,\perp}_{aa(=bb)}({\bm q},i \omega_n=0)$ on the symmetry axes is shown. (a) shows $\chi^{\parallel}$, while (b) shows the anisotropy $\chi^{\parallel} - \chi^{\perp}$ for $U=1.8$, $x=0.2$, and $T=0.02$.
(c)-(d) Momentum dependence of the band dispersion (right panels) and Fermi surfaces (left panels) for $\alpha=0.1$ and $0.3$. 
The arrows show the spin texture on a sublattice.
\label{fig:band_FS_chi}}
\end{center}
\end{figure*}

\begin{figure}[ht]
\begin{center}
\includegraphics[width=82mm]{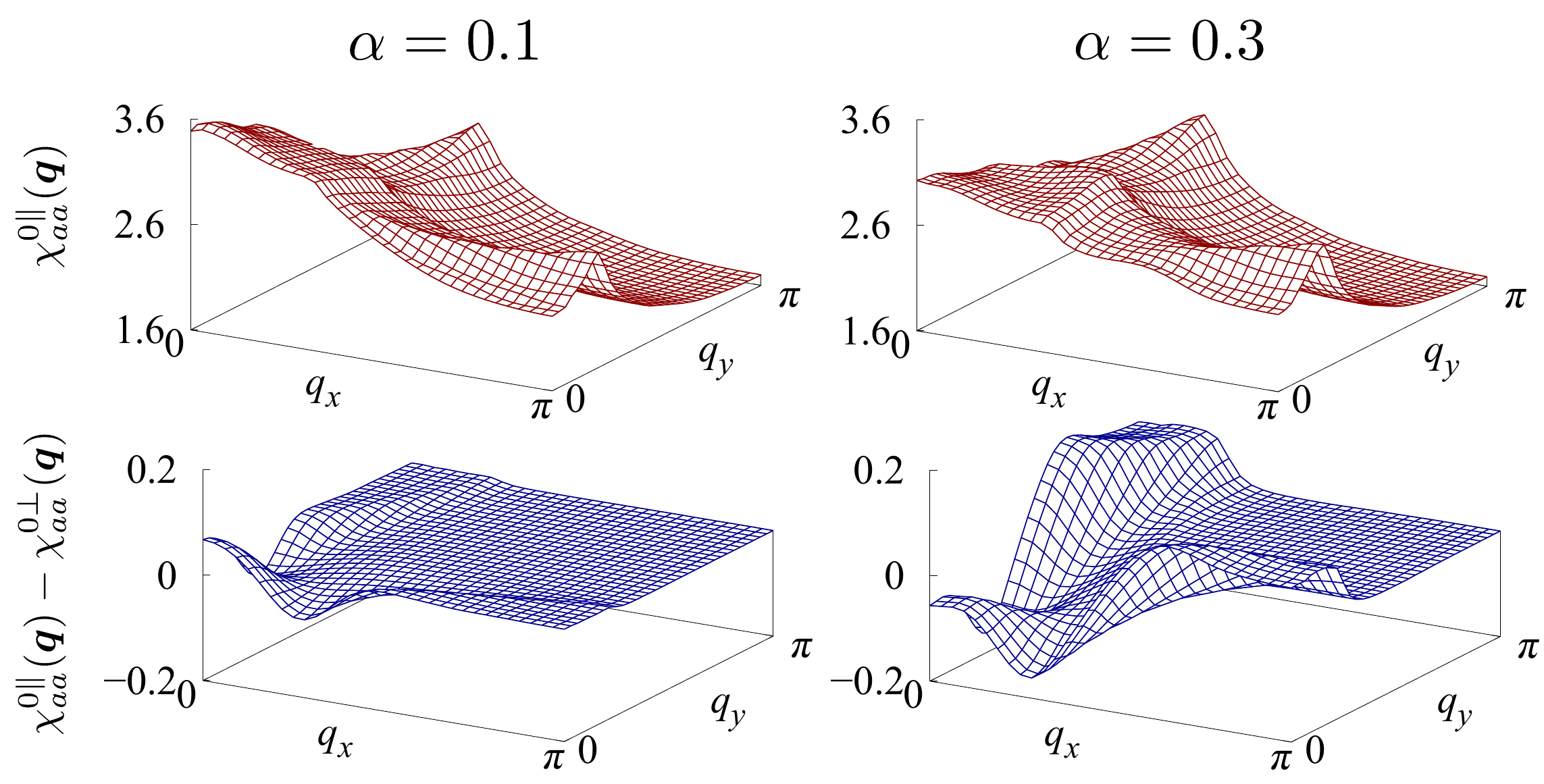}	
\caption{
Momentum dependence of the magnetic susceptibility for $U=0$ (irreducible bubble susceptibility).
\label{fig:chi0}}
\end{center}
\end{figure}

\begin{figure*}[ht]
\begin{center}
\includegraphics[width=165mm]{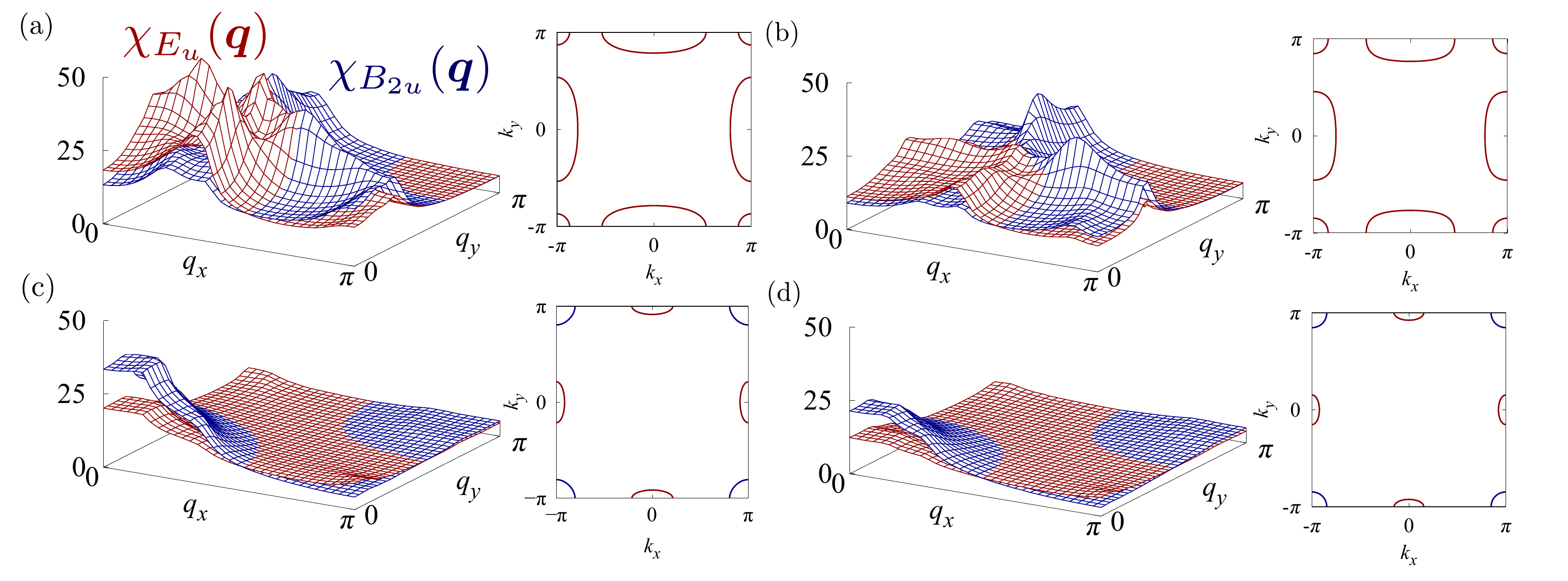}	
\caption{
Momentum dependence of the multipole susceptibilities $\chi_{Eu}(\bm q)$ and $\chi_{B2u}(\bm q)$ for $U=2.2$. (a) $\alpha=0.7$ and $x=0.2$, (b) $\alpha=1$ and $x=0.2$, (c) $\alpha=0.7$ and $x=0$, and (d) $\alpha=1$ and $x=0$. The right panels show the Fermi surfaces.
\label{fig:chi}}
\end{center}
\end{figure*}

\section{Magnetic Fluctuation}
\label{sec:chis}
First, we study the magnetic fluctuation.
When ASOC is absent, the magnetic anisotropy does not exist, and therefore we have $\chi^\parallel = \chi^\perp$. Then, the two sublattices are equivalent, and the model is equivalent to the ordinary single-sublattice Hubbard model which has been studied for a long time \cite{Yanase_review_sc}. Near half-filling, the AFM fluctuation with a wave vector around $\q=(\pi,\pi)$ is enhanced. On the other hand, when the ASOC is turned on, the two sublattices are nonequivalent, and the Brillouin zone is folded. In the folded Brillouin zone, the wave vector is $\q=\bm 0$, indicating a ferroic multipole fluctuation.
In our model, the nesting of the Fermi surface gives rise to the magnetic correlation parallel to the $c$ axis at $\q=\bm 0$ for $\alpha<0.2$ as shown in Fig.~\ref{fig:band_FS_chi}(a).
The anisotropy of the Ising-like magnetic fluctuation is compatible with the magnetic structure in BaMn$_2$As$_2$ \cite{Singh_BaMn2As2,Watanabe_mpole}, which possesses a weak spin-orbit coupling.
For $\alpha>0.2$, the Ising-like magnetic fluctuation is significantly suppressed by ASOC, and the incommensurate magnetic correlation perpendicular to the $c$ axis becomes predominant. 
This in-plane magnetic anisotropy is consistent with the 5$d$ transition metal oxide Sr$_2$IrO$_4$ having strong spin-orbit coupling \cite{Zhao_Sr2IrO4,Matteo_Sr2IrO4,Sumita_Ir,B.J.Kim_Sr2IrO4_2008,Y.K.Kim_Sr2IrO4_2014,Yan_Sr2IrO4,Battisti,Y.K.Kim_Sr2IrO4_2016,Meng,B.J.Kim_Sr2IrO4_2009,Boseggia,Clancy,Chetan,Watanabe_Ir,Huang_Sr2IrO4,Crawford}.
Thus, the model captures qualitative properties of these materials although a significantly simplified model is adopted. 
The suppression of magnetic fluctuation by the ASOC is a generic feature \cite{Yanase_CePt3Si_2}, and it has been confirmed by a NMR experiment in CeCoIn$_5$ superlattices \cite{Yamanaka_NMR}.
A qualitatively same conclusion is obtained from noninteracting magnetic susceptibility (see Fig.~\ref{fig:chi0}).

The ASOC dependence of the magnetic anisotropy may be attributed to the Fermi surface around the $M$ point. Figures \ref{fig:band_FS_chi}(c) and \ref{fig:band_FS_chi}(d) show the band structure and Fermi surfaces with a spin texture for $\alpha=0.1$ and $0.3$, respectively.
We find a Lifshitz transition at $\alpha \sim 0.2$, involving the change of the spin texture on the Fermi surface. 
Since the intersublattice hopping disappears on the $X$-$M$ line as ensured by nonsymmorphic crystal symmetry, we can define the spin texture on each sublattice. As we show in the left panels of Figs.~\ref{fig:band_FS_chi}(c) and \ref{fig:band_FS_chi}(d), the spin texture is antiparallel (parallel) between the two Fermi surfaces for $\alpha=0.1$ ($\alpha=0.3$). The change in magnetic anisotropy around $\q=\bm 0$ coincides with the Lifshitz transition.
Note that for the doping level $x=0$, the ASOC-induced Lifshitz transition does not occur up to $\alpha=1$, and then, the anisotropy is always $\chi^\parallel > \chi^\perp$. 
Thus, it is implied that the change in the magnetic anisotropy is related to the Lifshitz transition. 
%Therefore, our simplified model does not necessarily exhibit the correct magnetic anisotropy pertaining to the experiment.

Next, we classify the magnetic fluctuations into augmented cluster multipole fluctuations on the basis of the group theory \cite{Watanabe_mpole}.  When the AFM transition of $\bm m\parallel c$ occurs, the crystal symmetry of $D_{4h}$ is reduced to the subgroup $D'_{2d}$, in which the twofold rotational symmetry axes are rotated $45^\circ$ from $D_{2d}$.
The IRs of $D_{4h}$ are also reduced to representations of $D'_{2d}$.
Since only the $B_{2u}$ IR contains the fully symmetric $A_1$ IR of $D'_{2d}$, the magnetic order belongs to the $B_{2u}$ IR of $D_{4h}$.
A basis function of the $B_{2u}$ IR is a linear combination of magnetic quadrupole and hexadecapole moments.
On the other hand, the AFM structure of $\bm m\perp c$ reduces $D_{4h}$ to $C_{2v}$.
The $E_u$ IR is the candidate of the order parameter.
A basis function of the $E_u$ IR contains magnetic quadrupole and toroidal moments.
Both $B_{2u}$ and $E_u$ IRs represent odd-parity orders, which spontaneously break global inversion symmetry.
Thus, odd-parity multipole fluctuations are enhanced in our model. 

To clarify the multipole fluctuations,  we calculate the odd-parity multipole susceptibility defined as $\chi_{B_{2u}(E_u)}= \sum_{mm'} \chi^{\parallel(\perp)}_{mm'} (\zeta^0_{mm'}  -\zeta^x_{mm'})$, where $\zeta^\mu$ is a Pauli matrix for sublattice degree of freedom \cite{note_chi}.
Here, the magnetic representation is systematically derived from $\Gamma_{\rm tot} = \Gamma_{\rm mag} \otimes \Gamma_{\rm sub} = A_{2g}(E_g) \otimes B_{1u} = B_{2u}(E_u)$, where $\Gamma_{\rm sub}$ is an induced representation of ($A_1$ of $D_{2d}$) $\uparrow$ $ D_{4h}$ except for the fully symmetric representation of $D_{4h}$. 
When $\chi_{B_{2u}(E_u)}$ diverges at $\q=\bm 0$, an odd-parity magnetic multipole order accompanied by inversion symmetry breaking occurs.
Indeed, the sublattice off-diagonal components $\chi_{ab}$ ($=\chi_{ba}$) are negatively enhanced, and therefore 
 $\chi_{B_{2u}(E_u)}$ diverges by increasing $U$, which is consistent with the analysis of the eigenvector of $\hat \chi^0(\q)\hat \Gamma^0$. 
Figures \ref{fig:chi}(a) and \ref{fig:chi}(b) show $\bm q$ dependence of $\chi_{B_{2u}(E_u)}$ for $x=0.2$.
Here, we adopt a large ASOC since we discuss unconventional superconductivity in this region later.
As we have shown in Fig.~\ref{fig:band_FS_chi}, for a large ASOC the wave vector $\q$ of the magnetic order is finite at $x=0.2$.
We call such incommensurate order the quadrupole density wave in a broad sense.
On the other hand, incommensurate susceptibility is not enhanced in the undoped case ($x=0$) [Figs.~\ref{fig:chi}(c) and \ref{fig:chi}(d)]. 
Then, the ferroic multipole fluctuation of $\chi_{B_{2u}}$ is predominant because of the absence of a specific nesting in Fermi surfaces.
These odd-parity fluctuations affect superconductivity, as we demonstrate in the next section.

\section{Superconductivity}
\label{sec:sc}

Here, we examine superconductivity. Although this work is based on a model motivated by BaMn$_2$As$_2$ and Sr$_2$IrO$_4$, the following results are qualitatively valid in a broad range of odd-parity magnetic multipole materials which have been recently identified \cite{Watanabe_mpole_2}.

\begin{figure}[h]
\begin{center}
\includegraphics[width=82mm]{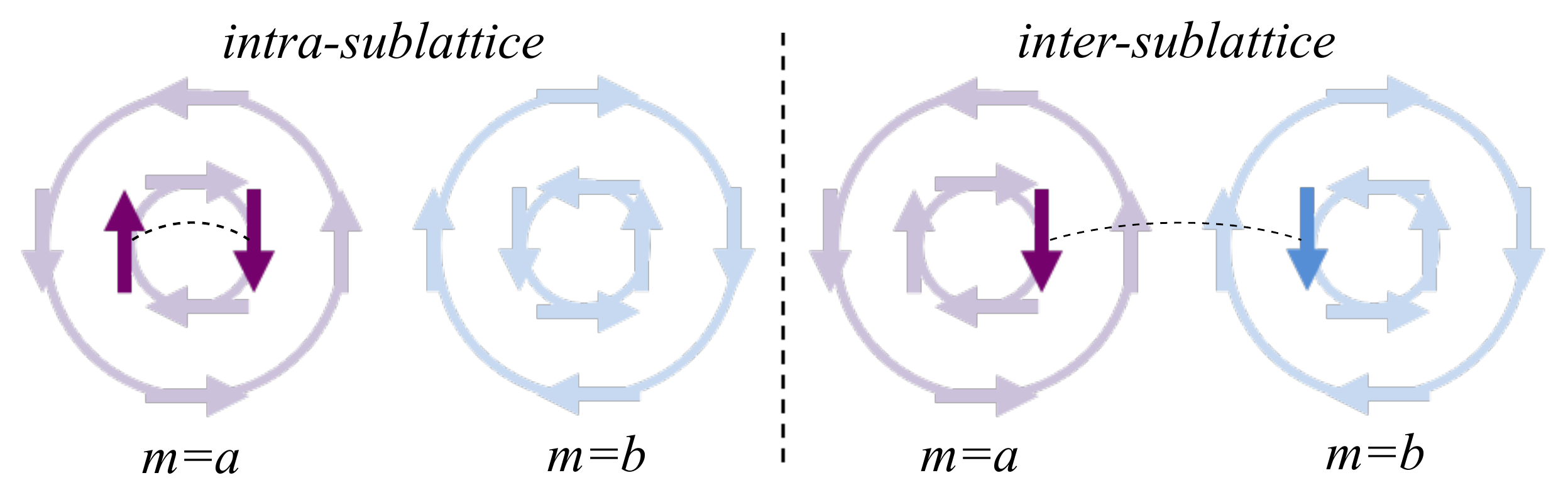}	
\caption{
Schematic figure of Fermi surfaces in a bilayer Rashba system. 
In the absence of intersublattice hybridization, the spin texture on each sublattice can be defined. In locally NCS systems, sublattices $a$ and $b$ have opposite spin textures.
Left panel: For intrasublattice pairing, Cooper pairs with total spin $S_z=0$, namely, the spin-singlet state and spin-triplet state with $\bm d (\k) \parallel \bm g (\k)$, are preferred, while other states are suppressed. Right panel: For intersublattice pairing, Cooper pairing with $S_z=1$ corresponding to the spin-triplet state with $\bm d (\k) \perp \bm g (\k)$ is stable. The spin-singlet pairing and spin-triplet pairing state with $\bm d (\k) \parallel \bm g (\k)$ are fragile by the ASOC.
\label{fig:rashba}}
\end{center}
\end{figure}

Before showing the numerical results of the Eliashberg equation, we discuss the effects of ASOC on superconductivity in locally NCS systems. 
%The antisymmetric spin-orbit interaction $\bm g(\k)$ belongs to $B_{1u}$ IR of $D_{4h}$ space point group. 
%Thus, the site symmetry of each sublattice is $D_{2d}$ without the space inversion symmetry.
The ASOC has two effects: (1) modulation of the one-particle Green's function, and  (2) that of the pairing interaction.
Considering effect (1), we may recognize that the stable superconducting state depends on whether the leading pairing channel is the intrasublattice pairing or intersublattice pairing  (see Fig.~\ref{fig:rashba} for an illustration) \cite{Fischer_noncentro}. 
This gives a selection rule summarized in Table~\ref{tb:select}.
The spin-singlet pairing state or spin-triplet pairing state with $\bm d (\k) \parallel \bm g (\k)$ are stable for intrasublattice pairing, while only the spin-triplet pairing state with $\bm d (\k) \perp \bm g (\k)$ is stable for intersublattice pairing. The other superconducting states are destabilized by sublattice-dependent ASOC. Although the selection rule for intrasublattice pairing is equivalent to that of globally NCS superconductors, the selection rule for intersublattice pairing is peculiar to locally NCS superconductors.
We may understand the selection rule with the help of the band structure in Fig.~\ref{fig:rashba}.
The effect of ASOC on the band structure is taken into account through the one-particle Green's function. On the other hand, 
effect (2) occurs through the modification of magnetic fluctuation, which has been investigated in Sec.~\ref{sec:chis}.
Later, we show that the modified magnetic fluctuation stabilizes odd-parity spin-triplet superconductivity.

\begin{table}[t]
\caption{Selection rules of the superconductivity in locally and globally NCS crystals \cite{Fischer_noncentro}.}
\label{tb:select}
\begin{tabular*}{1.0\columnwidth}{@{\extracolsep{\fill}}lcc}
\hline\hline
&\\[-10pt] globally NCS crystal & \multicolumn{2}{c}{locally NCS crystal} \\[-10pt] & \\ \hline
intrasublattice & intrasublattice & intersublattice \\
singlet, $\bm d (\k) \parallel \bm g (\k)$ & singlet, $\bm d (\k) \parallel \bm g (\k)$ & $\bm d (\k) \perp \bm g (\k)$ \\
%parity violation & globally NCS crystal & \multicolumn{2}{c}{locally NCS crystal} \\ \hline
%pairing channel & intrasublattice & intrasublattice & intersublattice \\
%stable symmetry & singlet, $\bm d (\k) \parallel \bm g (\k)$ & singlet, $\bm d (\k) \parallel \bm g (\k)$ & $\bm d (\k) \perp \bm g (\k)$ \\
\hline\hline 
\end{tabular*}
\end{table}

\begin{table*}[t]
\caption{List of gap functions for the $B_{1g}$, $B_{2g}$, $A_{1u}$, and $B_{1u}$ superconducting states. 
Intrasublattice component $d^{\mu 0}(\k)$, parity-mixed component $d^{\mu z}(\k)$, and intersublattice component $d^{\mu x}(\k)$ are listed.
The last column shows the leading component.}
\label{tb:gap}
\begin{tabular*}{1.0\textwidth}{@{\extracolsep{\fill}}lcccccc}
\hline\hline
&\\[-10pt] & Intrasublattice & & Parity mixed & & Intersublattice & Leading component \\[-10pt] & \\ \hline
$B_{1g}$ & $\cos k_x-\cos k_y$ & $\zeta^0\bar \sigma^0$ &  $\sin k_x \hat {\bm x} + \sin k_y \hat {\bm y}$ & $\zeta^z\bar \sigma^x,\zeta^z\bar \sigma^y$ & $\zeta^x\bar \sigma^0$ & $\zeta^0\bar \sigma^0$\\
$B_{2g}$ & $\sin k_x \sin k_y$ & $\zeta^0\bar \sigma^0$ & $\sin k_y \hat {\bm x} - \sin k_x \hat {\bm y}$ & $\zeta^z\bar \sigma^x,\zeta^z\bar \sigma^y$ & $\zeta^x\bar \sigma^0$ & $\zeta^x\bar \sigma^0$ \\
$A_{1u}$ & $\sin k_x \hat {\bm x} + \sin k_y \hat {\bm y}$ & $\zeta^0\bar \sigma^x,\zeta^0\bar \sigma^y$  & $\cos k_x-\cos k_y$ & $\zeta^z\bar \sigma^0$ & $\zeta^x\bar \sigma^x,\zeta^x\bar \sigma^y$ & $\zeta^0\bar \sigma^x,\zeta^0\bar \sigma^y$ \\
$B_{1u}$ & $\sin k_x \hat {\bm x} - \sin k_y \hat {\bm y}$ & $\zeta^0\bar \sigma^x,\zeta^0\bar \sigma^y$ & $\cos k_x + \cos k_y$ & $\zeta^z\bar \sigma^0$ & $\zeta^x\bar \sigma^x,\zeta^x\bar \sigma^y$ & $\zeta^0\bar \sigma^x,\zeta^0\bar \sigma^y$ \\
\hline\hline 
\end{tabular*}
\end{table*}

\begin{figure*}[ht]
\begin{center}
\includegraphics[width=165mm]{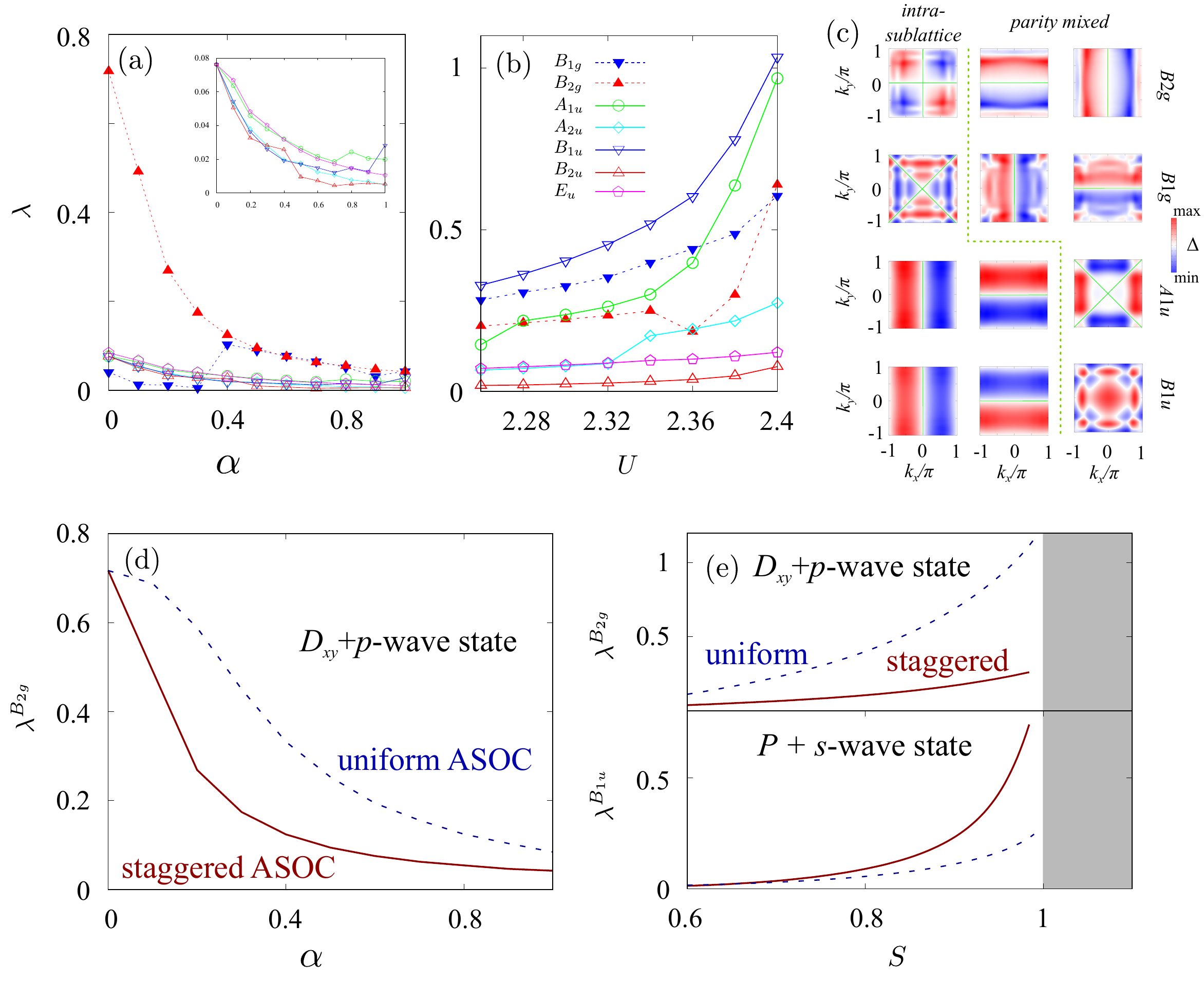}	
\caption{(a) ASOC dependence of eigenvalues of the Eliashberg equation $\lambda$ for $U=1.6$ and $T=0.02$. The $B_{1g} (d_{x^2-y^2}+p$-wave state$), B_{2g} (d_{xy}+p$-wave state$), A_{1u} (p+d_{x^2-y^2}$-wave state$), A_{2u} (p+d_{xy}$-wave state$), B_{1u} (p+s$-wave state$), B _{2u} (p+g$-wave state$), E _{u} (p+d_{zx,yz}$-wave state$)$ IRs are depicted.
(b) Coulomb interaction dependence of $\lambda$ for $\alpha=0.9$.
(c)  Intra-sublattice and parity-mixed components of gap functions for $U=1.4$ and $\alpha=0.1$. 
For the spin-triplet components, $d^{x0} (\k)$ and $d^{y0} (\k)$ are shown for the $A_{1u}$ and $B_{1u}$ IRs, while $d^{xz} (\k)$ and $d^{yz} (\k)$ are shown for the $B_{2g}$ and $B_{1g}$ IRs.  
(d) ASOC dependence of $\lambda^{B_{2g}}$ for the $B_{2g}$ IR at $U=1.6$. The solid line shows the result of our model introduced in Sec.~\ref{sec:sc} for the locally NCS crystal. For comparison, we show the dashed line obtained for the model containing a uniform ASOC instead of the staggered ASOC. The latter corresponds to the globally NCS crystal.
(e) Stoner factor $S=[\hat \Gamma \hat \chi (\q)]_{\rm max}$ dependence of $\lambda^{B_{2g}}$ (upper panel) and $\lambda^{B_{1u}}$ (lower panel) for $\alpha=0.9$. We again compare the locally NCS model (solid lines) with the globally NCS model (dashed lines). 
\label{fig:lambda_gap}}
\end{center}
\end{figure*}

Now we show the numerical results. Within the RPA theory, we obtained four stable superconducting states: $B_{2g}$, $B_{1u}$, $B_{1g}$, and $A_{1u}$ IRs.
The leading pair amplitude and other admixed components of the superconducting states are summarized in Table~\ref{tb:gap}.
As we have shown in Sec.~\ref{sec:gapstruct}, parity mixing does not appear in intersublattice components. Thus, Table~\ref{tb:gap} illustrates the intrasublattice component, parity-mixed intrasublattice component, and intersublattice component.  
For convenience we describe the order parameter of superconductivity in a standard manner, $\hat{\Delta}(\k,i\pi T) = \sum _{\mu\nu} d^{\mu\nu}(\k) \bar \sigma^{\mu}_{ss'} \zeta^{\nu}_{mm'}$, where $\sigma^\mu$ and $\zeta^\nu$ are the Pauli matrix for spin and sublattice degrees of freedom, respectively. We introduced $\bar \sigma^{\mu}_{ss'} = [\sigma^{\mu} i \sigma^y]_{ss'}$ for $\mu = 0, x, y, z$. This notation is used in Table~\ref{tb:gap}. 

%In the absence of the ASOC the $B_{2g}$ state is the most stable. This 
The $B_{2g}$ state corresponds to the spin-singlet $d_{x^2-y^2}$-wave pairing state in the well-studied single-sublattice Hubbard model. Since the {\it x} and {\it y} axes in the two-sublattice model are rotated $45^\circ$, the predominant component of the order parameter is an intersublattice spin-singlet component $d^{0x} (\k)$ of $d_{xy}$-wave symmetry. 
Consistent with many theoretical works on the single-sublattice Hubbard model near half filling \cite{Yanase_review_sc}, the $B_{2g}$ state is stable at $\alpha=0$. 
%This $B_{2g}$ state is dominant in the whole of the IRs for $\alpha=0$ eV. 
However, when the staggered ASOC is turned on, the eigenvalue of the Eliashberg equation $\lambda$ for the $B_{2g}$ state is steeply suppressed [Fig.~\ref{fig:lambda_gap}(a)]. 
This is mainly because intersublattice spin-singlet pairing is ruled out by the selection rule of locally NCS superconductors, in striking contrast to the globally NCS superconductors (see Table~\ref{tb:select}). 

To examine the effect of the staggered ASOC, we solved the Eliashberg equation for a similar model containing a uniform ASOC instead of the staggered ASOC.
The eigenvalues $\lambda$ of the two models are compared in Fig.~\ref{fig:lambda_gap}(d).
Consistent with the selection rule, the local parity violation more significantly suppresses $d$-wave superconductivity than does the global parity violation.
In both cases, superconductivity is suppressed by ASOC owing to the suppressed magnetic fluctuation.
In addition, the staggered ASOC causes pair breaking through the modulation of the one-particle Green's function. 
%The order parameter of the $B_{2g}$ IR is represented as $d^{00} (\k) + d^{0x} (\k) + d^{xz} (\k) + d^{yz} (\k)$.
The dominant pairing component $d^{0x} (\k)$, which is incompatible with the selection rule in Table~\ref{tb:select}, decreases in the same manner as $\lambda^{B_{2g}}$ by increasing $\alpha$.
%It indicates that this ingredient is a main resource of the $\lambda$-suppression.
Instead of that, an intrasublattice spin-singlet component compatible with the selection rule monotonically increases as $d^{00} (\k) \simeq d^{00}(\k)|_{\alpha=0} + A\alpha \sin k_x\sin k_y$. 
Owing to the parity mixing by the ASOC, an admixed staggered spin-triplet component, 
$\left(d^{xz}(\k), d^{yz}(\k)\right) \simeq B \alpha (\sin k_y, -\sin k_x)$, appears.
The momentum dependence of these components is shown in Fig.~\ref{fig:lambda_gap}(c).

%and $d^{xz} (\k) + d^{yz} (\k) \sim B\alpha(\sin k_y \hat {\bm x} - \sin k_x \hat {\bm y})$.

Although $d$-wave superconductivity is stable in a broad range near the AFM critical point, it is significantly suppressed in locally NCS crystals with a large spin-orbit coupling. Thus, we have a chance to see another exotic superconducting state. 
Candidates are the $B_{1u}$ and $A_{1u}$ states which show large eigenvalues of the Eliashberg equation. The other odd-parity IRs are less stable than these states. Both $B_{1u}$ and $A_{1u}$ states satisfy the condition $\bm d(\k) \parallel \bm g(\k)$, in a part of $\k$ space, compatible with the selection rule for the intrasublattice pairing. 
However, $\lambda$ of all the odd-parity IRs moderately decrease as increasing $\alpha$ [inset of Fig.~\ref{fig:lambda_gap}(a)] due to the suppression of magnetic fluctuations.
Thus, by looking at the $\alpha$ dependence of $\lambda$ we cannot determine which superconducting states are preferred. 
To examine the superconductivity in a large ASOC region, we calculated the $U$ dependence of $\lambda$ and investigated which superconducting states are stabilized in the vicinity of the magnetic critical point. 
Figure~\ref{fig:lambda_gap}(b) shows that the $B_{1u}$ state is predominant for $x=0.2$ and the eigenvalue of the Eliashberg equation reaches $\lambda = 1$ at $U\simeq 2.4$.
The second most stable superconducting state is the $A_{1u}$ IR. This state is the most  stable in the undoped system $x=0$. 
%has the secondary dominant eigenvalue.
Thus, odd-parity superconductivity may be realized in a large ASOC region by the magnetic multipole fluctuations.

In order to further elucidate an essential role of sublattice-dependent ASOC, we again compare our model to the model containing a uniform ASOC.
Figure~\ref{fig:lambda_gap}(e) compares the $U$ dependence of eigenvalues $\lambda$ for the $B_{2g}$ state and $B_{1u}$ state.
As shown in the lower panel, the $B_{1u}$ superconducting state is more stable in our model than in the globally NCS model.
In other words, the staggered ASOC favors the $B_{1u}$ state more significantly than does the uniform ASOC.
Because the magnetic fluctuation is identified as odd-parity magnetic multipole fluctuation only in the locally NCS model, it is implied that the modification of magnetic fluctuation by the staggered ASOC leads to odd-parity multipole fluctuation and favors odd-parity superconductivity.
Note that the $B_{1u}$ state is compatible with the selection rule in Table \ref{tb:select} for both models.
In contrast, the even-parity  $B_{2g}$ state is suppressed by the staggered ASOC [upper panel of Fig.~\ref{fig:lambda_gap}(e)]. 
%Interestingly, these states are stabilized by the local parity violation [Fig.~\ref{fig:lambda_gap}(e)].
%This is a remarkable difference in the globally NCS crystal.

%The order parameter of the $B_{1u}$ IR is represented as $d^{\mu\nu}(\k) + d^{0z} (\k)$ for $\mu=x,y$ and $\nu=0,x$, and given by the 
As we show in Table~\ref{tb:gap}, the leading order parameter of the $B_{1u}$ state is an intrasublattice spin-triplet pairing  
$d^{x0} (\k)$ and $d^{y0} (\k)$, namely, $(\sin k_x \bar \sigma^x - \sin k_y \bar \sigma^y) \zeta^0$. An admixed staggered spin-singlet component is $d^{0z} (\k) \simeq \delta + \cos k_x+\cos k_y$. 
On the other hand, the leading order parameter of the $A_{1u}$ state is $(\sin k_x \bar \sigma^x + \sin k_y \bar \sigma^y) \zeta^0$, and the induced component is $d^{0z} (\k) \simeq \cos k_x-\cos k_y$ [see Fig.~\ref{fig:lambda_gap}(c)].

Spin-triplet superconductors are known to be a platform of topological superconductivity, which has been one of the main subjects of modern condensed matter physics.
The spin-triplet superconductivity clarified in this work is also identified as topological superconducting states. 
According to the criterion for time-reversal-invariant topological superconductivity in two dimension \cite{M.Sato2010_TSC}, both $B_{1u}$ and $A_{1u}$ states are $Z_2$-topological superconducting states in class DIII, because the number of Fermi surfaces enclosing time-reversal-invariant momentum ($\Gamma$, $X$, and $M$ points) is odd. 

\begin{figure}[ht]
\begin{center}
\includegraphics[width=77mm]{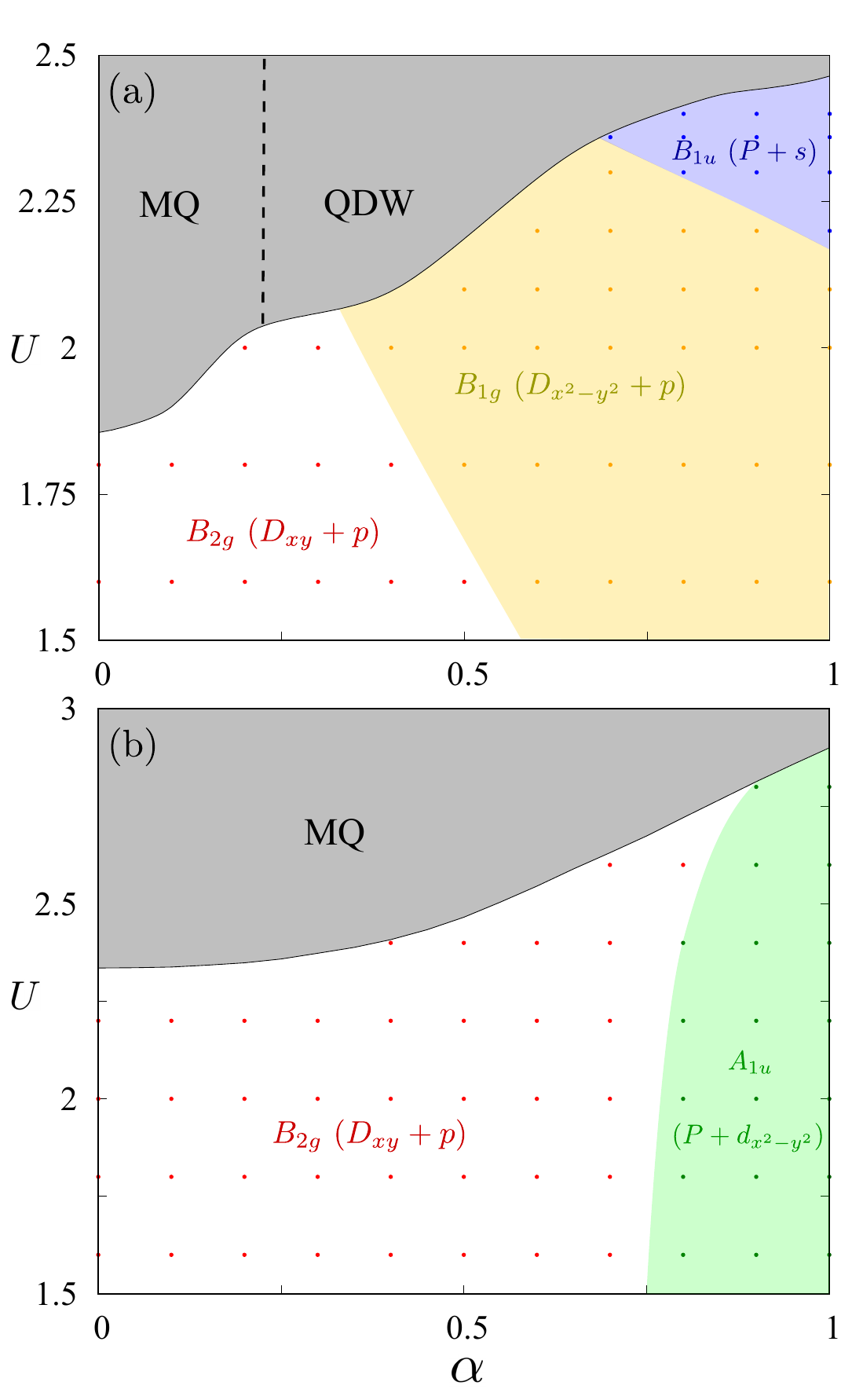}	
\caption{Phase diagram for the Coulomb interaction $U$ and staggered ASOC $\alpha$ at $T=0.02$. (a) $x=0.2$ and (b) $x=0$. The magnetic quadrupole (MQ) state and quadrupole density wave (QDW) state. In the paramagnetic state, the $B_{2g}$, $B_{1g}$, $B_{1u}$, and $A_{1u}$ superconducting states are illustrated. Capital and lowercase letters represent predominant and parity-mixed (subdominant) components of superconducting order parameter, respectively.
\label{fig:phase}}
\end{center}
\end{figure}

Figure~\ref{fig:phase} shows the phase diagram as a function of $\alpha$ and $U$. From Fig.~\ref{fig:phase}(a) for $x=0.2$, we identify the stable odd-parity $B_{1u}$ state for $\alpha>0.7$, while the $B_{2g}$ state is stabilized for $\alpha<0.3$.
The magnetic instability for $\alpha<0.2$ is the $B_{2u}$ magnetic quadrupole and hexadecapole order, which is monotonically suppressed by ASOC.
The magnetic instability for $\alpha=0.7$ is the $E_{u}$ magnetic quadrupole density wave with an incommensurate period. 
In an intermediate ASOC region, the $B_{1g}$ state represented by the predominant intrasublattice spin-singlet pairing $d^{00} (\k) \sim \cos k_x-\cos k_y$ is stable. %admixed with a sublattice-odd spin-triplet pairing $d^{\mu z} (\k) \sim \sin k_x \hat {\bm x} + \sin k_y \hat {\bm y}$.
This state is stabilized by the incommensurate magnetic fluctuation with a small wave vector $\q\sim(\pm1.14,\pm1.14)$. As is usually done by magnetic fluctuation, the sign change of the gap function between the Fermi surface connected by the wave vector is favored. The $B_{1g}$ superconducting state is compatible with this condition and also with the selection rule for locally NCS superconductors (Table~\ref{tb:select}).

In the undoped case, $x=0$, we obtain a similar but simpler phase diagram [Fig.~\ref{fig:phase}(b)]. 
In a large ASOC region, the odd-parity $A_{1u}$ superconducting state is realized near the ferroic odd-parity magnetic multipole state. The magnetic wave vector is always $\q =\bm 0$. Thus, incommensurate magnetic fluctuation is not a necessary condition for odd-parity superconductivity. Irrespective of the wave vector of multipole fluctuations, the odd-parity superconducting states are stabilized in a large ASOC region. In contrast, $B_{1g}$ superconducting state requires the incommensurate fluctuation, and it disappears in the phase diagram for $x=0$.

%The order parameter of the $B_{1g}$ IR is represented as the predominant singlet pairing $d^{00} (\k) \sim \cos k_x-\cos k_y$ admixed with a sublattice-odd triplet pairing $d^{\mu z} (\k) \sim \sin k_x \hat {\bm x} + \sin k_y \hat {\bm y}$.
%This state appears in the region between $B_{2g}$ and $B_{1u}$-states [Fig.~\ref{fig:lambda_gap}(f)].
%The incommensurate magnetic fluctuation induces the sign change of the gap function between the Fermi surface at $\q\sim(\pm1.14,\pm1.14)$.
%Therefore, this state can not be stabilized for undoped region $x=0$, where the ferromagnetic quadrupole fluctuation is dominant.
%Although the effect of the symmetric spin-orbit coupling is not included in our calculation, the issue is beyond the scope of this paper.

\begin{figure}[ht]
\begin{center}
\includegraphics[width=82mm]{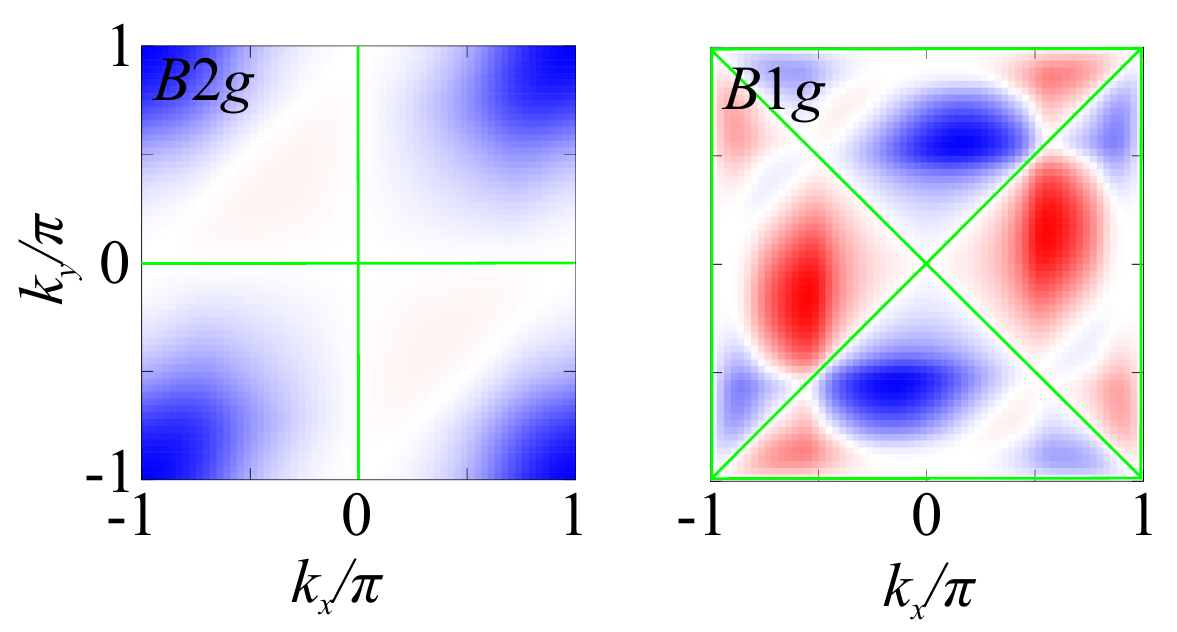}
\caption{
Intersublattice components of the gap function for $U=1.4$ and $\alpha=0.1$. We show the real part for (a) $B_{2g}$ IR and (b) $B_{1g}$ IR. 
\label{fig:gap_inter}}
\end{center}
\end{figure}

Finally, we comment on a peculiar momentum dependence of the gap function protected by nonsymmorphic space-group symmetry. 
From Eq.~(\ref{eq:GxFGx^T_ky_pi}) and a similar equation for $G_y=\{\sigma_y | \tau_x\}$, the intersublattice spin-singlet gap function shows an unusual nodal/gapped structure. 
As shown in Fig.~\ref{fig:gap_inter}, $d^{0x} (\k)$ for the $B_{1g}$ IR shows nodal lines at $k_{x,y}=\pm\pi$, while it is gapped for the $B_{2g}$ IR.
These nodal/gapped structures at the Brillouin zone boundaries are opposite to those in symmorphic crystals. 
Note that Fig.~\ref{fig:gap_inter} does not show fourfold rotation symmetry, because it depicts a real part of the gap function. The superconducting gap is fourfold symmetric in accordance with the symmetry of the system.
The gap functions numerically obtained in this paper satisfy the symmetry conditions discussed in Sec.~\ref{sec:gapstruct}. %revealed in Sec.~\ref{sec:chis}.
%protected by the nonsymmorphic symmetry.

%\begin{figure*}[ht]
%\begin{center}
%\includegraphics[width=165mm]{./lambda_phase_gap.pdf}	
%\caption{
%(a) ASOC dependence of $\lambda^{B_{2g}}$ ($d_{xy}+p$-wave state) for $U=1.6$ eV. Dashed (Solid) line shows uniform (staggered) ASOC, which corresponds to the globally (locally) NCS crystal.
%(b) Coulomb interaction dependence of $\lambda^{B_{1u}}$ ($p+s$-wave state) for $\alpha=0.9$ eV.
%(c) Phase diagram of Coulomb interaction vs. staggered ASOC for $x=0.2$.
%(d) Inter-sublattice components of the gap function for $U=1.4$ eV and $\alpha=0.1$ eV.
%\label{fig:lambda_phase_gap}}
%\end{center}
%\end{figure*}

%\begin{figure}[ht]
%\begin{center}
%\includegraphics[width=82mm]{./lambda_GISB-LISB.pdf}	
%\caption{
%ASOC dependence of $\lambda^{B_{2g}}$ ($d_{xy}+p$-wave state) for $U=1.6$ eV. Dashed (Solid) line shows uniform (staggered) ASOC, which corresponds to the globally (locally) NCS crystal.
%\label{fig:lambda_GISB-LISB}}
%\end{center}
%\end{figure}
%
%\begin{figure}[ht]
%\begin{center}
%\includegraphics[width=82mm]{./phase_u_aasoc.pdf}	
%\caption{
%(a) Phase diagram of Coulomb interaction vs. ASOC for $x=0.2$.
%\label{fig:phase_u_aasoc}}
%\end{center}
%\end{figure}

\section{Summary and Discussion}
\label{sec:sum}
In this paper, we have investigated the superconductivity induced by odd-parity magnetic multipole fluctuations in a locally NCS crystal. The obtained results are summarized below.

First, we have revealed the symmetry properties of the superconductivity with sublattice degrees of freedom.
The general representation for pair amplitudes including nonsymmorphic operations was derived.
After introducing a space group of a specific crystal structure, an unconventional gapped/nodal structure protected by the glide symmetry was shown. 
%which is concerned with recent classification theories\cite{Norman,Micklitz,Yanase_UPt3_Weyl,Kobayashi_TSC_2016,Nomoto_UPt3,Nomoto_UCoGe,Sumita_UPt3} beyond the Sigrist-Ueda method \cite{Sigrist_Ueda}.
For glide-even superconducting states, the intersublattice pair amplitude possesses a node on a Brillouin zone boundary, while those are gapped for glide-odd superconducting states.
On the other hand, the intrasublattice pair amplitude shows local parity mixing in a staggered form with respect to the sublattices. The admixture of spin-singlet and spin-triplet pairings is classified by the point group of local atomic sites. These symmetry analyses are consistent with the following numerical results.

Next, we have studied a two-dimensional two-sublattice model with an on-site Coulomb interaction term and a $D_{2d}$-type staggered ASOC term. 
The magnetic fluctuation is suppressed by increasing the ASOC, consistent with a theoretical study of CePt$_{3}$Si \cite{Yanase_CePt3Si_2} and an experiment for CeCoIn$_5$ superlattices \cite{Yamanaka_NMR}.
The ASOC term also induces an anisotropy in magnetic fluctuation.
From the classification of multipole order parameters, the antiferromagnetism of $\bm m\parallel c$ is classified into the odd-parity magnetic multipole order belonging to the $B_{2u}$ IR.
In the same way, the $\bm m\perp c$ AFM state is classified into the $E_{u}$ IR.
Both $B_{2u}$ and $E_{u}$ IRs represent odd-parity magnetic multipole orders accompanied by spontaneous global inversion symmetry breaking. 
In our model, these odd-parity multipole fluctuations are enhanced in the vicinity of the magnetic critical point.

Superconducting instability has been analyzed by solving the Eliashberg equation with the use of RPA.
We have demonstrated the selection rules of locally NCS superconductors \cite{Fischer_noncentro}.
Since the intersublattice spin-singlet pairing is ruled out by the selection rule, the $B_{2g}$ state, which corresponds to the well-studied $d_{x^2-y^2}$-wave superconducting state in the single-sublattice Hubbard model, is rapidly suppressed by turning on the ASOC.
Intriguingly, this behavior is in sharp contrast with globally NCS superconductors. 
When $d$-wave superconductivity is suppressed in a large ASOC region, the odd-parity superconductivity is stabilized by enhanced odd-parity multipole fluctuations. 
%Then, we can expect the exotic superconductivity for a large ASOC region in locally NCS system.
We found that the $B_{1u}$ or $A_{1u}$ state is stable. 
From the criterion for time-reversal-invariant topological superconductivity, both the $B_{1u}$ and $A_{1u}$ states are identified as the nontrivial $Z_2$ topological superconductivity in the DIII class.
Thus, our results may open a different platform of odd-parity topological superconductivity.

%where the odd-parity magnetic fluctuation is largely moderated by sublattice-dependent ASOC.
Here, we note that the staggered ASOC arising from the local parity violation in the crystal structure plays an essential role in stabilizing odd-parity superconductivity.
From a comparison with the globally NCS model, we have shown that the modification of magnetic fluctuation by the staggered ASOC significantly enhances odd-parity superconductivity.
Such an enhancement is not caused by the {\it sublattice-independent} ASOC in the globally NCS system. The modified magnetic fluctuation in the locally NCS system is regarded as odd-parity magnetic multipole fluctuations. Therefore, we conclude that odd-parity superconductivity is stabilized by odd-parity multipole fluctuations. 
%The locally NCS crystals favor the odd-parity superconductivity by the spin-orbit coupling, which suppresses the $d$-wave superconductivity and enhances the $p$-wave superconductivity.  

%The odd-parity magnetic fluctuation is enhanced in locally NCS crystal, not in globally NCS one.
%Comparison calculation between the locally and globally NCS crystal shows that the odd-parity superconductivity is not stabilized by the sublattice-{\it independent} ASOC. 
%Then, the local parity violation accompanied by the odd-parity fluctuation promotes the odd-parity superconductivity.

In the long-standing studies of unconventional superconductivity, spin-triplet superconductivity has attracted interest. Furthermore, renewed interest has been stimulated because odd-parity superconductivity may be topologically nontrivial. However, only limited examples, such as Sr$_2$RuO$_4$ \cite{Mackenzie_Sr2RuO4}, UPt$_3$ \cite{Joynt_UPt3}, UCoGe \cite{Huy_UCoGe}, and so on, are known as strong candidates for spin-triplet superconductors. This is because conditions for spin-triplet pairing are quite unfavorable in most materials.
Our paper has uncovered a different pairing mechanism favorable for spin-triplet pairing. The local parity violation in crystal structures, large spin-orbit coupling, and enhanced magnetic multipole fluctuations are conditions for spin-triplet superconductivity proposed in this paper.

Finally, we discuss candidate materials of odd-parity fluctuation and superconductivity.
First, Sr$_2$IrO$_4$ is a layered perovskite 5$d$ transition metal oxide and possesses a K$_2$NiF$_4$-type structure as does La$_2$CuO$_4$.
Many similarities to the high-temperature cuprate superconductors have been recognized, and thus it is expected to be a superconductor from both the experimental \cite{Y.K.Kim_Sr2IrO4_2016} and theoretical \cite{Watanabe_Ir,Sumita_Ir} sides.
AFM moments align to the $a$ axis with a small canted moment along the $b$ axis and show stacking patterns: $-++-$, $-+-+$, and $++++$.
From the viewpoint of multipole order, the $-++-$ pattern is magnetic octupole order preserving space inversion symmetry, while the $-+-+$ pattern is odd-parity magnetic quadrupole order \cite{Matteo_Sr2IrO4}.
Another candidate is BaMn$_2$As$_2$ crystallizing in a locally NCS ThCr$_2$Si$_2$-type structure, which is isostructural to the 122 systems of iron-based superconductors.
Undoped BaMn$_2$As$_2$ shows the $G$-type AFM order at $T_N=625$ K \cite{Singh_BaMn2As2}. The magnetic structure is classified into the odd-parity magnetic quadrupole and hexadecapole orders \cite{Watanabe_mpole}.
Many related materials show the same odd-parity magnetic order, and some of them may be superconducting \cite{Watanabe_mpole_2}.
A further experimental search is desired.
Hole-doped (Ba$_{1-x}$K$_x$)Mn$_2$As$_2$ realizes the metallic state.
However, superconductivity has not been observed up to now. 
A fascinating material is CrAs \cite{Watanabe_CrAs,Selte_CrAs,Kotegawa_CrAs_2014,Wu_CrAs,Kotegawa_CrAs_2015,Keller,Shen,Niu,Guo_CrAs}.
The space group is No.~62, $Pnma$ ($D_{2h}$ point group) lacking local inversion symmetry at the Cr sites. CrAs shows a first-order helical magnetic transition at $T_N\sim265$ K \cite{Watanabe_CrAs,Selte_CrAs}. When the helical magnetic order is suppressed by applied pressure, superconductivity occurs \cite{Kotegawa_CrAs_2014}. The phase diagram implies superconductivity induced by magnetic fluctuation.
The wave vector of the helical magnetism is incommensurate, $\q=(0,0,q_c)$ with $q_c\sim0.354$.
Thus, a local parity violation and odd-parity magnetic fluctuation may promote odd-parity superconductivity in CrAs.
Indeed, a recent experiment suggests spin-triplet superconductivity \cite{Guo_CrAs}.
As for odd-parity {\it electric} multipole fluctuation, SrTiO$_3$ \cite{Rischau_STO} and Cd$_2$Re$_2$O$_7$ \cite{Yamaura_Cd2Re2O7_2017} show superconductivity in the vicinity of the nonmagnetic order accompanied by global inversion symmetry breaking.
Recently, Ref.~\onlinecite{Kozii} theoretically proposed that odd-parity {\it electric} fluctuation may induce the odd-parity superconductivity.
More research of multipole materials will shed light on odd-parity superconductivity in condensed matter.
%, the antiferromagnetic 5$d$ electron system in ThCr$_2$Si$_2$-type structure is compatible and might be found in the recently listed Table \cite{Watanabe_mpole_2}.

\section*{Acknowledgments}
The authors are grateful to S. Sumita, H. Watanabe, Y. Yanagi, and A. Daido for fruitful discussions and comments. This work was supported by a Grant-in-Aid for Scientific Research on Innovative Areas ``J-Physics'' (Grant No.~JP15H05884) and ``Topological Materials Science'' (Grants No.~JP16H00991 and JP18H04225) from Japan Society for Promotion of Science (JSPS) and by JSPS KAKENHI (Grants No.~JP15K05164, No.~JP15H05745, and No.~JP18H01178).


\begin{thebibliography}{106}%
\makeatletter
\providecommand \@ifxundefined [1]{%
 \@ifx{#1\undefined}
}%
\providecommand \@ifnum [1]{%
 \ifnum #1\expandafter \@firstoftwo
 \else \expandafter \@secondoftwo
 \fi
}%
\providecommand \@ifx [1]{%
 \ifx #1\expandafter \@firstoftwo
 \else \expandafter \@secondoftwo
 \fi
}%
\providecommand \natexlab [1]{#1}%
\providecommand \enquote  [1]{``#1''}%
\providecommand \bibnamefont  [1]{#1}%
\providecommand \bibfnamefont [1]{#1}%
\providecommand \citenamefont [1]{#1}%
\providecommand \href@noop [0]{\@secondoftwo}%
\providecommand \href [0]{\begingroup \@sanitize@url \@href}%
\providecommand \@href[1]{\@@startlink{#1}\@@href}%
\providecommand \@@href[1]{\endgroup#1\@@endlink}%
\providecommand \@sanitize@url [0]{\catcode `\\12\catcode `\$12\catcode
  `\&12\catcode `\#12\catcode `\^12\catcode `\_12\catcode `\%12\relax}%
\providecommand \@@startlink[1]{}%
\providecommand \@@endlink[0]{}%
\providecommand \url  [0]{\begingroup\@sanitize@url \@url }%
\providecommand \@url [1]{\endgroup\@href {#1}{\urlprefix }}%
\providecommand \urlprefix  [0]{URL }%
\providecommand \Eprint [0]{\href }%
\providecommand \doibase [0]{http://dx.doi.org/}%
\providecommand \selectlanguage [0]{\@gobble}%
\providecommand \bibinfo  [0]{\@secondoftwo}%
\providecommand \bibfield  [0]{\@secondoftwo}%
\providecommand \translation [1]{[#1]}%
\providecommand \BibitemOpen [0]{}%
\providecommand \bibitemStop [0]{}%
\providecommand \bibitemNoStop [0]{.\EOS\space}%
\providecommand \EOS [0]{\spacefactor3000\relax}%
\providecommand \BibitemShut  [1]{\csname bibitem#1\endcsname}%
\let\auto@bib@innerbib\@empty
%</preamble>
\bibitem [{\citenamefont {Sigrist}\ \emph {et~al.}(2014)\citenamefont
  {Sigrist}, \citenamefont {Agterberg}, \citenamefont {Fischer}, \citenamefont
  {Goryo}, \citenamefont {Loder}, \citenamefont {Rhim}, \citenamefont
  {Maruyama}, \citenamefont {Yanase}, \citenamefont {Yoshida},\ and\
  \citenamefont {Youn}}]{Sigrist_LNCS}%
  \BibitemOpen
  \bibfield  {author} {\bibinfo {author} {\bibfnamefont {M.}~\bibnamefont
  {Sigrist}}, \bibinfo {author} {\bibfnamefont {D.~F.}\ \bibnamefont
  {Agterberg}}, \bibinfo {author} {\bibfnamefont {M.~H.}\ \bibnamefont
  {Fischer}}, \bibinfo {author} {\bibfnamefont {J.}~\bibnamefont {Goryo}},
  \bibinfo {author} {\bibfnamefont {F.}~\bibnamefont {Loder}}, \bibinfo
  {author} {\bibfnamefont {S.-H.}\ \bibnamefont {Rhim}}, \bibinfo {author}
  {\bibfnamefont {D.}~\bibnamefont {Maruyama}}, \bibinfo {author}
  {\bibfnamefont {Y.}~\bibnamefont {Yanase}}, \bibinfo {author} {\bibfnamefont
  {T.}~\bibnamefont {Yoshida}}, \ and\ \bibinfo {author} {\bibfnamefont
  {S.~J.}\ \bibnamefont {Youn}},\ }\href {\doibase 10.7566/JPSJ.83.061014}
  {\bibfield  {journal} {\bibinfo  {journal} {J. Phys. Soc. Jpn.}\ }\textbf
  {\bibinfo {volume} {83}},\ \bibinfo {pages} {061014} (\bibinfo {year}
  {2014})}\BibitemShut {NoStop}%
\bibitem [{\citenamefont {Maruyama}\ \emph {et~al.}(2012)\citenamefont
  {Maruyama}, \citenamefont {Sigrist},\ and\ \citenamefont
  {Yanase}}]{Maruyama_LNCS}%
  \BibitemOpen
  \bibfield  {author} {\bibinfo {author} {\bibfnamefont {D.}~\bibnamefont
  {Maruyama}}, \bibinfo {author} {\bibfnamefont {M.}~\bibnamefont {Sigrist}}, \
  and\ \bibinfo {author} {\bibfnamefont {Y.}~\bibnamefont {Yanase}},\ }\href
  {\doibase 10.1143/JPSJ.81.034702} {\bibfield  {journal} {\bibinfo  {journal}
  {J. Phys. Soc. Jpn.}\ }\textbf {\bibinfo {volume} {81}},\ \bibinfo {pages}
  {034702} (\bibinfo {year} {2012})}\BibitemShut {NoStop}%
\bibitem [{\citenamefont {Fischer}\ \emph {et~al.}(2011)\citenamefont
  {Fischer}, \citenamefont {Loder},\ and\ \citenamefont
  {Sigrist}}]{Fischer_noncentro}%
  \BibitemOpen
  \bibfield  {author} {\bibinfo {author} {\bibfnamefont {M.~H.}\ \bibnamefont
  {Fischer}}, \bibinfo {author} {\bibfnamefont {F.}~\bibnamefont {Loder}}, \
  and\ \bibinfo {author} {\bibfnamefont {M.}~\bibnamefont {Sigrist}},\ }\href
  {\doibase 10.1103/PhysRevB.84.184533} {\bibfield  {journal} {\bibinfo
  {journal} {Phys. Rev. B}\ }\textbf {\bibinfo {volume} {84}},\ \bibinfo
  {pages} {184533} (\bibinfo {year} {2011})}\BibitemShut {NoStop}%
\bibitem [{\citenamefont {Nakosai}\ \emph {et~al.}(2012)\citenamefont
  {Nakosai}, \citenamefont {Tanaka},\ and\ \citenamefont {Nagaosa}}]{Nakosai}%
  \BibitemOpen
  \bibfield  {author} {\bibinfo {author} {\bibfnamefont {S.}~\bibnamefont
  {Nakosai}}, \bibinfo {author} {\bibfnamefont {Y.}~\bibnamefont {Tanaka}}, \
  and\ \bibinfo {author} {\bibfnamefont {N.}~\bibnamefont {Nagaosa}},\ }\href
  {\doibase 10.1103/PhysRevLett.108.147003} {\bibfield  {journal} {\bibinfo
  {journal} {Phys. Rev. Lett.}\ }\textbf {\bibinfo {volume} {108}},\ \bibinfo
  {pages} {147003} (\bibinfo {year} {2012})}\BibitemShut {NoStop}%
\bibitem [{\citenamefont {Yoshida}\ \emph {et~al.}(2012)\citenamefont
  {Yoshida}, \citenamefont {Sigrist},\ and\ \citenamefont
  {Yanase}}]{Yoshida_PDW}%
  \BibitemOpen
  \bibfield  {author} {\bibinfo {author} {\bibfnamefont {T.}~\bibnamefont
  {Yoshida}}, \bibinfo {author} {\bibfnamefont {M.}~\bibnamefont {Sigrist}}, \
  and\ \bibinfo {author} {\bibfnamefont {Y.}~\bibnamefont {Yanase}},\ }\href
  {\doibase 10.1103/PhysRevB.86.134514} {\bibfield  {journal} {\bibinfo
  {journal} {Phys. Rev. B}\ }\textbf {\bibinfo {volume} {86}},\ \bibinfo
  {pages} {134514} (\bibinfo {year} {2012})}\BibitemShut {NoStop}%
\bibitem [{\citenamefont {Yoshida}\ \emph {et~al.}(2013)\citenamefont
  {Yoshida}, \citenamefont {Sigrist},\ and\ \citenamefont
  {Yanase}}]{Yoshida_CS}%
  \BibitemOpen
  \bibfield  {author} {\bibinfo {author} {\bibfnamefont {T.}~\bibnamefont
  {Yoshida}}, \bibinfo {author} {\bibfnamefont {M.}~\bibnamefont {Sigrist}}, \
  and\ \bibinfo {author} {\bibfnamefont {Y.}~\bibnamefont {Yanase}},\ }\href
  {\doibase 10.7566/JPSJ.82.074714} {\bibfield  {journal} {\bibinfo  {journal}
  {J. Phys. Soc. Jpn.}\ }\textbf {\bibinfo {volume} {82}},\ \bibinfo {pages}
  {074714} (\bibinfo {year} {2013})}\BibitemShut {NoStop}%
\bibitem [{\citenamefont {Yoshida}\ \emph {et~al.}(2015)\citenamefont
  {Yoshida}, \citenamefont {Sigrist},\ and\ \citenamefont
  {Yanase}}]{Yoshida_TCSC}%
  \BibitemOpen
  \bibfield  {author} {\bibinfo {author} {\bibfnamefont {T.}~\bibnamefont
  {Yoshida}}, \bibinfo {author} {\bibfnamefont {M.}~\bibnamefont {Sigrist}}, \
  and\ \bibinfo {author} {\bibfnamefont {Y.}~\bibnamefont {Yanase}},\ }\href
  {\doibase 10.1103/PhysRevLett.115.027001} {\bibfield  {journal} {\bibinfo
  {journal} {Phys. Rev. Lett.}\ }\textbf {\bibinfo {volume} {115}},\ \bibinfo
  {pages} {027001} (\bibinfo {year} {2015})}\BibitemShut {NoStop}%
\bibitem [{\citenamefont {Yoshida}\ \emph {et~al.}(2017)\citenamefont
  {Yoshida}, \citenamefont {Daido}, \citenamefont {Yanase},\ and\ \citenamefont
  {Kawakami}}]{Yoshida_CeCoIn5_Z8}%
  \BibitemOpen
  \bibfield  {author} {\bibinfo {author} {\bibfnamefont {T.}~\bibnamefont
  {Yoshida}}, \bibinfo {author} {\bibfnamefont {A.}~\bibnamefont {Daido}},
  \bibinfo {author} {\bibfnamefont {Y.}~\bibnamefont {Yanase}}, \ and\ \bibinfo
  {author} {\bibfnamefont {N.}~\bibnamefont {Kawakami}},\ }\href {\doibase
  10.1103/PhysRevLett.118.147001} {\bibfield  {journal} {\bibinfo  {journal}
  {Phys. Rev. Lett.}\ }\textbf {\bibinfo {volume} {118}},\ \bibinfo {pages}
  {147001} (\bibinfo {year} {2017})}\BibitemShut {NoStop}%
\bibitem [{Bau()}]{Bauer_Sigrist}%
  \BibitemOpen
  \href@noop {} {}\bibinfo {note} {\textit{Non-Centrosymmetric Superconductors:
  Introduction and Overview}, edited by E. Bauer and M. Sigrist, Lecture Notes
  in Physics Vol. 847 (Springer, Berlin, 2012)}\BibitemShut {NoStop}%
\bibitem [{\citenamefont {Settai}\ \emph {et~al.}(2007)\citenamefont {Settai},
  \citenamefont {Takeuchi},\ and\ \citenamefont {\=Onuki}}]{Settai_review}%
  \BibitemOpen
  \bibfield  {author} {\bibinfo {author} {\bibfnamefont {R.}~\bibnamefont
  {Settai}}, \bibinfo {author} {\bibfnamefont {T.}~\bibnamefont {Takeuchi}}, \
  and\ \bibinfo {author} {\bibfnamefont {Y.}~\bibnamefont {\=Onuki}},\ }\href
  {\doibase 10.1143/JPSJ.76.051003} {\bibfield  {journal} {\bibinfo  {journal}
  {J. Phys. Soc. Jpn.}\ }\textbf {\bibinfo {volume} {76}},\ \bibinfo {pages}
  {051003} (\bibinfo {year} {2007})}\BibitemShut {NoStop}%
\bibitem [{\citenamefont {Edelstein}(1995)}]{Edelstein_2}%
  \BibitemOpen
  \bibfield  {author} {\bibinfo {author} {\bibfnamefont {V.~M.}\ \bibnamefont
  {Edelstein}},\ }\href {\doibase 10.1103/PhysRevLett.75.2004} {\bibfield
  {journal} {\bibinfo  {journal} {Phys. Rev. Lett.}\ }\textbf {\bibinfo
  {volume} {75}},\ \bibinfo {pages} {2004} (\bibinfo {year}
  {1995})}\BibitemShut {NoStop}%
\bibitem [{\citenamefont {Gor'kov}\ and\ \citenamefont
  {Rashba}(2001)}]{Gorkov}%
  \BibitemOpen
  \bibfield  {author} {\bibinfo {author} {\bibfnamefont {L.~P.}\ \bibnamefont
  {Gor'kov}}\ and\ \bibinfo {author} {\bibfnamefont {E.~I.}\ \bibnamefont
  {Rashba}},\ }\href {\doibase 10.1103/PhysRevLett.87.037004} {\bibfield
  {journal} {\bibinfo  {journal} {Phys. Rev. Lett.}\ }\textbf {\bibinfo
  {volume} {87}},\ \bibinfo {pages} {037004} (\bibinfo {year}
  {2001})}\BibitemShut {NoStop}%
\bibitem [{\citenamefont {Frigeri}\ \emph {et~al.}(2004)\citenamefont
  {Frigeri}, \citenamefont {Agterberg}, \citenamefont {Koga},\ and\
  \citenamefont {Sigrist}}]{Frigeri_CePt3Si}%
  \BibitemOpen
  \bibfield  {author} {\bibinfo {author} {\bibfnamefont {P.~A.}\ \bibnamefont
  {Frigeri}}, \bibinfo {author} {\bibfnamefont {D.~F.}\ \bibnamefont
  {Agterberg}}, \bibinfo {author} {\bibfnamefont {A.}~\bibnamefont {Koga}}, \
  and\ \bibinfo {author} {\bibfnamefont {M.}~\bibnamefont {Sigrist}},\ }\href
  {\doibase 10.1103/PhysRevLett.92.097001} {\bibfield  {journal} {\bibinfo
  {journal} {Phys. Rev. Lett.}\ }\textbf {\bibinfo {volume} {92}},\ \bibinfo
  {pages} {097001} (\bibinfo {year} {2004})}\BibitemShut {NoStop}%
\bibitem [{\citenamefont {Samokhin}(2004)}]{Samokhin}%
  \BibitemOpen
  \bibfield  {author} {\bibinfo {author} {\bibfnamefont {K.~V.}\ \bibnamefont
  {Samokhin}},\ }\href {\doibase 10.1103/PhysRevB.70.104521} {\bibfield
  {journal} {\bibinfo  {journal} {Phys. Rev. B}\ }\textbf {\bibinfo {volume}
  {70}},\ \bibinfo {pages} {104521} (\bibinfo {year} {2004})}\BibitemShut
  {NoStop}%
\bibitem [{\citenamefont {Yanase}\ and\ \citenamefont
  {Sigrist}(2007)}]{Yanase_CePt3Si_1}%
  \BibitemOpen
  \bibfield  {author} {\bibinfo {author} {\bibfnamefont {Y.}~\bibnamefont
  {Yanase}}\ and\ \bibinfo {author} {\bibfnamefont {M.}~\bibnamefont
  {Sigrist}},\ }\href {\doibase 10.1143/JPSJ.76.043712} {\bibfield  {journal}
  {\bibinfo  {journal} {J. Phys. Soc. Jpn.}\ }\textbf {\bibinfo {volume}
  {76}},\ \bibinfo {pages} {043712} (\bibinfo {year} {2007})}\BibitemShut
  {NoStop}%
\bibitem [{\citenamefont {Yanase}\ and\ \citenamefont
  {Sigrist}(2008)}]{Yanase_CePt3Si_2}%
  \BibitemOpen
  \bibfield  {author} {\bibinfo {author} {\bibfnamefont {Y.}~\bibnamefont
  {Yanase}}\ and\ \bibinfo {author} {\bibfnamefont {M.}~\bibnamefont
  {Sigrist}},\ }\href {\doibase 10.1143/JPSJ.77.124711} {\bibfield  {journal}
  {\bibinfo  {journal} {J. Phys. Soc. Jpn.}\ }\textbf {\bibinfo {volume}
  {77}},\ \bibinfo {pages} {124711} (\bibinfo {year} {2008})}\BibitemShut
  {NoStop}%
\bibitem [{\citenamefont {Yokoyama}\ \emph {et~al.}(2007)\citenamefont
  {Yokoyama}, \citenamefont {Onari},\ and\ \citenamefont {Tanaka}}]{Yokoyama}%
  \BibitemOpen
  \bibfield  {author} {\bibinfo {author} {\bibfnamefont {T.}~\bibnamefont
  {Yokoyama}}, \bibinfo {author} {\bibfnamefont {S.}~\bibnamefont {Onari}}, \
  and\ \bibinfo {author} {\bibfnamefont {Y.}~\bibnamefont {Tanaka}},\ }\href
  {\doibase 10.1103/PhysRevB.75.172511} {\bibfield  {journal} {\bibinfo
  {journal} {Phys. Rev. B}\ }\textbf {\bibinfo {volume} {75}},\ \bibinfo
  {pages} {172511} (\bibinfo {year} {2007})}\BibitemShut {NoStop}%
\bibitem [{\citenamefont {Tada}\ \emph {et~al.}(2009)\citenamefont {Tada},
  \citenamefont {Kawakami},\ and\ \citenamefont {Fujimoto}}]{Tada}%
  \BibitemOpen
  \bibfield  {author} {\bibinfo {author} {\bibfnamefont {Y.}~\bibnamefont
  {Tada}}, \bibinfo {author} {\bibfnamefont {N.}~\bibnamefont {Kawakami}}, \
  and\ \bibinfo {author} {\bibfnamefont {S.}~\bibnamefont {Fujimoto}},\ }\href
  {http://stacks.iop.org/1367-2630/11/i=5/a=055070} {\bibfield  {journal}
  {\bibinfo  {journal} {New Journal of Physics}\ }\textbf {\bibinfo {volume}
  {11}},\ \bibinfo {pages} {055070} (\bibinfo {year} {2009})}\BibitemShut
  {NoStop}%
\bibitem [{\citenamefont {Agterberg}\ and\ \citenamefont
  {Kaur}(2007)}]{Agterberg_helical}%
  \BibitemOpen
  \bibfield  {author} {\bibinfo {author} {\bibfnamefont {D.~F.}\ \bibnamefont
  {Agterberg}}\ and\ \bibinfo {author} {\bibfnamefont {R.~P.}\ \bibnamefont
  {Kaur}},\ }\href {\doibase 10.1103/PhysRevB.75.064511} {\bibfield  {journal}
  {\bibinfo  {journal} {Phys. Rev. B}\ }\textbf {\bibinfo {volume} {75}},\
  \bibinfo {pages} {064511} (\bibinfo {year} {2007})}\BibitemShut {NoStop}%
\bibitem [{\citenamefont {Fujimoto}(2006)}]{Fujimoto_1}%
  \BibitemOpen
  \bibfield  {author} {\bibinfo {author} {\bibfnamefont {S.}~\bibnamefont
  {Fujimoto}},\ }\href {\doibase 10.1143/JPSJ.75.083704} {\bibfield  {journal}
  {\bibinfo  {journal} {J. Phys. Soc. Jpn.}\ }\textbf {\bibinfo {volume}
  {75}},\ \bibinfo {pages} {083704} (\bibinfo {year} {2006})}\BibitemShut
  {NoStop}%
\bibitem [{\citenamefont {Fujimoto}(2007)}]{Fujimoto_2}%
  \BibitemOpen
  \bibfield  {author} {\bibinfo {author} {\bibfnamefont {S.}~\bibnamefont
  {Fujimoto}},\ }\href {\doibase 10.1143/JPSJ.76.034712} {\bibfield  {journal}
  {\bibinfo  {journal} {J. Phys. Soc. Jpn.}\ }\textbf {\bibinfo {volume}
  {76}},\ \bibinfo {pages} {034712} (\bibinfo {year} {2007})}\BibitemShut
  {NoStop}%
\bibitem [{\citenamefont {Brydon}\ \emph {et~al.}(2016)\citenamefont {Brydon},
  \citenamefont {Wang}, \citenamefont {Weinert},\ and\ \citenamefont
  {Agterberg}}]{Brydon_YPtBi}%
  \BibitemOpen
  \bibfield  {author} {\bibinfo {author} {\bibfnamefont {P.~M.~R.}\
  \bibnamefont {Brydon}}, \bibinfo {author} {\bibfnamefont {L.}~\bibnamefont
  {Wang}}, \bibinfo {author} {\bibfnamefont {M.}~\bibnamefont {Weinert}}, \
  and\ \bibinfo {author} {\bibfnamefont {D.~F.}\ \bibnamefont {Agterberg}},\
  }\href {\doibase 10.1103/PhysRevLett.116.177001} {\bibfield  {journal}
  {\bibinfo  {journal} {Phys. Rev. Lett.}\ }\textbf {\bibinfo {volume} {116}},\
  \bibinfo {pages} {177001} (\bibinfo {year} {2016})}\BibitemShut {NoStop}%
\bibitem [{\citenamefont {Bauer}\ \emph {et~al.}(2004)\citenamefont {Bauer},
  \citenamefont {Hilscher}, \citenamefont {Michor}, \citenamefont {Paul},
  \citenamefont {Scheidt}, \citenamefont {Gribanov}, \citenamefont {Seropegin},
  \citenamefont {No\"el}, \citenamefont {Sigrist},\ and\ \citenamefont
  {Rogl}}]{Bauer_CePt3Si_1}%
  \BibitemOpen
  \bibfield  {author} {\bibinfo {author} {\bibfnamefont {E.}~\bibnamefont
  {Bauer}}, \bibinfo {author} {\bibfnamefont {G.}~\bibnamefont {Hilscher}},
  \bibinfo {author} {\bibfnamefont {H.}~\bibnamefont {Michor}}, \bibinfo
  {author} {\bibfnamefont {C.}~\bibnamefont {Paul}}, \bibinfo {author}
  {\bibfnamefont {E.~W.}\ \bibnamefont {Scheidt}}, \bibinfo {author}
  {\bibfnamefont {A.}~\bibnamefont {Gribanov}}, \bibinfo {author}
  {\bibfnamefont {Y.}~\bibnamefont {Seropegin}}, \bibinfo {author}
  {\bibfnamefont {H.}~\bibnamefont {No\"el}}, \bibinfo {author} {\bibfnamefont
  {M.}~\bibnamefont {Sigrist}}, \ and\ \bibinfo {author} {\bibfnamefont
  {P.}~\bibnamefont {Rogl}},\ }\href {\doibase 10.1103/PhysRevLett.92.027003}
  {\bibfield  {journal} {\bibinfo  {journal} {Phys. Rev. Lett.}\ }\textbf
  {\bibinfo {volume} {92}},\ \bibinfo {pages} {027003} (\bibinfo {year}
  {2004})}\BibitemShut {NoStop}%
\bibitem [{\citenamefont {Sugitani}\ \emph {et~al.}(2006)\citenamefont
  {Sugitani}, \citenamefont {Okuda}, \citenamefont {Shishido}, \citenamefont
  {Yamada}, \citenamefont {Thamizhavel}, \citenamefont {Yamamoto},
  \citenamefont {Matsuda}, \citenamefont {Haga}, \citenamefont {Takeuchi},
  \citenamefont {Settai},\ and\ \citenamefont {\=Onuki}}]{Sugitani_CeIrSi3}%
  \BibitemOpen
  \bibfield  {author} {\bibinfo {author} {\bibfnamefont {I.}~\bibnamefont
  {Sugitani}}, \bibinfo {author} {\bibfnamefont {Y.}~\bibnamefont {Okuda}},
  \bibinfo {author} {\bibfnamefont {H.}~\bibnamefont {Shishido}}, \bibinfo
  {author} {\bibfnamefont {T.}~\bibnamefont {Yamada}}, \bibinfo {author}
  {\bibfnamefont {A.}~\bibnamefont {Thamizhavel}}, \bibinfo {author}
  {\bibfnamefont {E.}~\bibnamefont {Yamamoto}}, \bibinfo {author}
  {\bibfnamefont {T.~D.}\ \bibnamefont {Matsuda}}, \bibinfo {author}
  {\bibfnamefont {Y.}~\bibnamefont {Haga}}, \bibinfo {author} {\bibfnamefont
  {T.}~\bibnamefont {Takeuchi}}, \bibinfo {author} {\bibfnamefont
  {R.}~\bibnamefont {Settai}}, \ and\ \bibinfo {author} {\bibfnamefont
  {Y.}~\bibnamefont {\=Onuki}},\ }\href {\doibase 10.1143/JPSJ.75.043703}
  {\bibfield  {journal} {\bibinfo  {journal} {J. Phys. Soc. Jpn.}\ }\textbf
  {\bibinfo {volume} {75}},\ \bibinfo {pages} {043703} (\bibinfo {year}
  {2006})}\BibitemShut {NoStop}%
\bibitem [{\citenamefont {Yanase}(2014)}]{Yanase_zigzag}%
  \BibitemOpen
  \bibfield  {author} {\bibinfo {author} {\bibfnamefont {Y.}~\bibnamefont
  {Yanase}},\ }\href {\doibase 10.7566/JPSJ.83.014703} {\bibfield  {journal}
  {\bibinfo  {journal} {J. Phys. Soc. Jpn.}\ }\textbf {\bibinfo {volume}
  {83}},\ \bibinfo {pages} {014703} (\bibinfo {year} {2014})}\BibitemShut
  {NoStop}%
\bibitem [{\citenamefont {\ifmmode~\check{Z}\else \v{Z}\fi{}elezn\'y}\ \emph
  {et~al.}(2014)\citenamefont {\ifmmode~\check{Z}\else \v{Z}\fi{}elezn\'y},
  \citenamefont {Gao}, \citenamefont {V\'yborn\'y}, \citenamefont {Zemen},
  \citenamefont {Ma\ifmmode~\check{s}\else \v{s}\fi{}ek}, \citenamefont
  {Manchon}, \citenamefont {Wunderlich}, \citenamefont {Sinova},\ and\
  \citenamefont {Jungwirth}}]{Zelezny_Mn2Au}%
  \BibitemOpen
  \bibfield  {author} {\bibinfo {author} {\bibfnamefont {J.}~\bibnamefont
  {\ifmmode~\check{Z}\else \v{Z}\fi{}elezn\'y}}, \bibinfo {author}
  {\bibfnamefont {H.}~\bibnamefont {Gao}}, \bibinfo {author} {\bibfnamefont
  {K.}~\bibnamefont {V\'yborn\'y}}, \bibinfo {author} {\bibfnamefont
  {J.}~\bibnamefont {Zemen}}, \bibinfo {author} {\bibfnamefont
  {J.}~\bibnamefont {Ma\ifmmode~\check{s}\else \v{s}\fi{}ek}}, \bibinfo
  {author} {\bibfnamefont {A.}~\bibnamefont {Manchon}}, \bibinfo {author}
  {\bibfnamefont {J.}~\bibnamefont {Wunderlich}}, \bibinfo {author}
  {\bibfnamefont {J.}~\bibnamefont {Sinova}}, \ and\ \bibinfo {author}
  {\bibfnamefont {T.}~\bibnamefont {Jungwirth}},\ }\href {\doibase
  10.1103/PhysRevLett.113.157201} {\bibfield  {journal} {\bibinfo  {journal}
  {Phys. Rev. Lett.}\ }\textbf {\bibinfo {volume} {113}},\ \bibinfo {pages}
  {157201} (\bibinfo {year} {2014})}\BibitemShut {NoStop}%
\bibitem [{\citenamefont {Wadley}\ \emph {et~al.}(2016)\citenamefont {Wadley},
  \citenamefont {Howells}, \citenamefont {{\v Z}elezn{\'y}}, \citenamefont
  {Andrews}, \citenamefont {Hills}, \citenamefont {Campion}, \citenamefont
  {Nov{\'a}k}, \citenamefont {Olejn{\'\i}k}, \citenamefont {Maccherozzi},
  \citenamefont {Dhesi}, \citenamefont {Martin}, \citenamefont {Wagner},
  \citenamefont {Wunderlich}, \citenamefont {Freimuth}, \citenamefont
  {Mokrousov}, \citenamefont {Kune{\v s}}, \citenamefont {Chauhan},
  \citenamefont {Grzybowski}, \citenamefont {Rushforth}, \citenamefont
  {Edmonds}, \citenamefont {Gallagher},\ and\ \citenamefont
  {Jungwirth}}]{Wadley}%
  \BibitemOpen
  \bibfield  {author} {\bibinfo {author} {\bibfnamefont {P.}~\bibnamefont
  {Wadley}}, \bibinfo {author} {\bibfnamefont {B.}~\bibnamefont {Howells}},
  \bibinfo {author} {\bibfnamefont {J.}~\bibnamefont {{\v Z}elezn{\'y}}},
  \bibinfo {author} {\bibfnamefont {C.}~\bibnamefont {Andrews}}, \bibinfo
  {author} {\bibfnamefont {V.}~\bibnamefont {Hills}}, \bibinfo {author}
  {\bibfnamefont {R.~P.}\ \bibnamefont {Campion}}, \bibinfo {author}
  {\bibfnamefont {V.}~\bibnamefont {Nov{\'a}k}}, \bibinfo {author}
  {\bibfnamefont {K.}~\bibnamefont {Olejn{\'\i}k}}, \bibinfo {author}
  {\bibfnamefont {F.}~\bibnamefont {Maccherozzi}}, \bibinfo {author}
  {\bibfnamefont {S.~S.}\ \bibnamefont {Dhesi}}, \bibinfo {author}
  {\bibfnamefont {S.~Y.}\ \bibnamefont {Martin}}, \bibinfo {author}
  {\bibfnamefont {T.}~\bibnamefont {Wagner}}, \bibinfo {author} {\bibfnamefont
  {J.}~\bibnamefont {Wunderlich}}, \bibinfo {author} {\bibfnamefont
  {F.}~\bibnamefont {Freimuth}}, \bibinfo {author} {\bibfnamefont
  {Y.}~\bibnamefont {Mokrousov}}, \bibinfo {author} {\bibfnamefont
  {J.}~\bibnamefont {Kune{\v s}}}, \bibinfo {author} {\bibfnamefont {J.~S.}\
  \bibnamefont {Chauhan}}, \bibinfo {author} {\bibfnamefont {M.~J.}\
  \bibnamefont {Grzybowski}}, \bibinfo {author} {\bibfnamefont {A.~W.}\
  \bibnamefont {Rushforth}}, \bibinfo {author} {\bibfnamefont {K.~W.}\
  \bibnamefont {Edmonds}}, \bibinfo {author} {\bibfnamefont {B.~L.}\
  \bibnamefont {Gallagher}}, \ and\ \bibinfo {author} {\bibfnamefont
  {T.}~\bibnamefont {Jungwirth}},\ }\href {\doibase 10.1126/science.aab1031}
  {\bibfield  {journal} {\bibinfo  {journal} {Science}\ }\textbf {\bibinfo
  {volume} {351}},\ \bibinfo {pages} {587} (\bibinfo {year}
  {2016})}\BibitemShut {NoStop}%
\bibitem [{\citenamefont {Watanabe}\ and\ \citenamefont
  {Yanase}(2017)}]{Watanabe_mpole}%
  \BibitemOpen
  \bibfield  {author} {\bibinfo {author} {\bibfnamefont {H.}~\bibnamefont
  {Watanabe}}\ and\ \bibinfo {author} {\bibfnamefont {Y.}~\bibnamefont
  {Yanase}},\ }\href {\doibase 10.1103/PhysRevB.96.064432} {\bibfield
  {journal} {\bibinfo  {journal} {Phys. Rev. B}\ }\textbf {\bibinfo {volume}
  {96}},\ \bibinfo {pages} {064432} (\bibinfo {year} {2017})}\BibitemShut
  {NoStop}%
\bibitem [{\citenamefont {Kuroki}\ \emph {et~al.}(2008)\citenamefont {Kuroki},
  \citenamefont {Onari}, \citenamefont {Arita}, \citenamefont {Usui},
  \citenamefont {Tanaka}, \citenamefont {Kontani},\ and\ \citenamefont
  {Aoki}}]{Kuroki_1}%
  \BibitemOpen
  \bibfield  {author} {\bibinfo {author} {\bibfnamefont {K.}~\bibnamefont
  {Kuroki}}, \bibinfo {author} {\bibfnamefont {S.}~\bibnamefont {Onari}},
  \bibinfo {author} {\bibfnamefont {R.}~\bibnamefont {Arita}}, \bibinfo
  {author} {\bibfnamefont {H.}~\bibnamefont {Usui}}, \bibinfo {author}
  {\bibfnamefont {Y.}~\bibnamefont {Tanaka}}, \bibinfo {author} {\bibfnamefont
  {H.}~\bibnamefont {Kontani}}, \ and\ \bibinfo {author} {\bibfnamefont
  {H.}~\bibnamefont {Aoki}},\ }\href {\doibase 10.1103/PhysRevLett.101.087004}
  {\bibfield  {journal} {\bibinfo  {journal} {Phys. Rev. Lett.}\ }\textbf
  {\bibinfo {volume} {101}},\ \bibinfo {pages} {087004} (\bibinfo {year}
  {2008})}\BibitemShut {NoStop}%
\bibitem [{\citenamefont {Kamihara}\ \emph {et~al.}(2008)\citenamefont
  {Kamihara}, \citenamefont {Watanabe}, \citenamefont {Hirano},\ and\
  \citenamefont {Hosono}}]{Kamihara_2}%
  \BibitemOpen
  \bibfield  {author} {\bibinfo {author} {\bibfnamefont {Y.}~\bibnamefont
  {Kamihara}}, \bibinfo {author} {\bibfnamefont {T.}~\bibnamefont {Watanabe}},
  \bibinfo {author} {\bibfnamefont {M.}~\bibnamefont {Hirano}}, \ and\ \bibinfo
  {author} {\bibfnamefont {H.}~\bibnamefont {Hosono}},\ }\href {\doibase
  10.1021/ja800073m} {\bibfield  {journal} {\bibinfo  {journal} {J. Am. Chem.
  Soc.}\ }\textbf {\bibinfo {volume} {130}},\ \bibinfo {pages} {3296} (\bibinfo
  {year} {2008})}\BibitemShut {NoStop}%
\bibitem [{\citenamefont {Mizukami}\ \emph {et~al.}(2011)\citenamefont
  {Mizukami}, \citenamefont {Shishido}, \citenamefont {Shibauchi},
  \citenamefont {Shimozawa}, \citenamefont {Yasumoto}, \citenamefont
  {Watanabe}, \citenamefont {Yamashita}, \citenamefont {Ikeda}, \citenamefont
  {Terashima}, \citenamefont {Kontani},\ and\ \citenamefont
  {Matsuda}}]{Mizukami_CeCoIn5}%
  \BibitemOpen
  \bibfield  {author} {\bibinfo {author} {\bibfnamefont {Y.}~\bibnamefont
  {Mizukami}}, \bibinfo {author} {\bibfnamefont {H.}~\bibnamefont {Shishido}},
  \bibinfo {author} {\bibfnamefont {T.}~\bibnamefont {Shibauchi}}, \bibinfo
  {author} {\bibfnamefont {M.}~\bibnamefont {Shimozawa}}, \bibinfo {author}
  {\bibfnamefont {S.}~\bibnamefont {Yasumoto}}, \bibinfo {author}
  {\bibfnamefont {D.}~\bibnamefont {Watanabe}}, \bibinfo {author}
  {\bibfnamefont {M.}~\bibnamefont {Yamashita}}, \bibinfo {author}
  {\bibfnamefont {H.}~\bibnamefont {Ikeda}}, \bibinfo {author} {\bibfnamefont
  {T.}~\bibnamefont {Terashima}}, \bibinfo {author} {\bibfnamefont
  {H.}~\bibnamefont {Kontani}}, \ and\ \bibinfo {author} {\bibfnamefont
  {Y.}~\bibnamefont {Matsuda}},\ }\href {http://dx.doi.org/10.1038/nphys2112}
  {\bibfield  {journal} {\bibinfo  {journal} {Nat. Phys.}\ }\textbf {\bibinfo
  {volume} {7}},\ \bibinfo {pages} {849} (\bibinfo {year} {2011})}\BibitemShut
  {NoStop}%
\bibitem [{\citenamefont {Sumita}\ and\ \citenamefont
  {Yanase}(2016)}]{Sumita_zigzag}%
  \BibitemOpen
  \bibfield  {author} {\bibinfo {author} {\bibfnamefont {S.}~\bibnamefont
  {Sumita}}\ and\ \bibinfo {author} {\bibfnamefont {Y.}~\bibnamefont
  {Yanase}},\ }\href {\doibase 10.1103/PhysRevB.93.224507} {\bibfield
  {journal} {\bibinfo  {journal} {Phys. Rev. B}\ }\textbf {\bibinfo {volume}
  {93}},\ \bibinfo {pages} {224507} (\bibinfo {year} {2016})}\BibitemShut
  {NoStop}%
\bibitem [{\citenamefont {Fu}(2015)}]{Fu_Multipole}%
  \BibitemOpen
  \bibfield  {author} {\bibinfo {author} {\bibfnamefont {L.}~\bibnamefont
  {Fu}},\ }\href {\doibase 10.1103/PhysRevLett.115.026401} {\bibfield
  {journal} {\bibinfo  {journal} {Phys. Rev. Lett.}\ }\textbf {\bibinfo
  {volume} {115}},\ \bibinfo {pages} {026401} (\bibinfo {year}
  {2015})}\BibitemShut {NoStop}%
\bibitem [{\citenamefont {Hitomi}\ and\ \citenamefont
  {Yanase}(2014)}]{Hitomi_Sr3Ru2O7}%
  \BibitemOpen
  \bibfield  {author} {\bibinfo {author} {\bibfnamefont {T.}~\bibnamefont
  {Hitomi}}\ and\ \bibinfo {author} {\bibfnamefont {Y.}~\bibnamefont
  {Yanase}},\ }\href {\doibase 10.7566/JPSJ.83.114704} {\bibfield  {journal}
  {\bibinfo  {journal} {J. Phys. Soc. Jpn.}\ }\textbf {\bibinfo {volume}
  {83}},\ \bibinfo {pages} {114704} (\bibinfo {year} {2014})}\BibitemShut
  {NoStop}%
\bibitem [{\citenamefont {Hitomi}\ and\ \citenamefont
  {Yanase}(2016)}]{Hitomi_bilayer}%
  \BibitemOpen
  \bibfield  {author} {\bibinfo {author} {\bibfnamefont {T.}~\bibnamefont
  {Hitomi}}\ and\ \bibinfo {author} {\bibfnamefont {Y.}~\bibnamefont
  {Yanase}},\ }\href {\doibase 10.7566/JPSJ.85.124702} {\bibfield  {journal}
  {\bibinfo  {journal} {J. Phys. Soc. Jpn.}\ }\textbf {\bibinfo {volume}
  {85}},\ \bibinfo {pages} {124702} (\bibinfo {year} {2016})}\BibitemShut
  {NoStop}%
\bibitem [{\citenamefont {Hayami}\ \emph
  {et~al.}(2014{\natexlab{a}})\citenamefont {Hayami}, \citenamefont
  {Kusunose},\ and\ \citenamefont {Motome}}]{Hayami_Toroidal}%
  \BibitemOpen
  \bibfield  {author} {\bibinfo {author} {\bibfnamefont {S.}~\bibnamefont
  {Hayami}}, \bibinfo {author} {\bibfnamefont {H.}~\bibnamefont {Kusunose}}, \
  and\ \bibinfo {author} {\bibfnamefont {Y.}~\bibnamefont {Motome}},\ }\href
  {\doibase 10.1103/PhysRevB.90.024432} {\bibfield  {journal} {\bibinfo
  {journal} {Phys. Rev. B}\ }\textbf {\bibinfo {volume} {90}},\ \bibinfo
  {pages} {024432} (\bibinfo {year} {2014}{\natexlab{a}})}\BibitemShut
  {NoStop}%
\bibitem [{\citenamefont {Hayami}\ \emph
  {et~al.}(2014{\natexlab{b}})\citenamefont {Hayami}, \citenamefont
  {Kusunose},\ and\ \citenamefont {Motome}}]{Hayami_valley}%
  \BibitemOpen
  \bibfield  {author} {\bibinfo {author} {\bibfnamefont {S.}~\bibnamefont
  {Hayami}}, \bibinfo {author} {\bibfnamefont {H.}~\bibnamefont {Kusunose}}, \
  and\ \bibinfo {author} {\bibfnamefont {Y.}~\bibnamefont {Motome}},\ }\href
  {\doibase 10.1103/PhysRevB.90.081115} {\bibfield  {journal} {\bibinfo
  {journal} {Phys. Rev. B}\ }\textbf {\bibinfo {volume} {90}},\ \bibinfo
  {pages} {081115} (\bibinfo {year} {2014}{\natexlab{b}})}\BibitemShut
  {NoStop}%
\bibitem [{\citenamefont {Hayami}\ \emph {et~al.}(2015)\citenamefont {Hayami},
  \citenamefont {Kusunose},\ and\ \citenamefont {Motome}}]{Hayami_zigzag}%
  \BibitemOpen
  \bibfield  {author} {\bibinfo {author} {\bibfnamefont {S.}~\bibnamefont
  {Hayami}}, \bibinfo {author} {\bibfnamefont {H.}~\bibnamefont {Kusunose}}, \
  and\ \bibinfo {author} {\bibfnamefont {Y.}~\bibnamefont {Motome}},\ }\href
  {\doibase 10.7566/JPSJ.84.064717} {\bibfield  {journal} {\bibinfo  {journal}
  {J. Phys. Soc. Jpn.}\ }\textbf {\bibinfo {volume} {84}},\ \bibinfo {pages}
  {064717} (\bibinfo {year} {2015})}\BibitemShut {NoStop}%
\bibitem [{\citenamefont {Sumita}\ \emph {et~al.}(2017)\citenamefont {Sumita},
  \citenamefont {Nomoto},\ and\ \citenamefont {Yanase}}]{Sumita_Ir}%
  \BibitemOpen
  \bibfield  {author} {\bibinfo {author} {\bibfnamefont {S.}~\bibnamefont
  {Sumita}}, \bibinfo {author} {\bibfnamefont {T.}~\bibnamefont {Nomoto}}, \
  and\ \bibinfo {author} {\bibfnamefont {Y.}~\bibnamefont {Yanase}},\ }\href
  {\doibase 10.1103/PhysRevLett.119.027001} {\bibfield  {journal} {\bibinfo
  {journal} {Phys. Rev. Lett.}\ }\textbf {\bibinfo {volume} {119}},\ \bibinfo
  {pages} {027001} (\bibinfo {year} {2017})}\BibitemShut {NoStop}%
\bibitem [{\citenamefont {Di~Matteo}\ and\ \citenamefont
  {Norman}(2016)}]{Matteo_Sr2IrO4}%
  \BibitemOpen
  \bibfield  {author} {\bibinfo {author} {\bibfnamefont {S.}~\bibnamefont
  {Di~Matteo}}\ and\ \bibinfo {author} {\bibfnamefont {M.~R.}\ \bibnamefont
  {Norman}},\ }\href {\doibase 10.1103/PhysRevB.94.075148} {\bibfield
  {journal} {\bibinfo  {journal} {Phys. Rev. B}\ }\textbf {\bibinfo {volume}
  {94}},\ \bibinfo {pages} {075148} (\bibinfo {year} {2016})}\BibitemShut
  {NoStop}%
\bibitem [{\citenamefont {Hayami}\ \emph {et~al.}(2018)\citenamefont {Hayami},
  \citenamefont {Kusunose},\ and\ \citenamefont {Motome}}]{Hayami_AOsO4}%
  \BibitemOpen
  \bibfield  {author} {\bibinfo {author} {\bibfnamefont {S.}~\bibnamefont
  {Hayami}}, \bibinfo {author} {\bibfnamefont {H.}~\bibnamefont {Kusunose}}, \
  and\ \bibinfo {author} {\bibfnamefont {Y.}~\bibnamefont {Motome}},\ }\href
  {\doibase 10.1103/PhysRevB.97.024414} {\bibfield  {journal} {\bibinfo
  {journal} {Phys. Rev. B}\ }\textbf {\bibinfo {volume} {97}},\ \bibinfo
  {pages} {024414} (\bibinfo {year} {2018})}\BibitemShut {NoStop}%
\bibitem [{\citenamefont {Yanagi}\ \emph {et~al.}(2018)\citenamefont {Yanagi},
  \citenamefont {Hayami},\ and\ \citenamefont {Kusunose}}]{Yanagi_Co4Nb2O9}%
  \BibitemOpen
  \bibfield  {author} {\bibinfo {author} {\bibfnamefont {Y.}~\bibnamefont
  {Yanagi}}, \bibinfo {author} {\bibfnamefont {S.}~\bibnamefont {Hayami}}, \
  and\ \bibinfo {author} {\bibfnamefont {H.}~\bibnamefont {Kusunose}},\ }\href
  {\doibase 10.1103/PhysRevB.97.020404} {\bibfield  {journal} {\bibinfo
  {journal} {Phys. Rev. B}\ }\textbf {\bibinfo {volume} {97}},\ \bibinfo
  {pages} {020404} (\bibinfo {year} {2018})}\BibitemShut {NoStop}%
\bibitem [{\citenamefont {Hayami}\ and\ \citenamefont
  {Kusunose}(2018)}]{Hayami_mpole_hyb}%
  \BibitemOpen
  \bibfield  {author} {\bibinfo {author} {\bibfnamefont {S.}~\bibnamefont
  {Hayami}}\ and\ \bibinfo {author} {\bibfnamefont {H.}~\bibnamefont
  {Kusunose}},\ }\href {\doibase 10.7566/JPSJ.87.033709} {\bibfield  {journal}
  {\bibinfo  {journal} {J. Phys. Soc. Jpn.}\ }\textbf {\bibinfo {volume}
  {87}},\ \bibinfo {pages} {033709} (\bibinfo {year} {2018})}\BibitemShut
  {NoStop}%
\bibitem [{\citenamefont {Kuramoto}\ \emph {et~al.}(2009)\citenamefont
  {Kuramoto}, \citenamefont {Kusunose},\ and\ \citenamefont
  {Kiss}}]{Kuramoto_multipole}%
  \BibitemOpen
  \bibfield  {author} {\bibinfo {author} {\bibfnamefont {Y.}~\bibnamefont
  {Kuramoto}}, \bibinfo {author} {\bibfnamefont {H.}~\bibnamefont {Kusunose}},
  \ and\ \bibinfo {author} {\bibfnamefont {A.}~\bibnamefont {Kiss}},\ }\href
  {\doibase 10.1143/JPSJ.78.072001} {\bibfield  {journal} {\bibinfo  {journal}
  {J. Phys. Soc. Jpn.}\ }\textbf {\bibinfo {volume} {78}},\ \bibinfo {pages}
  {072001} (\bibinfo {year} {2009})}\BibitemShut {NoStop}%
\bibitem [{\citenamefont {Haule}\ and\ \citenamefont
  {Kotliar}(2009)}]{Haule_URu2Si2}%
  \BibitemOpen
  \bibfield  {author} {\bibinfo {author} {\bibfnamefont {K.}~\bibnamefont
  {Haule}}\ and\ \bibinfo {author} {\bibfnamefont {G.}~\bibnamefont
  {Kotliar}},\ }\href@noop {} {\bibfield  {journal} {\bibinfo  {journal} {Nat.
  Phys.}\ }\textbf {\bibinfo {volume} {5}},\ \bibinfo {pages} {796} (\bibinfo
  {year} {2009})}\BibitemShut {NoStop}%
\bibitem [{\citenamefont {Kusunose}\ and\ \citenamefont
  {Harima}(2011)}]{Kusunose_URu2Si2}%
  \BibitemOpen
  \bibfield  {author} {\bibinfo {author} {\bibfnamefont {H.}~\bibnamefont
  {Kusunose}}\ and\ \bibinfo {author} {\bibfnamefont {H.}~\bibnamefont
  {Harima}},\ }\href {\doibase 10.1143/JPSJ.80.084702} {\bibfield  {journal}
  {\bibinfo  {journal} {J. Phys. Soc. Jpn.}\ }\textbf {\bibinfo {volume}
  {80}},\ \bibinfo {pages} {084702} (\bibinfo {year} {2011})}\BibitemShut
  {NoStop}%
\bibitem [{\citenamefont {Ikeda}\ \emph {et~al.}(2012)\citenamefont {Ikeda},
  \citenamefont {Suzuki}, \citenamefont {Arita}, \citenamefont {Takimoto},
  \citenamefont {Shibauchi},\ and\ \citenamefont {Matsuda}}]{Ikeda_URu2Si2}%
  \BibitemOpen
  \bibfield  {author} {\bibinfo {author} {\bibfnamefont {H.}~\bibnamefont
  {Ikeda}}, \bibinfo {author} {\bibfnamefont {M.-T.}\ \bibnamefont {Suzuki}},
  \bibinfo {author} {\bibfnamefont {R.}~\bibnamefont {Arita}}, \bibinfo
  {author} {\bibfnamefont {T.}~\bibnamefont {Takimoto}}, \bibinfo {author}
  {\bibfnamefont {T.}~\bibnamefont {Shibauchi}}, \ and\ \bibinfo {author}
  {\bibfnamefont {Y.}~\bibnamefont {Matsuda}},\ }\href
  {http://dx.doi.org/10.1038/nphys2330} {\bibfield  {journal} {\bibinfo
  {journal} {Nat. Phys.}\ }\textbf {\bibinfo {volume} {8}},\ \bibinfo {pages}
  {528} (\bibinfo {year} {2012})}\BibitemShut {NoStop}%
\bibitem [{\citenamefont {Singh}\ \emph {et~al.}(2009)\citenamefont {Singh},
  \citenamefont {Green}, \citenamefont {Huang}, \citenamefont {Kreyssig},
  \citenamefont {McQueeney}, \citenamefont {Johnston},\ and\ \citenamefont
  {Goldman}}]{Singh_BaMn2As2}%
  \BibitemOpen
  \bibfield  {author} {\bibinfo {author} {\bibfnamefont {Y.}~\bibnamefont
  {Singh}}, \bibinfo {author} {\bibfnamefont {M.~A.}\ \bibnamefont {Green}},
  \bibinfo {author} {\bibfnamefont {Q.}~\bibnamefont {Huang}}, \bibinfo
  {author} {\bibfnamefont {A.}~\bibnamefont {Kreyssig}}, \bibinfo {author}
  {\bibfnamefont {R.~J.}\ \bibnamefont {McQueeney}}, \bibinfo {author}
  {\bibfnamefont {D.~C.}\ \bibnamefont {Johnston}}, \ and\ \bibinfo {author}
  {\bibfnamefont {A.~I.}\ \bibnamefont {Goldman}},\ }\href {\doibase
  10.1103/PhysRevB.80.100403} {\bibfield  {journal} {\bibinfo  {journal} {Phys.
  Rev. B}\ }\textbf {\bibinfo {volume} {80}},\ \bibinfo {pages} {100403}
  (\bibinfo {year} {2009})}\BibitemShut {NoStop}%
\bibitem [{\citenamefont {Zhao}\ \emph {et~al.}(2015)\citenamefont {Zhao},
  \citenamefont {Torchinsky}, \citenamefont {Chu}, \citenamefont {Ivanov},
  \citenamefont {Lifshitz}, \citenamefont {Flint}, \citenamefont {Qi},
  \citenamefont {Cao},\ and\ \citenamefont {Hsieh}}]{Zhao_Sr2IrO4}%
  \BibitemOpen
  \bibfield  {author} {\bibinfo {author} {\bibfnamefont {L.}~\bibnamefont
  {Zhao}}, \bibinfo {author} {\bibfnamefont {D.~H.}\ \bibnamefont
  {Torchinsky}}, \bibinfo {author} {\bibfnamefont {H.}~\bibnamefont {Chu}},
  \bibinfo {author} {\bibfnamefont {V.}~\bibnamefont {Ivanov}}, \bibinfo
  {author} {\bibfnamefont {R.}~\bibnamefont {Lifshitz}}, \bibinfo {author}
  {\bibfnamefont {R.}~\bibnamefont {Flint}}, \bibinfo {author} {\bibfnamefont
  {T.}~\bibnamefont {Qi}}, \bibinfo {author} {\bibfnamefont {G.}~\bibnamefont
  {Cao}}, \ and\ \bibinfo {author} {\bibfnamefont {D.}~\bibnamefont {Hsieh}},\
  }\href {http://dx.doi.org/10.1038/nphys3517} {\bibfield  {journal} {\bibinfo
  {journal} {Nat. Phys.}\ }\textbf {\bibinfo {volume} {12}},\ \bibinfo {pages}
  {32} (\bibinfo {year} {2015})}\BibitemShut {NoStop}%
\bibitem [{\citenamefont {Kim}\ \emph {et~al.}(2008)\citenamefont {Kim},
  \citenamefont {Jin}, \citenamefont {Moon}, \citenamefont {Kim}, \citenamefont
  {Park}, \citenamefont {Leem}, \citenamefont {Yu}, \citenamefont {Noh},
  \citenamefont {Kim}, \citenamefont {Oh}, \citenamefont {Park}, \citenamefont
  {Durairaj}, \citenamefont {Cao},\ and\ \citenamefont
  {Rotenberg}}]{B.J.Kim_Sr2IrO4_2008}%
  \BibitemOpen
  \bibfield  {author} {\bibinfo {author} {\bibfnamefont {B.~J.}\ \bibnamefont
  {Kim}}, \bibinfo {author} {\bibfnamefont {H.}~\bibnamefont {Jin}}, \bibinfo
  {author} {\bibfnamefont {S.~J.}\ \bibnamefont {Moon}}, \bibinfo {author}
  {\bibfnamefont {J.-Y.}\ \bibnamefont {Kim}}, \bibinfo {author} {\bibfnamefont
  {B.-G.}\ \bibnamefont {Park}}, \bibinfo {author} {\bibfnamefont {C.~S.}\
  \bibnamefont {Leem}}, \bibinfo {author} {\bibfnamefont {J.}~\bibnamefont
  {Yu}}, \bibinfo {author} {\bibfnamefont {T.~W.}\ \bibnamefont {Noh}},
  \bibinfo {author} {\bibfnamefont {C.}~\bibnamefont {Kim}}, \bibinfo {author}
  {\bibfnamefont {S.-J.}\ \bibnamefont {Oh}}, \bibinfo {author} {\bibfnamefont
  {J.-H.}\ \bibnamefont {Park}}, \bibinfo {author} {\bibfnamefont
  {V.}~\bibnamefont {Durairaj}}, \bibinfo {author} {\bibfnamefont
  {G.}~\bibnamefont {Cao}}, \ and\ \bibinfo {author} {\bibfnamefont
  {E.}~\bibnamefont {Rotenberg}},\ }\href {\doibase
  10.1103/PhysRevLett.101.076402} {\bibfield  {journal} {\bibinfo  {journal}
  {Phys. Rev. Lett.}\ }\textbf {\bibinfo {volume} {101}},\ \bibinfo {pages}
  {076402} (\bibinfo {year} {2008})}\BibitemShut {NoStop}%
\bibitem [{\citenamefont {Kim}\ \emph {et~al.}(2014)\citenamefont {Kim},
  \citenamefont {Krupin}, \citenamefont {Denlinger}, \citenamefont {Bostwick},
  \citenamefont {Rotenberg}, \citenamefont {Zhao}, \citenamefont {Mitchell},
  \citenamefont {Allen},\ and\ \citenamefont {Kim}}]{Y.K.Kim_Sr2IrO4_2014}%
  \BibitemOpen
  \bibfield  {author} {\bibinfo {author} {\bibfnamefont {Y.~K.}\ \bibnamefont
  {Kim}}, \bibinfo {author} {\bibfnamefont {O.}~\bibnamefont {Krupin}},
  \bibinfo {author} {\bibfnamefont {J.~D.}\ \bibnamefont {Denlinger}}, \bibinfo
  {author} {\bibfnamefont {A.}~\bibnamefont {Bostwick}}, \bibinfo {author}
  {\bibfnamefont {E.}~\bibnamefont {Rotenberg}}, \bibinfo {author}
  {\bibfnamefont {Q.}~\bibnamefont {Zhao}}, \bibinfo {author} {\bibfnamefont
  {J.~F.}\ \bibnamefont {Mitchell}}, \bibinfo {author} {\bibfnamefont {J.~W.}\
  \bibnamefont {Allen}}, \ and\ \bibinfo {author} {\bibfnamefont {B.~J.}\
  \bibnamefont {Kim}},\ }\href {\doibase 10.1126/science.1251151} {\bibfield
  {journal} {\bibinfo  {journal} {Science}\ }\textbf {\bibinfo {volume}
  {345}},\ \bibinfo {pages} {187} (\bibinfo {year} {2014})}\BibitemShut
  {NoStop}%
\bibitem [{\citenamefont {Yan}\ \emph {et~al.}(2015)\citenamefont {Yan},
  \citenamefont {Ren}, \citenamefont {Xu}, \citenamefont {Xie}, \citenamefont
  {Tao}, \citenamefont {Choi}, \citenamefont {Lee}, \citenamefont {Choi},
  \citenamefont {Zhang},\ and\ \citenamefont {Feng}}]{Yan_Sr2IrO4}%
  \BibitemOpen
  \bibfield  {author} {\bibinfo {author} {\bibfnamefont {Y.~J.}\ \bibnamefont
  {Yan}}, \bibinfo {author} {\bibfnamefont {M.~Q.}\ \bibnamefont {Ren}},
  \bibinfo {author} {\bibfnamefont {H.~C.}\ \bibnamefont {Xu}}, \bibinfo
  {author} {\bibfnamefont {B.~P.}\ \bibnamefont {Xie}}, \bibinfo {author}
  {\bibfnamefont {R.}~\bibnamefont {Tao}}, \bibinfo {author} {\bibfnamefont
  {H.~Y.}\ \bibnamefont {Choi}}, \bibinfo {author} {\bibfnamefont
  {N.}~\bibnamefont {Lee}}, \bibinfo {author} {\bibfnamefont {Y.~J.}\
  \bibnamefont {Choi}}, \bibinfo {author} {\bibfnamefont {T.}~\bibnamefont
  {Zhang}}, \ and\ \bibinfo {author} {\bibfnamefont {D.~L.}\ \bibnamefont
  {Feng}},\ }\href {\doibase 10.1103/PhysRevX.5.041018} {\bibfield  {journal}
  {\bibinfo  {journal} {Phys. Rev. X}\ }\textbf {\bibinfo {volume} {5}},\
  \bibinfo {pages} {041018} (\bibinfo {year} {2015})}\BibitemShut {NoStop}%
\bibitem [{\citenamefont {Battisti}\ \emph {et~al.}(2016)\citenamefont
  {Battisti}, \citenamefont {Bastiaans}, \citenamefont {Fedoseev},
  \citenamefont {de~la Torre}, \citenamefont {Iliopoulos}, \citenamefont
  {Tamai}, \citenamefont {Hunter}, \citenamefont {Perry}, \citenamefont
  {Zaanen}, \citenamefont {Baumberger},\ and\ \citenamefont
  {Allan}}]{Battisti}%
  \BibitemOpen
  \bibfield  {author} {\bibinfo {author} {\bibfnamefont {I.}~\bibnamefont
  {Battisti}}, \bibinfo {author} {\bibfnamefont {K.~M.}\ \bibnamefont
  {Bastiaans}}, \bibinfo {author} {\bibfnamefont {V.}~\bibnamefont {Fedoseev}},
  \bibinfo {author} {\bibfnamefont {A.}~\bibnamefont {de~la Torre}}, \bibinfo
  {author} {\bibfnamefont {N.}~\bibnamefont {Iliopoulos}}, \bibinfo {author}
  {\bibfnamefont {A.}~\bibnamefont {Tamai}}, \bibinfo {author} {\bibfnamefont
  {E.~C.}\ \bibnamefont {Hunter}}, \bibinfo {author} {\bibfnamefont {R.~S.}\
  \bibnamefont {Perry}}, \bibinfo {author} {\bibfnamefont {J.}~\bibnamefont
  {Zaanen}}, \bibinfo {author} {\bibfnamefont {F.}~\bibnamefont {Baumberger}},
  \ and\ \bibinfo {author} {\bibfnamefont {M.~P.}\ \bibnamefont {Allan}},\
  }\href {http://dx.doi.org/10.1038/nphys3894} {\bibfield  {journal} {\bibinfo
  {journal} {Nat. Phys.}\ }\textbf {\bibinfo {volume} {13}},\ \bibinfo {pages}
  {21} (\bibinfo {year} {2016})}\BibitemShut {NoStop}%
\bibitem [{\citenamefont {Kim}\ \emph {et~al.}(2016)\citenamefont {Kim},
  \citenamefont {Sung}, \citenamefont {Denlinger},\ and\ \citenamefont
  {Kim}}]{Y.K.Kim_Sr2IrO4_2016}%
  \BibitemOpen
  \bibfield  {author} {\bibinfo {author} {\bibfnamefont {Y.~K.}\ \bibnamefont
  {Kim}}, \bibinfo {author} {\bibfnamefont {N.~H.}\ \bibnamefont {Sung}},
  \bibinfo {author} {\bibfnamefont {J.~D.}\ \bibnamefont {Denlinger}}, \ and\
  \bibinfo {author} {\bibfnamefont {B.~J.}\ \bibnamefont {Kim}},\ }\href
  {\doibase 10.1038/nphys3503} {\bibfield  {journal} {\bibinfo  {journal} {Nat.
  Phys.}\ }\textbf {\bibinfo {volume} {12}},\ \bibinfo {pages} {37} (\bibinfo
  {year} {2016})}\BibitemShut {NoStop}%
\bibitem [{\citenamefont {Meng}\ \emph {et~al.}(2014)\citenamefont {Meng},
  \citenamefont {Kim},\ and\ \citenamefont {Kee}}]{Meng}%
  \BibitemOpen
  \bibfield  {author} {\bibinfo {author} {\bibfnamefont {Z.~Y.}\ \bibnamefont
  {Meng}}, \bibinfo {author} {\bibfnamefont {Y.~B.}\ \bibnamefont {Kim}}, \
  and\ \bibinfo {author} {\bibfnamefont {H.-Y.}\ \bibnamefont {Kee}},\ }\href
  {\doibase 10.1103/PhysRevLett.113.177003} {\bibfield  {journal} {\bibinfo
  {journal} {Phys. Rev. Lett.}\ }\textbf {\bibinfo {volume} {113}},\ \bibinfo
  {pages} {177003} (\bibinfo {year} {2014})}\BibitemShut {NoStop}%
\bibitem [{\citenamefont {Kim}\ \emph {et~al.}(2009)\citenamefont {Kim},
  \citenamefont {Ohsumi}, \citenamefont {Komesu}, \citenamefont {Sakai},
  \citenamefont {Morita}, \citenamefont {Takagi},\ and\ \citenamefont
  {Arima}}]{B.J.Kim_Sr2IrO4_2009}%
  \BibitemOpen
  \bibfield  {author} {\bibinfo {author} {\bibfnamefont {B.~J.}\ \bibnamefont
  {Kim}}, \bibinfo {author} {\bibfnamefont {H.}~\bibnamefont {Ohsumi}},
  \bibinfo {author} {\bibfnamefont {T.}~\bibnamefont {Komesu}}, \bibinfo
  {author} {\bibfnamefont {S.}~\bibnamefont {Sakai}}, \bibinfo {author}
  {\bibfnamefont {T.}~\bibnamefont {Morita}}, \bibinfo {author} {\bibfnamefont
  {H.}~\bibnamefont {Takagi}}, \ and\ \bibinfo {author} {\bibfnamefont
  {T.}~\bibnamefont {Arima}},\ }\href {\doibase 10.1126/science.1167106}
  {\bibfield  {journal} {\bibinfo  {journal} {Science}\ }\textbf {\bibinfo
  {volume} {323}},\ \bibinfo {pages} {1329} (\bibinfo {year}
  {2009})}\BibitemShut {NoStop}%
\bibitem [{\citenamefont {Boseggia}\ \emph {et~al.}(2013)\citenamefont
  {Boseggia}, \citenamefont {Walker}, \citenamefont {Vale}, \citenamefont
  {Springell}, \citenamefont {Feng}, \citenamefont {Perry}, \citenamefont
  {Sala}, \citenamefont {R{\o}nnow}, \citenamefont {Collins},\ and\
  \citenamefont {McMorrow}}]{Boseggia}%
  \BibitemOpen
  \bibfield  {author} {\bibinfo {author} {\bibfnamefont {S.}~\bibnamefont
  {Boseggia}}, \bibinfo {author} {\bibfnamefont {H.~C.}\ \bibnamefont
  {Walker}}, \bibinfo {author} {\bibfnamefont {J.}~\bibnamefont {Vale}},
  \bibinfo {author} {\bibfnamefont {R.}~\bibnamefont {Springell}}, \bibinfo
  {author} {\bibfnamefont {Z.}~\bibnamefont {Feng}}, \bibinfo {author}
  {\bibfnamefont {R.~S.}\ \bibnamefont {Perry}}, \bibinfo {author}
  {\bibfnamefont {M.~M.}\ \bibnamefont {Sala}}, \bibinfo {author}
  {\bibfnamefont {H.~M.}\ \bibnamefont {R{\o}nnow}}, \bibinfo {author}
  {\bibfnamefont {S.~P.}\ \bibnamefont {Collins}}, \ and\ \bibinfo {author}
  {\bibfnamefont {D.~F.}\ \bibnamefont {McMorrow}},\ }\href
  {http://stacks.iop.org/0953-8984/25/i=42/a=422202} {\bibfield  {journal}
  {\bibinfo  {journal} {J. Phys.: Condens. Matter}\ }\textbf {\bibinfo {volume}
  {25}},\ \bibinfo {pages} {422202} (\bibinfo {year} {2013})}\BibitemShut
  {NoStop}%
\bibitem [{\citenamefont {Clancy}\ \emph {et~al.}(2014)\citenamefont {Clancy},
  \citenamefont {Lupascu}, \citenamefont {Gretarsson}, \citenamefont {Islam},
  \citenamefont {Hu}, \citenamefont {Casa}, \citenamefont {Nelson},
  \citenamefont {LaMarra}, \citenamefont {Cao},\ and\ \citenamefont
  {Kim}}]{Clancy}%
  \BibitemOpen
  \bibfield  {author} {\bibinfo {author} {\bibfnamefont {J.~P.}\ \bibnamefont
  {Clancy}}, \bibinfo {author} {\bibfnamefont {A.}~\bibnamefont {Lupascu}},
  \bibinfo {author} {\bibfnamefont {H.}~\bibnamefont {Gretarsson}}, \bibinfo
  {author} {\bibfnamefont {Z.}~\bibnamefont {Islam}}, \bibinfo {author}
  {\bibfnamefont {Y.~F.}\ \bibnamefont {Hu}}, \bibinfo {author} {\bibfnamefont
  {D.}~\bibnamefont {Casa}}, \bibinfo {author} {\bibfnamefont {C.~S.}\
  \bibnamefont {Nelson}}, \bibinfo {author} {\bibfnamefont {S.~C.}\
  \bibnamefont {LaMarra}}, \bibinfo {author} {\bibfnamefont {G.}~\bibnamefont
  {Cao}}, \ and\ \bibinfo {author} {\bibfnamefont {Y.-J.}\ \bibnamefont
  {Kim}},\ }\href {\doibase 10.1103/PhysRevB.89.054409} {\bibfield  {journal}
  {\bibinfo  {journal} {Phys. Rev. B}\ }\textbf {\bibinfo {volume} {89}},\
  \bibinfo {pages} {054409} (\bibinfo {year} {2014})}\BibitemShut {NoStop}%
\bibitem [{\citenamefont {Dhital}\ \emph {et~al.}(2013)\citenamefont {Dhital},
  \citenamefont {Hogan}, \citenamefont {Yamani}, \citenamefont {de~la Cruz},
  \citenamefont {Chen}, \citenamefont {Khadka}, \citenamefont {Ren},\ and\
  \citenamefont {Wilson}}]{Chetan}%
  \BibitemOpen
  \bibfield  {author} {\bibinfo {author} {\bibfnamefont {C.}~\bibnamefont
  {Dhital}}, \bibinfo {author} {\bibfnamefont {T.}~\bibnamefont {Hogan}},
  \bibinfo {author} {\bibfnamefont {Z.}~\bibnamefont {Yamani}}, \bibinfo
  {author} {\bibfnamefont {C.}~\bibnamefont {de~la Cruz}}, \bibinfo {author}
  {\bibfnamefont {X.}~\bibnamefont {Chen}}, \bibinfo {author} {\bibfnamefont
  {S.}~\bibnamefont {Khadka}}, \bibinfo {author} {\bibfnamefont
  {Z.}~\bibnamefont {Ren}}, \ and\ \bibinfo {author} {\bibfnamefont {S.~D.}\
  \bibnamefont {Wilson}},\ }\href {\doibase 10.1103/PhysRevB.87.144405}
  {\bibfield  {journal} {\bibinfo  {journal} {Phys. Rev. B}\ }\textbf {\bibinfo
  {volume} {87}},\ \bibinfo {pages} {144405} (\bibinfo {year}
  {2013})}\BibitemShut {NoStop}%
\bibitem [{\citenamefont {Watanabe}\ \emph {et~al.}(2013)\citenamefont
  {Watanabe}, \citenamefont {Shirakawa},\ and\ \citenamefont
  {Yunoki}}]{Watanabe_Ir}%
  \BibitemOpen
  \bibfield  {author} {\bibinfo {author} {\bibfnamefont {H.}~\bibnamefont
  {Watanabe}}, \bibinfo {author} {\bibfnamefont {T.}~\bibnamefont {Shirakawa}},
  \ and\ \bibinfo {author} {\bibfnamefont {S.}~\bibnamefont {Yunoki}},\ }\href
  {\doibase 10.1103/PhysRevLett.110.027002} {\bibfield  {journal} {\bibinfo
  {journal} {Phys. Rev. Lett.}\ }\textbf {\bibinfo {volume} {110}},\ \bibinfo
  {pages} {027002} (\bibinfo {year} {2013})}\BibitemShut {NoStop}%
\bibitem [{\citenamefont {Huang}\ \emph {et~al.}(1994)\citenamefont {Huang},
  \citenamefont {Soubeyroux}, \citenamefont {Chmaissem}, \citenamefont {Sora},
  \citenamefont {Santoro}, \citenamefont {Cava}, \citenamefont {Krajewski},\
  and\ \citenamefont {Peck}}]{Huang_Sr2IrO4}%
  \BibitemOpen
  \bibfield  {author} {\bibinfo {author} {\bibfnamefont {Q.}~\bibnamefont
  {Huang}}, \bibinfo {author} {\bibfnamefont {J.}~\bibnamefont {Soubeyroux}},
  \bibinfo {author} {\bibfnamefont {O.}~\bibnamefont {Chmaissem}}, \bibinfo
  {author} {\bibfnamefont {I.}~\bibnamefont {Sora}}, \bibinfo {author}
  {\bibfnamefont {A.}~\bibnamefont {Santoro}}, \bibinfo {author} {\bibfnamefont
  {R.}~\bibnamefont {Cava}}, \bibinfo {author} {\bibfnamefont {J.}~\bibnamefont
  {Krajewski}}, \ and\ \bibinfo {author} {\bibfnamefont {W.}~\bibnamefont
  {Peck}},\ }\href {\doibase https://doi.org/10.1006/jssc.1994.1316} {\bibfield
   {journal} {\bibinfo  {journal} {J. Solid State Chem.}\ }\textbf {\bibinfo
  {volume} {112}},\ \bibinfo {pages} {355 } (\bibinfo {year}
  {1994})}\BibitemShut {NoStop}%
\bibitem [{\citenamefont {Crawford}\ \emph {et~al.}(1994)\citenamefont
  {Crawford}, \citenamefont {Subramanian}, \citenamefont {Harlow},
  \citenamefont {Fernandez-Baca}, \citenamefont {Wang},\ and\ \citenamefont
  {Johnston}}]{Crawford}%
  \BibitemOpen
  \bibfield  {author} {\bibinfo {author} {\bibfnamefont {M.~K.}\ \bibnamefont
  {Crawford}}, \bibinfo {author} {\bibfnamefont {M.~A.}\ \bibnamefont
  {Subramanian}}, \bibinfo {author} {\bibfnamefont {R.~L.}\ \bibnamefont
  {Harlow}}, \bibinfo {author} {\bibfnamefont {J.~A.}\ \bibnamefont
  {Fernandez-Baca}}, \bibinfo {author} {\bibfnamefont {Z.~R.}\ \bibnamefont
  {Wang}}, \ and\ \bibinfo {author} {\bibfnamefont {D.~C.}\ \bibnamefont
  {Johnston}},\ }\href {\doibase 10.1103/PhysRevB.49.9198} {\bibfield
  {journal} {\bibinfo  {journal} {Phys. Rev. B}\ }\textbf {\bibinfo {volume}
  {49}},\ \bibinfo {pages} {9198} (\bibinfo {year} {1994})}\BibitemShut
  {NoStop}%
\bibitem [{\citenamefont {Dodge}(2017)}]{Steven_Cd2Re2O7}%
  \BibitemOpen
  \bibfield  {author} {\bibinfo {author} {\bibfnamefont {J.~S.}\ \bibnamefont
  {Dodge}},\ }\href {\doibase 10.1126/science.aam8369} {\bibfield  {journal}
  {\bibinfo  {journal} {Science}\ }\textbf {\bibinfo {volume} {356}},\ \bibinfo
  {pages} {246} (\bibinfo {year} {2017})}\BibitemShut {NoStop}%
\bibitem [{\citenamefont {Harter}\ \emph {et~al.}(2017)\citenamefont {Harter},
  \citenamefont {Zhao}, \citenamefont {Yan}, \citenamefont {Mandrus},\ and\
  \citenamefont {Hsieh}}]{Harter_Cd2Re2O7}%
  \BibitemOpen
  \bibfield  {author} {\bibinfo {author} {\bibfnamefont {J.~W.}\ \bibnamefont
  {Harter}}, \bibinfo {author} {\bibfnamefont {Z.~Y.}\ \bibnamefont {Zhao}},
  \bibinfo {author} {\bibfnamefont {J.-Q.}\ \bibnamefont {Yan}}, \bibinfo
  {author} {\bibfnamefont {D.~G.}\ \bibnamefont {Mandrus}}, \ and\ \bibinfo
  {author} {\bibfnamefont {D.}~\bibnamefont {Hsieh}},\ }\href {\doibase
  10.1126/science.aad1188} {\bibfield  {journal} {\bibinfo  {journal}
  {Science}\ }\textbf {\bibinfo {volume} {356}},\ \bibinfo {pages} {295}
  (\bibinfo {year} {2017})}\BibitemShut {NoStop}%
\bibitem [{\citenamefont {Di~Matteo}\ and\ \citenamefont
  {Norman}(2017)}]{Matteo_Cd2Re2O7}%
  \BibitemOpen
  \bibfield  {author} {\bibinfo {author} {\bibfnamefont {S.}~\bibnamefont
  {Di~Matteo}}\ and\ \bibinfo {author} {\bibfnamefont {M.~R.}\ \bibnamefont
  {Norman}},\ }\href {\doibase 10.1103/PhysRevB.96.115156} {\bibfield
  {journal} {\bibinfo  {journal} {Phys. Rev. B}\ }\textbf {\bibinfo {volume}
  {96}},\ \bibinfo {pages} {115156} (\bibinfo {year} {2017})}\BibitemShut
  {NoStop}%
\bibitem [{\citenamefont {Harima}(2002)}]{Harima_Cd2Re2O7}%
  \BibitemOpen
  \bibfield  {author} {\bibinfo {author} {\bibfnamefont {H.}~\bibnamefont
  {Harima}},\ }\href {\doibase https://doi.org/10.1016/S0022-3697(02)00058-6}
  {\bibfield  {journal} {\bibinfo  {journal} {Journal of Physics and Chemistry
  of Solids}\ }\textbf {\bibinfo {volume} {63}},\ \bibinfo {pages} {1035 }
  (\bibinfo {year} {2002})},\ \bibinfo {note} {proceedings of the 8th ISSP
  International Symposium}\BibitemShut {NoStop}%
\bibitem [{\citenamefont {Yamaura}\ \emph {et~al.}(2017)\citenamefont
  {Yamaura}, \citenamefont {Takeda}, \citenamefont {Ikeda}, \citenamefont
  {Hirao}, \citenamefont {Ohishi}, \citenamefont {Kobayashi},\ and\
  \citenamefont {Hiroi}}]{Yamaura_Cd2Re2O7_2017}%
  \BibitemOpen
  \bibfield  {author} {\bibinfo {author} {\bibfnamefont {J.-i.}\ \bibnamefont
  {Yamaura}}, \bibinfo {author} {\bibfnamefont {K.}~\bibnamefont {Takeda}},
  \bibinfo {author} {\bibfnamefont {Y.}~\bibnamefont {Ikeda}}, \bibinfo
  {author} {\bibfnamefont {N.}~\bibnamefont {Hirao}}, \bibinfo {author}
  {\bibfnamefont {Y.}~\bibnamefont {Ohishi}}, \bibinfo {author} {\bibfnamefont
  {T.~C.}\ \bibnamefont {Kobayashi}}, \ and\ \bibinfo {author} {\bibfnamefont
  {Z.}~\bibnamefont {Hiroi}},\ }\href {\doibase 10.1103/PhysRevB.95.020102}
  {\bibfield  {journal} {\bibinfo  {journal} {Phys. Rev. B}\ }\textbf {\bibinfo
  {volume} {95}},\ \bibinfo {pages} {020102} (\bibinfo {year}
  {2017})}\BibitemShut {NoStop}%
\bibitem [{\citenamefont {Yamaura}\ and\ \citenamefont
  {Hiroi}(2002)}]{Yamaura_Cd2Re2O7_2002}%
  \BibitemOpen
  \bibfield  {author} {\bibinfo {author} {\bibfnamefont {J.-I.}\ \bibnamefont
  {Yamaura}}\ and\ \bibinfo {author} {\bibfnamefont {Z.}~\bibnamefont
  {Hiroi}},\ }\href {\doibase 10.1143/JPSJ.71.2598} {\bibfield  {journal}
  {\bibinfo  {journal} {J. Phys. Soc. Jpn.}\ }\textbf {\bibinfo {volume}
  {71}},\ \bibinfo {pages} {2598} (\bibinfo {year} {2002})}\BibitemShut
  {NoStop}%
\bibitem [{\citenamefont {Hiroi}\ \emph {et~al.}(2002)\citenamefont {Hiroi},
  \citenamefont {Yamaura}, \citenamefont {Muraoka},\ and\ \citenamefont
  {Hanawa}}]{Hiroi_Cd2Re2O7}%
  \BibitemOpen
  \bibfield  {author} {\bibinfo {author} {\bibfnamefont {Z.}~\bibnamefont
  {Hiroi}}, \bibinfo {author} {\bibfnamefont {J.-I.}\ \bibnamefont {Yamaura}},
  \bibinfo {author} {\bibfnamefont {Y.}~\bibnamefont {Muraoka}}, \ and\
  \bibinfo {author} {\bibfnamefont {M.}~\bibnamefont {Hanawa}},\ }\href
  {\doibase 10.1143/JPSJ.71.1634} {\bibfield  {journal} {\bibinfo  {journal}
  {J. Phys. Soc. Jpn.}\ }\textbf {\bibinfo {volume} {71}},\ \bibinfo {pages}
  {1634} (\bibinfo {year} {2002})}\BibitemShut {NoStop}%
\bibitem [{\citenamefont {Hanawa}\ \emph {et~al.}(2002)\citenamefont {Hanawa},
  \citenamefont {Yamaura}, \citenamefont {Muraoka}, \citenamefont {Sakai},\
  and\ \citenamefont {Hiroi}}]{Hanawa_Cd2Re2O7}%
  \BibitemOpen
  \bibfield  {author} {\bibinfo {author} {\bibfnamefont {M.}~\bibnamefont
  {Hanawa}}, \bibinfo {author} {\bibfnamefont {J.}~\bibnamefont {Yamaura}},
  \bibinfo {author} {\bibfnamefont {Y.}~\bibnamefont {Muraoka}}, \bibinfo
  {author} {\bibfnamefont {F.}~\bibnamefont {Sakai}}, \ and\ \bibinfo {author}
  {\bibfnamefont {Z.}~\bibnamefont {Hiroi}},\ }\href {\doibase
  https://doi.org/10.1016/S0022-3697(02)00090-2} {\bibfield  {journal}
  {\bibinfo  {journal} {Journal of Physics and Chemistry of Solids}\ }\textbf
  {\bibinfo {volume} {63}},\ \bibinfo {pages} {1027 } (\bibinfo {year}
  {2002})},\ \bibinfo {note} {proceedings of the 8th ISSP International
  Symposium}\BibitemShut {NoStop}%
\bibitem [{\citenamefont {Sakai}\ \emph {et~al.}(2001)\citenamefont {Sakai},
  \citenamefont {Yoshimura}, \citenamefont {Ohno}, \citenamefont {Kato},
  \citenamefont {Kambe}, \citenamefont {Walstedt}, \citenamefont {Matsuda},
  \citenamefont {Haga},\ and\ \citenamefont {\=Onuki}}]{Sakai_Cd2Re2O7}%
  \BibitemOpen
  \bibfield  {author} {\bibinfo {author} {\bibfnamefont {H.}~\bibnamefont
  {Sakai}}, \bibinfo {author} {\bibfnamefont {K.}~\bibnamefont {Yoshimura}},
  \bibinfo {author} {\bibfnamefont {H.}~\bibnamefont {Ohno}}, \bibinfo {author}
  {\bibfnamefont {H.}~\bibnamefont {Kato}}, \bibinfo {author} {\bibfnamefont
  {S.}~\bibnamefont {Kambe}}, \bibinfo {author} {\bibfnamefont {R.~E.}\
  \bibnamefont {Walstedt}}, \bibinfo {author} {\bibfnamefont {T.~D.}\
  \bibnamefont {Matsuda}}, \bibinfo {author} {\bibfnamefont {Y.}~\bibnamefont
  {Haga}}, \ and\ \bibinfo {author} {\bibfnamefont {Y.}~\bibnamefont
  {\=Onuki}},\ }\href {http://stacks.iop.org/0953-8984/13/i=33/a=105}
  {\bibfield  {journal} {\bibinfo  {journal} {J. Phys.: Condens. Matter}\
  }\textbf {\bibinfo {volume} {13}},\ \bibinfo {pages} {L785} (\bibinfo {year}
  {2001})}\BibitemShut {NoStop}%
\bibitem [{\citenamefont {Bednorz}\ and\ \citenamefont
  {M\"uller}(1984)}]{Bednorz_STO}%
  \BibitemOpen
  \bibfield  {author} {\bibinfo {author} {\bibfnamefont {J.~G.}\ \bibnamefont
  {Bednorz}}\ and\ \bibinfo {author} {\bibfnamefont {K.~A.}\ \bibnamefont
  {M\"uller}},\ }\href {\doibase 10.1103/PhysRevLett.52.2289} {\bibfield
  {journal} {\bibinfo  {journal} {Phys. Rev. Lett.}\ }\textbf {\bibinfo
  {volume} {52}},\ \bibinfo {pages} {2289} (\bibinfo {year}
  {1984})}\BibitemShut {NoStop}%
\bibitem [{\citenamefont {Schooley}\ \emph {et~al.}(1964)\citenamefont
  {Schooley}, \citenamefont {Hosler},\ and\ \citenamefont {Cohen}}]{Schooley}%
  \BibitemOpen
  \bibfield  {author} {\bibinfo {author} {\bibfnamefont {J.~F.}\ \bibnamefont
  {Schooley}}, \bibinfo {author} {\bibfnamefont {W.~R.}\ \bibnamefont
  {Hosler}}, \ and\ \bibinfo {author} {\bibfnamefont {M.~L.}\ \bibnamefont
  {Cohen}},\ }\href {\doibase 10.1103/PhysRevLett.12.474} {\bibfield  {journal}
  {\bibinfo  {journal} {Phys. Rev. Lett.}\ }\textbf {\bibinfo {volume} {12}},\
  \bibinfo {pages} {474} (\bibinfo {year} {1964})}\BibitemShut {NoStop}%
\bibitem [{\citenamefont {Rischau}\ \emph {et~al.}(2017)\citenamefont
  {Rischau}, \citenamefont {Lin}, \citenamefont {Grams}, \citenamefont {Finck},
  \citenamefont {Harms}, \citenamefont {Engelmayer}, \citenamefont {Lorenz},
  \citenamefont {Gallais}, \citenamefont {Fauqu{\'e}}, \citenamefont
  {Hemberger},\ and\ \citenamefont {Behnia}}]{Rischau_STO}%
  \BibitemOpen
  \bibfield  {author} {\bibinfo {author} {\bibfnamefont {C.~W.}\ \bibnamefont
  {Rischau}}, \bibinfo {author} {\bibfnamefont {X.}~\bibnamefont {Lin}},
  \bibinfo {author} {\bibfnamefont {C.~P.}\ \bibnamefont {Grams}}, \bibinfo
  {author} {\bibfnamefont {D.}~\bibnamefont {Finck}}, \bibinfo {author}
  {\bibfnamefont {S.}~\bibnamefont {Harms}}, \bibinfo {author} {\bibfnamefont
  {J.}~\bibnamefont {Engelmayer}}, \bibinfo {author} {\bibfnamefont
  {T.}~\bibnamefont {Lorenz}}, \bibinfo {author} {\bibfnamefont
  {Y.}~\bibnamefont {Gallais}}, \bibinfo {author} {\bibfnamefont
  {B.}~\bibnamefont {Fauqu{\'e}}}, \bibinfo {author} {\bibfnamefont
  {J.}~\bibnamefont {Hemberger}}, \ and\ \bibinfo {author} {\bibfnamefont
  {K.}~\bibnamefont {Behnia}},\ }\href {http://dx.doi.org/10.1038/nphys4085}
  {\bibfield  {journal} {\bibinfo  {journal} {Nat. Phys.}\ }\textbf {\bibinfo
  {volume} {13}},\ \bibinfo {pages} {643} (\bibinfo {year} {2017})}\BibitemShut
  {NoStop}%
\bibitem [{\citenamefont {Watanabe}\ and\ \citenamefont
  {Yanase}()}]{Watanabe_mpole_2}%
  \BibitemOpen
  \bibfield  {author} {\bibinfo {author} {\bibfnamefont {H.}~\bibnamefont
  {Watanabe}}\ and\ \bibinfo {author} {\bibfnamefont {Y.}~\bibnamefont
  {Yanase}},\ }\href {http://arxiv.org/abs/1805.10828} {\bibinfo  {journal}
  {ArXiv e-prints}\ ,\ \bibinfo {pages} {arXiv:1805.10828}}\BibitemShut
  {NoStop}%
\bibitem [{\citenamefont {Kozii}\ and\ \citenamefont {Fu}(2015)}]{Kozii}%
  \BibitemOpen
\bibfield  {journal} {  }\bibfield  {author} {\bibinfo {author} {\bibfnamefont
  {V.}~\bibnamefont {Kozii}}\ and\ \bibinfo {author} {\bibfnamefont
  {L.}~\bibnamefont {Fu}},\ }\href {\doibase 10.1103/PhysRevLett.115.207002}
  {\bibfield  {journal} {\bibinfo  {journal} {Phys. Rev. Lett.}\ }\textbf
  {\bibinfo {volume} {115}},\ \bibinfo {pages} {207002} (\bibinfo {year}
  {2015})}\BibitemShut {NoStop}%
\bibitem [{\citenamefont {Wang}\ \emph {et~al.}(2016)\citenamefont {Wang},
  \citenamefont {Cho}, \citenamefont {Hughes},\ and\ \citenamefont
  {Fradkin}}]{Wang_Cd2Re2O7}%
  \BibitemOpen
  \bibfield  {author} {\bibinfo {author} {\bibfnamefont {Y.}~\bibnamefont
  {Wang}}, \bibinfo {author} {\bibfnamefont {G.~Y.}\ \bibnamefont {Cho}},
  \bibinfo {author} {\bibfnamefont {T.~L.}\ \bibnamefont {Hughes}}, \ and\
  \bibinfo {author} {\bibfnamefont {E.}~\bibnamefont {Fradkin}},\ }\href
  {\doibase 10.1103/PhysRevB.93.134512} {\bibfield  {journal} {\bibinfo
  {journal} {Phys. Rev. B}\ }\textbf {\bibinfo {volume} {93}},\ \bibinfo
  {pages} {134512} (\bibinfo {year} {2016})}\BibitemShut {NoStop}%
\bibitem [{\citenamefont {Edge}\ \emph {et~al.}(2015)\citenamefont {Edge},
  \citenamefont {Kedem}, \citenamefont {Aschauer}, \citenamefont {Spaldin},\
  and\ \citenamefont {Balatsky}}]{Edge}%
  \BibitemOpen
  \bibfield  {author} {\bibinfo {author} {\bibfnamefont {J.~M.}\ \bibnamefont
  {Edge}}, \bibinfo {author} {\bibfnamefont {Y.}~\bibnamefont {Kedem}},
  \bibinfo {author} {\bibfnamefont {U.}~\bibnamefont {Aschauer}}, \bibinfo
  {author} {\bibfnamefont {N.~A.}\ \bibnamefont {Spaldin}}, \ and\ \bibinfo
  {author} {\bibfnamefont {A.~V.}\ \bibnamefont {Balatsky}},\ }\href {\doibase
  10.1103/PhysRevLett.115.247002} {\bibfield  {journal} {\bibinfo  {journal}
  {Phys. Rev. Lett.}\ }\textbf {\bibinfo {volume} {115}},\ \bibinfo {pages}
  {247002} (\bibinfo {year} {2015})}\BibitemShut {NoStop}%
\bibitem [{\citenamefont {Takimoto}\ and\ \citenamefont
  {Thalmeier}(2009)}]{Takimoto_CePt3Si_2}%
  \BibitemOpen
  \bibfield  {author} {\bibinfo {author} {\bibfnamefont {T.}~\bibnamefont
  {Takimoto}}\ and\ \bibinfo {author} {\bibfnamefont {P.}~\bibnamefont
  {Thalmeier}},\ }\href {\doibase 10.1143/JPSJ.78.103703} {\bibfield  {journal}
  {\bibinfo  {journal} {J. Phys. Soc. Jpn.}\ }\textbf {\bibinfo {volume}
  {78}},\ \bibinfo {pages} {103703} (\bibinfo {year} {2009})}\BibitemShut
  {NoStop}%
\bibitem [{\citenamefont {Sigrist}\ and\ \citenamefont
  {Ueda}(1991)}]{Sigrist_Ueda}%
  \BibitemOpen
  \bibfield  {author} {\bibinfo {author} {\bibfnamefont {M.}~\bibnamefont
  {Sigrist}}\ and\ \bibinfo {author} {\bibfnamefont {K.}~\bibnamefont {Ueda}},\
  }\href {\doibase 10.1103/RevModPhys.63.239} {\bibfield  {journal} {\bibinfo
  {journal} {Rev. Mod. Phys.}\ }\textbf {\bibinfo {volume} {63}},\ \bibinfo
  {pages} {239} (\bibinfo {year} {1991})}\BibitemShut {NoStop}%
\bibitem [{\citenamefont {Nomoto}\ \emph {et~al.}(2016)\citenamefont {Nomoto},
  \citenamefont {Hattori},\ and\ \citenamefont {Ikeda}}]{Nomoto_MSC}%
  \BibitemOpen
  \bibfield  {author} {\bibinfo {author} {\bibfnamefont {T.}~\bibnamefont
  {Nomoto}}, \bibinfo {author} {\bibfnamefont {K.}~\bibnamefont {Hattori}}, \
  and\ \bibinfo {author} {\bibfnamefont {H.}~\bibnamefont {Ikeda}},\ }\href
  {\doibase 10.1103/PhysRevB.94.174513} {\bibfield  {journal} {\bibinfo
  {journal} {Phys. Rev. B}\ }\textbf {\bibinfo {volume} {94}},\ \bibinfo
  {pages} {174513} (\bibinfo {year} {2016})}\BibitemShut {NoStop}%
\bibitem [{\citenamefont {Heid}(1995)}]{Heid}%
  \BibitemOpen
  \bibfield  {author} {\bibinfo {author} {\bibfnamefont {R.}~\bibnamefont
  {Heid}},\ }\href {\doibase 10.1007/s002570050003} {\bibfield  {journal}
  {\bibinfo  {journal} {Z. Phys. B: Condens. Matter}\ }\textbf {\bibinfo
  {volume} {99}},\ \bibinfo {pages} {15} (\bibinfo {year} {1995})}\BibitemShut
  {NoStop}%
\bibitem [{\citenamefont {Yanase}(2016)}]{Yanase_UPt3_Weyl}%
  \BibitemOpen
  \bibfield  {author} {\bibinfo {author} {\bibfnamefont {Y.}~\bibnamefont
  {Yanase}},\ }\href {\doibase 10.1103/PhysRevB.94.174502} {\bibfield
  {journal} {\bibinfo  {journal} {Phys. Rev. B}\ }\textbf {\bibinfo {volume}
  {94}},\ \bibinfo {pages} {174502} (\bibinfo {year} {2016})}\BibitemShut
  {NoStop}%
\bibitem [{\citenamefont {Nomoto}\ and\ \citenamefont
  {Ikeda}(2016)}]{Nomoto_UPt3}%
  \BibitemOpen
  \bibfield  {author} {\bibinfo {author} {\bibfnamefont {T.}~\bibnamefont
  {Nomoto}}\ and\ \bibinfo {author} {\bibfnamefont {H.}~\bibnamefont {Ikeda}},\
  }\href {\doibase 10.1103/PhysRevLett.117.217002} {\bibfield  {journal}
  {\bibinfo  {journal} {Phys. Rev. Lett.}\ }\textbf {\bibinfo {volume} {117}},\
  \bibinfo {pages} {217002} (\bibinfo {year} {2016})}\BibitemShut {NoStop}%
\bibitem [{\citenamefont {Nomoto}\ and\ \citenamefont
  {Ikeda}(2017)}]{Nomoto_UCoGe}%
  \BibitemOpen
  \bibfield  {author} {\bibinfo {author} {\bibfnamefont {T.}~\bibnamefont
  {Nomoto}}\ and\ \bibinfo {author} {\bibfnamefont {H.}~\bibnamefont {Ikeda}},\
  }\href {\doibase 10.7566/JPSJ.86.023703} {\bibfield  {journal} {\bibinfo
  {journal} {J. Phys. Soc. Jpn.}\ }\textbf {\bibinfo {volume} {86}},\ \bibinfo
  {pages} {023703} (\bibinfo {year} {2017})}\BibitemShut {NoStop}%
\bibitem [{\citenamefont {Norman}(1995)}]{Norman}%
  \BibitemOpen
  \bibfield  {author} {\bibinfo {author} {\bibfnamefont {M.~R.}\ \bibnamefont
  {Norman}},\ }\href {\doibase 10.1103/PhysRevB.52.15093} {\bibfield  {journal}
  {\bibinfo  {journal} {Phys. Rev. B}\ }\textbf {\bibinfo {volume} {52}},\
  \bibinfo {pages} {15093} (\bibinfo {year} {1995})}\BibitemShut {NoStop}%
\bibitem [{\citenamefont {Micklitz}\ and\ \citenamefont
  {Norman}(2009)}]{Micklitz}%
  \BibitemOpen
  \bibfield  {author} {\bibinfo {author} {\bibfnamefont {T.}~\bibnamefont
  {Micklitz}}\ and\ \bibinfo {author} {\bibfnamefont {M.~R.}\ \bibnamefont
  {Norman}},\ }\href {\doibase 10.1103/PhysRevB.80.100506} {\bibfield
  {journal} {\bibinfo  {journal} {Phys. Rev. B}\ }\textbf {\bibinfo {volume}
  {80}},\ \bibinfo {pages} {100506} (\bibinfo {year} {2009})}\BibitemShut
  {NoStop}%
\bibitem [{\citenamefont {Kobayashi}\ \emph {et~al.}(2016)\citenamefont
  {Kobayashi}, \citenamefont {Yanase},\ and\ \citenamefont
  {Sato}}]{Kobayashi_TSC_2016}%
  \BibitemOpen
  \bibfield  {author} {\bibinfo {author} {\bibfnamefont {S.}~\bibnamefont
  {Kobayashi}}, \bibinfo {author} {\bibfnamefont {Y.}~\bibnamefont {Yanase}}, \
  and\ \bibinfo {author} {\bibfnamefont {M.}~\bibnamefont {Sato}},\ }\href
  {\doibase 10.1103/PhysRevB.94.134512} {\bibfield  {journal} {\bibinfo
  {journal} {Phys. Rev. B}\ }\textbf {\bibinfo {volume} {94}},\ \bibinfo
  {pages} {134512} (\bibinfo {year} {2016})}\BibitemShut {NoStop}%
\bibitem [{\citenamefont {Sumita}\ and\ \citenamefont
  {Yanase}(2018)}]{Sumita_UPt3}%
  \BibitemOpen
  \bibfield  {author} {\bibinfo {author} {\bibfnamefont {S.}~\bibnamefont
  {Sumita}}\ and\ \bibinfo {author} {\bibfnamefont {Y.}~\bibnamefont
  {Yanase}},\ }\href {\doibase 10.1103/PhysRevB.97.134512} {\bibfield
  {journal} {\bibinfo  {journal} {Phys. Rev. B}\ }\textbf {\bibinfo {volume}
  {97}},\ \bibinfo {pages} {134512} (\bibinfo {year} {2018})}\BibitemShut
  {NoStop}%
\bibitem [{\citenamefont {Kobayashi}\ \emph {et~al.}(2018)\citenamefont
  {Kobayashi}, \citenamefont {Sumita}, \citenamefont {Yanase},\ and\
  \citenamefont {Sato}}]{Kobayashi_TSC_2018}%
  \BibitemOpen
  \bibfield  {author} {\bibinfo {author} {\bibfnamefont {S.}~\bibnamefont
  {Kobayashi}}, \bibinfo {author} {\bibfnamefont {S.}~\bibnamefont {Sumita}},
  \bibinfo {author} {\bibfnamefont {Y.}~\bibnamefont {Yanase}}, \ and\ \bibinfo
  {author} {\bibfnamefont {M.}~\bibnamefont {Sato}},\ }\href {\doibase
  10.1103/PhysRevB.97.180504} {\bibfield  {journal} {\bibinfo  {journal} {Phys.
  Rev. B}\ }\textbf {\bibinfo {volume} {97}},\ \bibinfo {pages} {180504}
  (\bibinfo {year} {2018})}\BibitemShut {NoStop}%
\bibitem [{\citenamefont {Yanase}\ \emph {et~al.}(2003)\citenamefont {Yanase},
  \citenamefont {Jujo}, \citenamefont {Nomura}, \citenamefont {Ikeda},
  \citenamefont {Hotta},\ and\ \citenamefont {Yamada}}]{Yanase_review_sc}%
  \BibitemOpen
  \bibfield  {author} {\bibinfo {author} {\bibfnamefont {Y.}~\bibnamefont
  {Yanase}}, \bibinfo {author} {\bibfnamefont {T.}~\bibnamefont {Jujo}},
  \bibinfo {author} {\bibfnamefont {T.}~\bibnamefont {Nomura}}, \bibinfo
  {author} {\bibfnamefont {H.}~\bibnamefont {Ikeda}}, \bibinfo {author}
  {\bibfnamefont {T.}~\bibnamefont {Hotta}}, \ and\ \bibinfo {author}
  {\bibfnamefont {K.}~\bibnamefont {Yamada}},\ }\href {\doibase
  https://doi.org/10.1016/j.physrep.2003.07.002} {\bibfield  {journal}
  {\bibinfo  {journal} {Phys. Rep.}\ }\textbf {\bibinfo {volume} {387}},\
  \bibinfo {pages} {1 } (\bibinfo {year} {2003})}\BibitemShut {NoStop}%
\bibitem [{\citenamefont {Yamanaka}\ \emph {et~al.}(2015)\citenamefont
  {Yamanaka}, \citenamefont {Shimozawa}, \citenamefont {Endo}, \citenamefont
  {Mizukami}, \citenamefont {Shishido}, \citenamefont {Terashima},
  \citenamefont {Shibauchi}, \citenamefont {Matsuda},\ and\ \citenamefont
  {Ishida}}]{Yamanaka_NMR}%
  \BibitemOpen
  \bibfield  {author} {\bibinfo {author} {\bibfnamefont {T.}~\bibnamefont
  {Yamanaka}}, \bibinfo {author} {\bibfnamefont {M.}~\bibnamefont {Shimozawa}},
  \bibinfo {author} {\bibfnamefont {R.}~\bibnamefont {Endo}}, \bibinfo {author}
  {\bibfnamefont {Y.}~\bibnamefont {Mizukami}}, \bibinfo {author}
  {\bibfnamefont {H.}~\bibnamefont {Shishido}}, \bibinfo {author}
  {\bibfnamefont {T.}~\bibnamefont {Terashima}}, \bibinfo {author}
  {\bibfnamefont {T.}~\bibnamefont {Shibauchi}}, \bibinfo {author}
  {\bibfnamefont {Y.}~\bibnamefont {Matsuda}}, \ and\ \bibinfo {author}
  {\bibfnamefont {K.}~\bibnamefont {Ishida}},\ }\href {\doibase
  10.1103/PhysRevB.92.241105} {\bibfield  {journal} {\bibinfo  {journal} {Phys.
  Rev. B}\ }\textbf {\bibinfo {volume} {92}},\ \bibinfo {pages} {241105}
  (\bibinfo {year} {2015})}\BibitemShut {NoStop}%
\bibitem [{not()}]{note_chi}%
  \BibitemOpen
  \href@noop {} {}\bibinfo {note} {Because nonsymmorphic space groups contain
  nonprimitive translation operations, there are several choices of an
  inversion center, giving an ambiguity of an atomic cluster. In the present
  case, we have two choices for the unit cell. One is shown in Fig. 1, and the
  other is obtained by the glide operation. $\chi_{B_{2u}(E_u)}(\bm q)$ at $\bm
  q \neq (0,0)$ depends on the choice of the unit cell. To avoid this
  ambiguity, we define bulk susceptibility by average over the choices of the
  unit cell, $\chi_{B_{2u}(E_u)}(\bm q) = (\chi^{(1)}_{B_{2u}(E_u)}(\bm q) +
  \chi^{(2)}_{B_{2u}(E_u)}(\bm q))/2$.}\BibitemShut {Stop}%
\bibitem [{\citenamefont {Sato}(2010)}]{M.Sato2010_TSC}%
  \BibitemOpen
  \bibfield  {author} {\bibinfo {author} {\bibfnamefont {M.}~\bibnamefont
  {Sato}},\ }\href {\doibase 10.1103/PhysRevB.81.220504} {\bibfield  {journal}
  {\bibinfo  {journal} {Phys. Rev. B}\ }\textbf {\bibinfo {volume} {81}},\
  \bibinfo {pages} {220504} (\bibinfo {year} {2010})}\BibitemShut {NoStop}%
\bibitem [{\citenamefont {Mackenzie}\ and\ \citenamefont
  {Maeno}(2003)}]{Mackenzie_Sr2RuO4}%
  \BibitemOpen
  \bibfield  {author} {\bibinfo {author} {\bibfnamefont {A.~P.}\ \bibnamefont
  {Mackenzie}}\ and\ \bibinfo {author} {\bibfnamefont {Y.}~\bibnamefont
  {Maeno}},\ }\href {\doibase 10.1103/RevModPhys.75.657} {\bibfield  {journal}
  {\bibinfo  {journal} {Rev. Mod. Phys.}\ }\textbf {\bibinfo {volume} {75}},\
  \bibinfo {pages} {657} (\bibinfo {year} {2003})}\BibitemShut {NoStop}%
\bibitem [{\citenamefont {Joynt}\ and\ \citenamefont
  {Taillefer}(2002)}]{Joynt_UPt3}%
  \BibitemOpen
  \bibfield  {author} {\bibinfo {author} {\bibfnamefont {R.}~\bibnamefont
  {Joynt}}\ and\ \bibinfo {author} {\bibfnamefont {L.}~\bibnamefont
  {Taillefer}},\ }\href {\doibase 10.1103/RevModPhys.74.235} {\bibfield
  {journal} {\bibinfo  {journal} {Rev. Mod. Phys.}\ }\textbf {\bibinfo {volume}
  {74}},\ \bibinfo {pages} {235} (\bibinfo {year} {2002})}\BibitemShut
  {NoStop}%
\bibitem [{\citenamefont {Huy}\ \emph {et~al.}(2007)\citenamefont {Huy},
  \citenamefont {Gasparini}, \citenamefont {de~Nijs}, \citenamefont {Huang},
  \citenamefont {Klaasse}, \citenamefont {Gortenmulder}, \citenamefont
  {de~Visser}, \citenamefont {Hamann}, \citenamefont {G\"orlach},\ and\
  \citenamefont {L\"ohneysen}}]{Huy_UCoGe}%
  \BibitemOpen
  \bibfield  {author} {\bibinfo {author} {\bibfnamefont {N.~T.}\ \bibnamefont
  {Huy}}, \bibinfo {author} {\bibfnamefont {A.}~\bibnamefont {Gasparini}},
  \bibinfo {author} {\bibfnamefont {D.~E.}\ \bibnamefont {de~Nijs}}, \bibinfo
  {author} {\bibfnamefont {Y.}~\bibnamefont {Huang}}, \bibinfo {author}
  {\bibfnamefont {J.~C.~P.}\ \bibnamefont {Klaasse}}, \bibinfo {author}
  {\bibfnamefont {T.}~\bibnamefont {Gortenmulder}}, \bibinfo {author}
  {\bibfnamefont {A.}~\bibnamefont {de~Visser}}, \bibinfo {author}
  {\bibfnamefont {A.}~\bibnamefont {Hamann}}, \bibinfo {author} {\bibfnamefont
  {T.}~\bibnamefont {G\"orlach}}, \ and\ \bibinfo {author} {\bibfnamefont
  {H.~v.}\ \bibnamefont {L\"ohneysen}},\ }\href {\doibase
  10.1103/PhysRevLett.99.067006} {\bibfield  {journal} {\bibinfo  {journal}
  {Phys. Rev. Lett.}\ }\textbf {\bibinfo {volume} {99}},\ \bibinfo {pages}
  {067006} (\bibinfo {year} {2007})}\BibitemShut {NoStop}%
\bibitem [{\citenamefont {Watanabe}\ \emph {et~al.}(1969)\citenamefont
  {Watanabe}, \citenamefont {Kazama}, \citenamefont {Yamaguchi},\ and\
  \citenamefont {Ohashi}}]{Watanabe_CrAs}%
  \BibitemOpen
  \bibfield  {author} {\bibinfo {author} {\bibfnamefont {H.}~\bibnamefont
  {Watanabe}}, \bibinfo {author} {\bibfnamefont {N.}~\bibnamefont {Kazama}},
  \bibinfo {author} {\bibfnamefont {Y.}~\bibnamefont {Yamaguchi}}, \ and\
  \bibinfo {author} {\bibfnamefont {M.}~\bibnamefont {Ohashi}},\ }\href
  {\doibase 10.1063/1.1657559} {\bibfield  {journal} {\bibinfo  {journal} {J.
  Appl. Phys.}\ }\textbf {\bibinfo {volume} {40}},\ \bibinfo {pages} {1128}
  (\bibinfo {year} {1969})}\BibitemShut {NoStop}%
\bibitem [{\citenamefont {Selte}\ \emph {et~al.}(1971)\citenamefont {Selte},
  \citenamefont {Kjekshus}, \citenamefont {Jamison}, \citenamefont {Andresen},\
  and\ \citenamefont {Engebretsen}}]{Selte_CrAs}%
  \BibitemOpen
  \bibfield  {author} {\bibinfo {author} {\bibfnamefont {K.}~\bibnamefont
  {Selte}}, \bibinfo {author} {\bibfnamefont {A.}~\bibnamefont {Kjekshus}},
  \bibinfo {author} {\bibfnamefont {W.~E.}\ \bibnamefont {Jamison}}, \bibinfo
  {author} {\bibfnamefont {A.}~\bibnamefont {Andresen}}, \ and\ \bibinfo
  {author} {\bibfnamefont {J.~E.}\ \bibnamefont {Engebretsen}},\ }\href
  {\doibase 10.3891/acta.chem.scand.25-1703} {\bibfield  {journal} {\bibinfo
  {journal} {Acta Chem. Scand.}\ }\textbf {\bibinfo {volume} {25}},\ \bibinfo
  {pages} {1703} (\bibinfo {year} {1971})}\BibitemShut {NoStop}%
\bibitem [{\citenamefont {Kotegawa}\ \emph {et~al.}(2014)\citenamefont
  {Kotegawa}, \citenamefont {Nakahara}, \citenamefont {Tou},\ and\
  \citenamefont {Sugawara}}]{Kotegawa_CrAs_2014}%
  \BibitemOpen
  \bibfield  {author} {\bibinfo {author} {\bibfnamefont {H.}~\bibnamefont
  {Kotegawa}}, \bibinfo {author} {\bibfnamefont {S.}~\bibnamefont {Nakahara}},
  \bibinfo {author} {\bibfnamefont {H.}~\bibnamefont {Tou}}, \ and\ \bibinfo
  {author} {\bibfnamefont {H.}~\bibnamefont {Sugawara}},\ }\href {\doibase
  10.7566/JPSJ.83.093702} {\bibfield  {journal} {\bibinfo  {journal} {J. Phys.
  Soc. Jpn.}\ }\textbf {\bibinfo {volume} {83}},\ \bibinfo {pages} {093702}
  (\bibinfo {year} {2014})}\BibitemShut {NoStop}%
\bibitem [{\citenamefont {Wu}\ \emph {et~al.}(2014)\citenamefont {Wu},
  \citenamefont {Cheng}, \citenamefont {Matsubayashi}, \citenamefont {Kong},
  \citenamefont {Lin}, \citenamefont {Jin}, \citenamefont {Wang}, \citenamefont
  {Uwatoko},\ and\ \citenamefont {Luo}}]{Wu_CrAs}%
  \BibitemOpen
  \bibfield  {author} {\bibinfo {author} {\bibfnamefont {W.}~\bibnamefont
  {Wu}}, \bibinfo {author} {\bibfnamefont {J.}~\bibnamefont {Cheng}}, \bibinfo
  {author} {\bibfnamefont {K.}~\bibnamefont {Matsubayashi}}, \bibinfo {author}
  {\bibfnamefont {P.}~\bibnamefont {Kong}}, \bibinfo {author} {\bibfnamefont
  {F.}~\bibnamefont {Lin}}, \bibinfo {author} {\bibfnamefont {C.}~\bibnamefont
  {Jin}}, \bibinfo {author} {\bibfnamefont {N.}~\bibnamefont {Wang}}, \bibinfo
  {author} {\bibfnamefont {Y.}~\bibnamefont {Uwatoko}}, \ and\ \bibinfo
  {author} {\bibfnamefont {J.}~\bibnamefont {Luo}},\ }\href
  {http://dx.doi.org/10.1038/ncomms6508} {\bibfield  {journal} {\bibinfo
  {journal} {Nat. Commun.}\ }\textbf {\bibinfo {volume} {5}},\ \bibinfo {pages}
  {5508} (\bibinfo {year} {2014})}\BibitemShut {NoStop}%
\bibitem [{\citenamefont {Kotegawa}\ \emph {et~al.}(2015)\citenamefont
  {Kotegawa}, \citenamefont {Nakahara}, \citenamefont {Akamatsu}, \citenamefont
  {Tou}, \citenamefont {Sugawara},\ and\ \citenamefont
  {Harima}}]{Kotegawa_CrAs_2015}%
  \BibitemOpen
  \bibfield  {author} {\bibinfo {author} {\bibfnamefont {H.}~\bibnamefont
  {Kotegawa}}, \bibinfo {author} {\bibfnamefont {S.}~\bibnamefont {Nakahara}},
  \bibinfo {author} {\bibfnamefont {R.}~\bibnamefont {Akamatsu}}, \bibinfo
  {author} {\bibfnamefont {H.}~\bibnamefont {Tou}}, \bibinfo {author}
  {\bibfnamefont {H.}~\bibnamefont {Sugawara}}, \ and\ \bibinfo {author}
  {\bibfnamefont {H.}~\bibnamefont {Harima}},\ }\href {\doibase
  10.1103/PhysRevLett.114.117002} {\bibfield  {journal} {\bibinfo  {journal}
  {Phys. Rev. Lett.}\ }\textbf {\bibinfo {volume} {114}},\ \bibinfo {pages}
  {117002} (\bibinfo {year} {2015})}\BibitemShut {NoStop}%
\bibitem [{\citenamefont {Keller}\ \emph {et~al.}(2015)\citenamefont {Keller},
  \citenamefont {White}, \citenamefont {Frontzek}, \citenamefont {Babkevich},
  \citenamefont {Susner}, \citenamefont {Sims}, \citenamefont {Sefat},
  \citenamefont {R\o{}nnow},\ and\ \citenamefont {R\"uegg}}]{Keller}%
  \BibitemOpen
  \bibfield  {author} {\bibinfo {author} {\bibfnamefont {L.}~\bibnamefont
  {Keller}}, \bibinfo {author} {\bibfnamefont {J.~S.}\ \bibnamefont {White}},
  \bibinfo {author} {\bibfnamefont {M.}~\bibnamefont {Frontzek}}, \bibinfo
  {author} {\bibfnamefont {P.}~\bibnamefont {Babkevich}}, \bibinfo {author}
  {\bibfnamefont {M.~A.}\ \bibnamefont {Susner}}, \bibinfo {author}
  {\bibfnamefont {Z.~C.}\ \bibnamefont {Sims}}, \bibinfo {author}
  {\bibfnamefont {A.~S.}\ \bibnamefont {Sefat}}, \bibinfo {author}
  {\bibfnamefont {H.~M.}\ \bibnamefont {R\o{}nnow}}, \ and\ \bibinfo {author}
  {\bibfnamefont {C.}~\bibnamefont {R\"uegg}},\ }\href {\doibase
  10.1103/PhysRevB.91.020409} {\bibfield  {journal} {\bibinfo  {journal} {Phys.
  Rev. B}\ }\textbf {\bibinfo {volume} {91}},\ \bibinfo {pages} {020409}
  (\bibinfo {year} {2015})}\BibitemShut {NoStop}%
\bibitem [{\citenamefont {Shen}\ \emph {et~al.}(2016)\citenamefont {Shen},
  \citenamefont {Wang}, \citenamefont {Hao}, \citenamefont {Pan}, \citenamefont
  {Feng}, \citenamefont {Huang}, \citenamefont {Harriger}, \citenamefont
  {Leao}, \citenamefont {Zhao}, \citenamefont {Chisnell}, \citenamefont {Lynn},
  \citenamefont {Cao}, \citenamefont {Hu},\ and\ \citenamefont {Zhao}}]{Shen}%
  \BibitemOpen
  \bibfield  {author} {\bibinfo {author} {\bibfnamefont {Y.}~\bibnamefont
  {Shen}}, \bibinfo {author} {\bibfnamefont {Q.}~\bibnamefont {Wang}}, \bibinfo
  {author} {\bibfnamefont {Y.}~\bibnamefont {Hao}}, \bibinfo {author}
  {\bibfnamefont {B.}~\bibnamefont {Pan}}, \bibinfo {author} {\bibfnamefont
  {Y.}~\bibnamefont {Feng}}, \bibinfo {author} {\bibfnamefont {Q.}~\bibnamefont
  {Huang}}, \bibinfo {author} {\bibfnamefont {L.~W.}\ \bibnamefont {Harriger}},
  \bibinfo {author} {\bibfnamefont {J.~B.}\ \bibnamefont {Leao}}, \bibinfo
  {author} {\bibfnamefont {Y.}~\bibnamefont {Zhao}}, \bibinfo {author}
  {\bibfnamefont {R.~M.}\ \bibnamefont {Chisnell}}, \bibinfo {author}
  {\bibfnamefont {J.~W.}\ \bibnamefont {Lynn}}, \bibinfo {author}
  {\bibfnamefont {H.}~\bibnamefont {Cao}}, \bibinfo {author} {\bibfnamefont
  {J.}~\bibnamefont {Hu}}, \ and\ \bibinfo {author} {\bibfnamefont
  {J.}~\bibnamefont {Zhao}},\ }\href {\doibase 10.1103/PhysRevB.93.060503}
  {\bibfield  {journal} {\bibinfo  {journal} {Phys. Rev. B}\ }\textbf {\bibinfo
  {volume} {93}},\ \bibinfo {pages} {060503} (\bibinfo {year}
  {2016})}\BibitemShut {NoStop}%
\bibitem [{\citenamefont {Niu}\ \emph {et~al.}(2017)\citenamefont {Niu},
  \citenamefont {Yu}, \citenamefont {Yip}, \citenamefont {Lim}, \citenamefont
  {Kotegawa}, \citenamefont {Matsuoka}, \citenamefont {Sugawara}, \citenamefont
  {Tou}, \citenamefont {Yanase},\ and\ \citenamefont {Goh}}]{Niu}%
  \BibitemOpen
  \bibfield  {author} {\bibinfo {author} {\bibfnamefont {Q.}~\bibnamefont
  {Niu}}, \bibinfo {author} {\bibfnamefont {W.~C.}\ \bibnamefont {Yu}},
  \bibinfo {author} {\bibfnamefont {K.~Y.}\ \bibnamefont {Yip}}, \bibinfo
  {author} {\bibfnamefont {Z.~L.}\ \bibnamefont {Lim}}, \bibinfo {author}
  {\bibfnamefont {H.}~\bibnamefont {Kotegawa}}, \bibinfo {author}
  {\bibfnamefont {E.}~\bibnamefont {Matsuoka}}, \bibinfo {author}
  {\bibfnamefont {H.}~\bibnamefont {Sugawara}}, \bibinfo {author}
  {\bibfnamefont {H.}~\bibnamefont {Tou}}, \bibinfo {author} {\bibfnamefont
  {Y.}~\bibnamefont {Yanase}}, \ and\ \bibinfo {author} {\bibfnamefont {S.~K.}\
  \bibnamefont {Goh}},\ }\href {http://dx.doi.org/10.1038/ncomms15358}
  {\bibfield  {journal} {\bibinfo  {journal} {Nat. Commun.}\ }\textbf {\bibinfo
  {volume} {8}},\ \bibinfo {pages} {15358} (\bibinfo {year}
  {2017})}\BibitemShut {NoStop}%
\bibitem [{\citenamefont {Guo}\ \emph {et~al.}(2018)\citenamefont {Guo},
  \citenamefont {Smidman}, \citenamefont {Shen}, \citenamefont {Wu},
  \citenamefont {Lin}, \citenamefont {Han}, \citenamefont {Chen}, \citenamefont
  {Wu}, \citenamefont {Wang}, \citenamefont {Jiang}, \citenamefont {Lu},
  \citenamefont {Hu}, \citenamefont {Luo},\ and\ \citenamefont
  {Yuan}}]{Guo_CrAs}%
  \BibitemOpen
  \bibfield  {author} {\bibinfo {author} {\bibfnamefont {C.~Y.}\ \bibnamefont
  {Guo}}, \bibinfo {author} {\bibfnamefont {M.}~\bibnamefont {Smidman}},
  \bibinfo {author} {\bibfnamefont {B.}~\bibnamefont {Shen}}, \bibinfo {author}
  {\bibfnamefont {W.}~\bibnamefont {Wu}}, \bibinfo {author} {\bibfnamefont
  {F.~K.}\ \bibnamefont {Lin}}, \bibinfo {author} {\bibfnamefont {X.~L.}\
  \bibnamefont {Han}}, \bibinfo {author} {\bibfnamefont {Y.}~\bibnamefont
  {Chen}}, \bibinfo {author} {\bibfnamefont {F.}~\bibnamefont {Wu}}, \bibinfo
  {author} {\bibfnamefont {Y.~F.}\ \bibnamefont {Wang}}, \bibinfo {author}
  {\bibfnamefont {W.~B.}\ \bibnamefont {Jiang}}, \bibinfo {author}
  {\bibfnamefont {X.}~\bibnamefont {Lu}}, \bibinfo {author} {\bibfnamefont
  {J.~P.}\ \bibnamefont {Hu}}, \bibinfo {author} {\bibfnamefont {J.~L.}\
  \bibnamefont {Luo}}, \ and\ \bibinfo {author} {\bibfnamefont {H.~Q.}\
  \bibnamefont {Yuan}},\ }\href {\doibase 10.1103/PhysRevB.98.024520}
  {\bibfield  {journal} {\bibinfo  {journal} {Phys. Rev. B}\ }\textbf {\bibinfo
  {volume} {98}},\ \bibinfo {pages} {024520} (\bibinfo {year}
  {2018})}\BibitemShut {NoStop}%
\end{thebibliography}
\end{document}